\documentclass[prd,onecolumn,notitlepage,eqsecnum,nofootinbib,floatfix,superscriptaddress]{revtex4-1}
\usepackage{amsmath, amsthm, amssymb, amsfonts, amsbsy,mathrsfs}

\usepackage{graphicx, epstopdf}
\DeclareGraphicsExtensions{.eps}
\graphicspath{{./images/}}

\usepackage{calligra,bm}
\usepackage{stmaryrd}

\usepackage[hidelinks,bookmarks=true]{hyperref} 
\hypersetup{pdfstartview=FitH,pdfhighlight=/O,colorlinks=false}

\begin{document}

\preprint{UTTG-04-17}

\title{The Weiss variation of the gravitational action}

\author{Justin C. Feng}
\author{Richard A. Matzner}
\affiliation{Theory Group, Department of Physics, University of Texas at Austin}

\date{\today}

\begin{abstract}
The Weiss variational principle in mechanics and classical field theory is a variational principle which allows displacements of the boundary. We review the Weiss variation in mechanics and classical field theory, and present a novel geometric derivation of the Weiss variation for the gravitational action: the Einstein-Hilbert action plus the Gibbons-Hawking-York boundary term. In particular, we use the first and second variation of area formulas (we present a derivation accessible to physicists in an appendix) to interpret and vary the Gibbons-Hawking-York boundary term. The Weiss variation for the gravitational action is in principle known to the Relativity community, but the variation of area approach formalizes the derivation, and facilitates the discussion of time evolution in General Relativity. A potentially useful feature of the formalism presented in this article is that it avoids an explicit 3+1 decomposition in the bulk spacetime.
\end{abstract}


\maketitle
\tableofcontents

\section{Introduction}

The Weiss variational principle is a variational principle which includes variations of the boundaries for the action integral \cite{SudarshanCM,MatznerShepleyCM}. In particular, the Weiss variation includes infinitesimal displacements of the endpoints for a mechanical action, or displacement of the spacetime boundaries for the action of a classical field theory. In general relativity (GR), the approach we present is based on and makes explicit that displacements of the boundary produce the contribution of the Gibbons-Hawking-York surface integral to the field equations. In mechanics, the Weiss variation may be used to identify a Hamiltonian without performing a Legendre transformation, and in classical field theory, one may use the Weiss variation to identify the Hamiltonian without performing a 3+1 split in the bulk. In particular, one may use the Weiss variation without an explicit 3+1 split in the bulk to extract the canonical variables and Hamiltonian for GR. The Weiss variation also provides a quick way to obtain the Hamilton-Jacobi equation directly from boundary terms. In this sense, the Weiss variation formalism provides a complementary addition to the existing canonical formalism for boundary terms in GR (see \cite{BrownYork1993,HawkingHorowitz1996,BrownLauYork1997,BoothMann1999,Booth2000T,BrownLauYork2002} and references therein).

In the gravitational case, variational principles are complicated by the fact that the Einstein-Hilbert action contains second-order derivatives of the metric tensor; the variation of the Einstein-Hilbert action will, as a result, contain boundary terms proportional to the derivatives of the metric tensor variations \cite{MTW}. One is then forced to hold both the metric tensor \textit{and} its derivatives fixed at the boundary. If the Gibbons-Hawking-York (GHY) boundary term is added to the Einstein-Hilbert action  \cite{York1972,GibbonsHawking1977,York1986}, and if the components of the metric tensor (only) are held fixed at the boundary, the variation of the GHY boundary term will cancel out boundary terms proportional to the derivatives of the metric tensor variations (also see \cite{Wald,Poisson}).

The Weiss variation of the gravitational action (the Einstein-Hilbert action plus the GHY boundary term) requires the variation of the GHY boundary term under displacements of the boundary. This is the principal insight that we bring to this problem. One may obtain the variation of the GHY term by brute force, but since the GHY boundary term is expressed in terms of geometric quantities (it is in fact the integral of the mean curvature of the boundary surface), it is natural to use a formalism in which the geometric meaning is apparent. Fortunately, the variation of area formulas \cite{Frankel} and \cite{ChowHamilton,Nitsche1989,Dierkesetal2010}, which describe the variation of the volume of a hypersurface under displacements, provide such a formalism. Upon comparison with the first variation of area formula, one may interpret the GHY boundary term as the first-order variation of ``area'' (the 3-volume) for the boundary under a unit displacement of the boundary surface in the direction of the unit normal vector--the GHY boundary term is a special case of the first variation of area formula. The second variation of area formula \cite{ChowHamilton,Nitsche1989,Dierkesetal2010} describes a variation in the first variation of area formula under a displacement (which may be independent of the first displacement) of the hypersurface. With it we obtain an expression for the variation of the GHY boundary term. 

The Weiss variation for the gravitational action is not a new result. Once we present the Weiss variation for a mechanical system, a reader familiar with the ADM canonical formulation of GR \cite{ADM62,Poisson,Gourgolhoun3+1} should be able to infer the Weiss variation for the gravitational action without much difficulty. Some terms in the Weiss variation can be inferred from results in the existing literature \cite{BrownYork1993,HawkingHorowitz1996,BrownLauYork1997,BrownLauYork2002} which examine metric variations of the GHY boundary term. The variation of gravitational action that results from boundary displacements is also well-known in the form of the Einstein-Hamilton-Jacobi equation \cite{Peres1962,Gerlach1969} (see also \cite{Rovelli2004} and references therein).\footnote{We discuss in Sec. \ref{Sec-Weissmech} the relationship between Hamilton-Jacobi theory and the Weiss variation.} What is new is our \textit{geometric} derivation of the Weiss variation for the gravitational action and variations due to explicit boundary displacements, which do not to our knowledge appear in the existing literature. Furthermore, our formalism avoids an explicit 3+1 decomposition of the bulk spacetime. It should be stressed that we claim no lack of generality in ignoring explicit boundary displacements; infinitesimal boundary displacements can in principle be absorbed into metric variations (in both the bulk and boundary metric). While our formalism may be useful for describing situations where a global 3+1 decomposition is inappropriate (for instance, spacetime manifolds which fail to be globally hyperbolic), we again do not suggest a lack of generality in employing a 3+1 decomposition for boundary variations, as the use of such a formalism for boundary variation only requires the existence of such a decomposition in a neighborhood of the boundary.

Below, we review the Weiss variational principle in mechanics and in classical field theory. Next, we briefly review some definitions and results in semi-Riemannian geometry, and present the variation of area formulas. We review the standard variation for the gravitational action (for spacetimes with and without spatial boundary) and extend it to include contributions from displacements of the action. Finally, we rewrite the variation of the resulting action in Weiss form.

We assume a 4-dimensional spacetime manifold \(\mathcal{M}\), with \(\textbf{U} \subset \mathcal{M}\) and \(\textbf{W} \subset \mathcal{M}\) being subsets of a spacetime manifold \(\mathcal{M}\) of nonzero volume. We use \(\textbf{W}\) and \(\textbf{U}\) to distinguish between spacetime regions with and without spatial boundary; \(\textbf{W}\) has spatial boundary, and \(\textbf{U}\) does not. We use the MTW \cite{MTW} signature \((-,+,+,+)\) for the metric tensor \(g_{\mu \nu}\), \(x\) will represent a points on the spacetime manifold \(\mathcal{M}\), and \(y\) will represent points on hypersurfaces (surfaces of codimension one) in \(\mathcal{M}\), \(\textbf{U}\) or \(\textbf{W}\) . Greek indices refer to coordinates on the spacetime manifold \(\mathcal{M}\), \(\textbf{U}\) and \(\textbf{W}\); coordinates on \(\mathcal{M}\) will be denoted \(x^\mu\) with \(x^0=t\) being the time coordinate. Lowercase Latin indices refer either to mechanical degrees of freedom or coordinates on hypersurfaces--the distinction should be apparent from the context. Capital Latin indices from the beginning of the alphabet will either refer to two-dimensional surfaces in \(\mathcal{M}\), or to the components of a generic classical field--again, the distinction should be apparent from the context. Thus, coordinates on hypersurfaces will be denoted \(y^i\), and coordinates on two-dimensional surfaces will be denoted \(z^A\).

\section{The Weiss variational principle}
\subsection{Mechanics}\label{Sec-Weissmech}
We begin by reviewing the \textit{Weiss variational principle} in mechanics, as discussed in \cite{SudarshanCM}, \cite{MatznerShepleyCM} and \cite{Weiss1936}. Mechanical systems are typically described by an action functional of the form:
\begin{equation} \label{3M-Action}
\begin{aligned}
S[q]:=\int^{t_2}_{t_1} L(q,\dot{q},t) \> dt \\
\end{aligned}
\end{equation}

\noindent where the quantities \(q^i\) form the degrees of freedom for the mechanical system in question. The functions \(q^i=q^i(t)\) describe paths in the manifold formed from \(q^i\) and \(t\). The primary feature of the Weiss variation is that endpoint variations are allowed--even displacements of the endpoints themselves. Here, we consider two paths described by the functions \(q^i(t)\) and \({q^\prime}^i(t)\), which differ infinitesimally in the following manner:
\begin{equation} \label{3M-WeissPaths}
\begin{aligned}
{q^\prime}^i(t)=q^i(t)+ \epsilon \> \eta^i(t)
\end{aligned}
\end{equation}

\noindent where \(\epsilon \ll 1\) is an infinitesimal parameter and \(\eta^i(t)\) is some function, which is \textit{not} assumed to vanish at the endpoints. The difference in the endpoints may be characterized by differences in the value of the time parameter for the endpoints, 
\begin{equation} \label{3M-WeissTimeParam}
\begin{aligned}
{t^\prime}_1 &=t_1+ \epsilon \> \tau_1\\
{t^\prime}_2 &=t_2+ \epsilon \> \tau_2\\
\end{aligned}
\end{equation}

\noindent The action for the path \({q^\prime}^i(t)\) which has endpoints \({t^\prime}_1\) and \({t^\prime}_2\) takes the following form, to first order in \(\epsilon\):
\begin{equation} \label{3M-PrimedAction}
\begin{aligned}
S[q^\prime]&= \int^{t_2^\prime}_{t_1^\prime} L(q^\prime(t),\dot{q}^\prime(t),t) dt =  \int^{t_2+\epsilon \> \tau_2}_{t_1 + \epsilon \> \tau_1} L(q(t)+ \epsilon \> \eta(t),\dot{q}(t)+\epsilon \> \dot{\eta}(t),t) dt \\
&=  \int^{t_2}_{t_1} \epsilon \left(\frac{\partial L}{\partial q^i} \>\eta^i(t)+ \frac{\partial L}{\partial \dot{q}^i} \> \dot{\eta}^i(t) \right) dt+\int^{t_2}_{t_1} L(q(t),\dot{q}(t),t) dt \\
&\>\>\>\>\>+\int^{t_2+\epsilon \> \tau_2}_{t_2} L(q(t),\dot{q}(t),t) dt  - \int^{t_1+ \epsilon \> \tau_1}_{t_1 } L(q(t),\dot{q}(t),t) dt \\
&= S[q]+\epsilon \int^{t_2}_{t_1} \left(\frac{\partial L}{\partial q^i}- \frac{d}{dt} \left(\frac{\partial L}{\partial \dot{q}^i} \right)\right)\eta^i(t) \> dt + \left( \frac{\partial L}{\partial \dot{q}^i} \> \epsilon \> \eta^i(t)+  L \> \Delta t\right) \biggr |^{t_2}_{t_1}\\
\end{aligned}
\end{equation}

\noindent where \(\Delta t=\Delta t (t)\) is a function that satisfies \(\Delta t (t_1)= \epsilon \> \tau_1\) and \(\Delta t (t_2)= \epsilon \> \tau_2\), and \(S[q]\) is the action for the path \(q^i(t)\) with the endpoints \({t}_1\) and \({t}_2\). It is convenient to rewrite the boundary/endpoint term in terms of the total displacement of the endpoints \(\Delta q^i_1\) and \(\Delta q^i_2\):
\begin{equation} \label{3M-EndpointDisplacementA}
\begin{aligned}
\Delta q^i_1&:={q^\prime}^i({t^\prime}_1)-q^i(t_1)=\epsilon (\eta^i(t_1) + \tau_1 \, \dot{q}^i(t_1))+O(\epsilon^2)\\
\Delta q^i_2&:={q^\prime}^i({t^\prime}_2)-q^i(t_2)=\epsilon (\eta^i(t_2) + \tau_2 \, \dot{q}^i(t_2))+O(\epsilon^2)\\
\end{aligned}
\end{equation}

\noindent The variation in the action, to first order in \(\epsilon\), becomes:
\begin{equation} \label{3M-WeissVariation}
\begin{aligned}
\delta S &=\epsilon \int^{t_2}_{t_1} \left(\frac{\partial L}{\partial q^i} - \frac{d}{dt} \left(\frac{\partial L}{\partial \dot{q}^i} \right)\right)\eta^i(t) \> dt + \left( p_i \> \Delta q^i - \left(p_i \> \dot{q}^i-L\right) \Delta t\right) \biggr |^{t_2}_{t_1}\\
\end{aligned}
\end{equation}

\noindent where \(\Delta q^i=\Delta q^i(t)\) satisfies \(\Delta q^i(t_1)=\Delta q^i_1\) and \(\Delta q^i(t_2)=\Delta q^i_2\), and we have defined the following:
\begin{equation} \label{3M-WeissMomDefn}
\begin{aligned}
p_i&:=\frac{\partial L}{\partial \dot{q}^i}\\
\end{aligned}
\end{equation}

\noindent Note that the quantity appearing in front of \(\Delta t\) in (\ref{3M-WeissVariation}) is in fact the Hamiltonian:
\begin{equation} \label{3M-WeissHamDefn}
\begin{aligned}
H:=p_i \> \dot{q}^i-L\\
\end{aligned}
\end{equation}

\noindent The Weiss variational principle states that the physical paths \(q^i(t)\) are those which have the property that general infinitesimal variations about \(q^i(t)\) produce \textit{only} boundary/endpoint contributions to lowest order in the variation parameters. Simply put, physical paths \(q^i(t)\) are those for which the first order variations about \(q^i(t)\) yield variations in the action of the form:
\begin{equation} \label{3M-WeissVariationPhysical}
\begin{aligned}
\delta S &= \left( p_i \> \Delta q^i - H \> \Delta t\right) \biggr |^{t_2}_{t_1}\\
\end{aligned}
\end{equation}

\noindent Upon comparing (\ref{3M-WeissVariationPhysical}) with (\ref{3M-WeissVariation}), (\ref{3M-WeissVariationPhysical}) implies that physical paths \(q^i(t)\) are those for which the following term vanishes:
\begin{equation} \label{3M-WeissVariationBulkVanish}
\begin{aligned}
\epsilon \int^{t_2}_{t_1} \left(\frac{\partial L}{\partial q^i} - \frac{d}{dt} \left(\frac{\partial L}{\partial \dot{q}^i} \right)\right)\eta^i(t) \> dt =0\\
\end{aligned}
\end{equation}

\noindent If we demand that the above equation is satisfied for general infinitesimal variations \(\delta q^i(t)= \epsilon \, \eta^i(t)\), we recover the Euler-Lagrange equations:
\begin{equation} \label{3M-Euler-Lagrange}
\begin{aligned}
\frac{\partial L}{\partial q^i} - \frac{d}{dt} \left(\frac{\partial L}{\partial \dot{q}^i} \right)=0 .
\end{aligned}
\end{equation}

\noindent We note that the Weiss variation (\ref{3M-WeissVariation}) allows one to identify the Hamiltonian without having to perform a Legendre transformation (cf. (\ref{3M-WeissHamDefn})). It may be argued that in doing so, one is essentially identifying N{\"o}ether currents, but here, no reference is made to symmetries and no transformation of the time parameter \(t\) has been performed; instead, one displaces the endpoints.\footnote{If the action is invariant under time translations and the displacement of the endpoints is chosen so they are consistent with a translation in time, then one recovers the result that the Hamiltonian is the N{\"o}ether current for time translation symmetry.} 

The Weiss variation also provides a quick way to obtain the Hamilton-Jacobi equation without the machinery of canonical transformations. The classical action \(S_c\) is defined as the value of the action evaluated on solutions to the Euler-Lagrange equations. If we know the solutions to the Euler-Lagrange equations for a given set of endpoint values \(q^i_1:=q^i(t_1)\), \(q^i_2:=q^i(t_2)\), the classical action may be written as a function of the endpoint values and endpoint times: \(S_c=S_c(t_1,q^i_1;t_2,q^i_2)\). We may hold \(t_1\) fixed, and upon comparing the resulting differential of the classical action \(dS_c=(\partial S_c/\partial q^i_1) \, dq^i_1+(\partial S_c/\partial q^i_2) \, dq^i_2+(\partial S_c/\partial t_2) \, dt \) with (\ref{3M-WeissVariationPhysical}), we recover the formula relating \(p_i\) to the derivatives of the action and the Hamilton-Jacobi equation:
\begin{equation} \label{3M-HamJacMom1}
\begin{aligned}
\frac{\partial S_c}{\partial q^i_1} &= p_i|_{t_1}\\
\end{aligned}
\end{equation}

\begin{equation} \label{3M-HamJacMom2}
\begin{aligned}
\frac{\partial S_c}{\partial q^i_2} &= p_i|_{t_2}\\
\end{aligned}
\end{equation}

\begin{equation} \label{3M-HamJac}
\begin{aligned}
\frac{\partial S_c}{\partial t_2} &=- H\left(\frac{\partial S_c}{\partial q^i_2} ,q^i_2,t_2\right)
\end{aligned}
\end{equation}

\noindent We do not include the derivative \({\partial S_c}/{\partial t_1}\), since \(t_1\) is held fixed. We stress that \(q^i_1\) is \textit{not} held fixed, so that we may construct (\ref{3M-HamJacMom1}); equation (\ref{3M-HamJacMom1}) is important because it ultimately allows us to obtain physical paths \(q^i(t)\) from solutions \(S_c=S_c(q^i_1,q^i_2,t_1,t_2)\) to the Hamilton-Jacobi equation (\ref{3M-HamJac}). Given \(S_c=S_c(q^i_1,q^i_2,t_1,t_2)\), equation (\ref{3M-HamJacMom1}) allows us to write down an \textit{algebraic}\footnote{Note that for some function \(S_c=S_c(q^i_1,q^i_2,t_2)\), the left-hand side of formula (\ref{3M-HamJacMom1}) is an explicit function of \(q^i_2\), \(t_2\) and \(q^i_1\).} equation relating \(q^i_2\) and \(t_2\) to the initial values \(q^i_1\) and \(p_i|_{t_1}\); note that equation (\ref{3M-HamJacMom2}) is insufficient\footnote{Equation (\ref{3M-HamJacMom2}) is used to construct the Hamilton-Jacobi equation (\ref{3M-HamJac}) itself; in particular, it is used to replace the momentum argument in the Hamiltonian with the derivative \({\partial S_c}/{\partial q^i_2}\) of the action.} for this, since it depends on the \textit{final} momentum \(p_i|_{t_2}\), rather than the initial momentum \(p_i|_{t_1}\). If the solutions of the Hamilton-Jacobi equation (\ref{3M-HamJac}) are known for \textit{all} values of \(q^i_1\), then we simply solve (\ref{3M-HamJacMom1}) for \(q^i_2\) to obtain the function \(q^i_2(t_2)\) for a given set of initial values \(q^i_1\) and \(p_i|_{t_1}\).

\subsection{Classical field theory}
Now consider the Weiss variation for a classical field theory in a region \(\textbf{W}\) with spatial boundary, \(\textbf{W} \subset \mathcal{M}\). Given a collection of fields \(\varphi^A(x)\), the index \(A\) being the field index (it may either serve as a coordinate index, a spinor index, an index to distinguish fields, or a combination), we begin by considering the following action functional:
\begin{equation} \label{3F-FieldActionNC}
S[\varphi^A]=\int_{\textbf{W}} \mathscr{L}(\varphi^A,\partial_\mu \varphi^A,x^\mu) \> d^4 x
\end{equation}

\noindent where \(\mathscr{L}\) is called the \textit{Lagrangian density}, which is a function \(\mathscr{L}(X^A,Y_\mu^A, x^\mu)\) of  \(X^A=\varphi^A(x)\), their first derivatives \(Y_\mu^A=\partial_\mu \varphi^A\), and \(x^\mu \in \mathbb{R}^4\). For clarity, we have chosen not to suppress the field and the Greek spacetime/\(\mathbb{R}^4\) indices in the arguments of \(\mathscr{L}\).\footnote{Also, since we shall later include the metric as an argument in the action functional, we choose not to suppress indices to avoid confusing the metric with its determinant--the symbol \(g\) is reserved for the determinant of the metric.}

We may identify one of the variables in \(\mathbb{R}^4\), which we will call \(x^0=t\), as a time variable, and the remaining variables \(y^i\) are interpreted as spatial variables. The volume element \(d^4x\) (we absorb any factors of \(\sqrt{|g|}\) into the Lagrangian density) may be split into spatial and temporal parts, so that \(d^4x=dt \> d^3y\). From the Lagrangian density, one may obtain the field Lagrangian\index{Lagrangian!Field Lagrangian} by isolating the spatial part of the integral, so that:
\begin{equation} \label{3F-FieldLagrangianNC}
L[\varphi^A,\dot{\varphi}^A;\Sigma_t]=\int_{\Sigma_t} \mathscr{L}(\varphi^A,\partial_\mu \varphi^A,x^\mu) \> d^3 y
\end{equation}

\noindent where \(\Sigma_t\) is a hypersurface of constant \(t\), and the semicolon in \(L[\varphi^A,\dot{\varphi}^A;\Sigma_t]\) denotes that it is a functional of functions defined on \(\Sigma_t\), in particular the functions \(\varphi^A|_{\Sigma_t}(y)\) and \(\dot{\varphi}^A|_{\Sigma_t}(y)\).

To obtain the variation of the action, we add an infinitesimal function \(\delta \varphi^A(x)\) to \(\varphi^A(x)\). To obtain a general variation, we infinitesimally distort the region \(\textbf{W}\); the infinitesimally distorted region will be denoted \(\textbf{W}^\prime\). The boundary \(\partial\textbf{W}\) of the region \(\textbf{W}\) may be defined parametrically by \(x^\mu(y)\), where \(y^i\) are coordinates on the boundary surface \(\partial\textbf{W}\). If \(\partial\textbf{W}^\prime\) is the boundary of the region \(\textbf{W}^\prime\), then we may describe the displaced boundary \(\partial\textbf{W}^\prime\) parametrically by \(x^\mu(y)+\delta x^\mu(y)\), where \(\delta x^\mu\) is an infinitesimal displacement of the boundary.

The varied action takes the form:
\begin{equation} \label{3F-VariedActionNC}
S[\varphi^A+\delta \varphi^A]=\int_{\textbf{W}^\prime} \mathscr{L} (\varphi^A+\delta \varphi^A,\partial_\mu (\varphi^A+\delta \varphi^A),x^\mu) \> d^4 x
\end{equation}

\noindent We may obtain an expression for the above valid to first order in \(\delta \varphi^A(x)\) and the boundary displacements \(\delta x^\mu\) by performing a Taylor expansion of the Lagrangian density about \(\varphi^A\):
\begin{equation} \label{3F-LagrangianExpandNC}
\begin{aligned}
\mathscr{L} (\varphi^A+\delta \varphi^A,\partial_\mu (\varphi^A+\delta \varphi^A),x^\mu)=\,&\mathscr{L}(\varphi^A,\partial_\mu \varphi^A,x^\mu)+\frac{\partial \mathscr{L}}{\partial \varphi^A} \> \delta \varphi^A  + \frac{\partial \mathscr{L}}{\partial (\partial_\mu \varphi^A)} \> \partial_\mu \delta \varphi^A
\end{aligned}
\end{equation}

\noindent where the following quantities are defined:
\begin{equation} \label{3F-LagrangianDerivativesNC}
\begin{aligned}
\frac{\partial \mathscr{L}}{\partial \varphi^A} := \frac{\partial \mathscr{L}(X^A,Y_\mu^A,x^\mu)}{\partial X^A}\biggr |_{X^A=\varphi^A(x), \> Y_\mu^A=\partial_\mu \varphi^A(x)}  \\
\frac{\partial \mathscr{L}}{\partial (\partial_\mu \varphi^A)} := \frac{\partial \mathscr{L}(X^A,Y_\mu^A,x^\mu)}{\partial Y_\mu^A}\biggr |_{X^A=\varphi^A(x), \> Y_\mu^A=\partial_\mu \varphi^A(x)}  
\end{aligned}
\end{equation}

\begin{figure} 
\begin{center}
\includegraphics[scale=1]{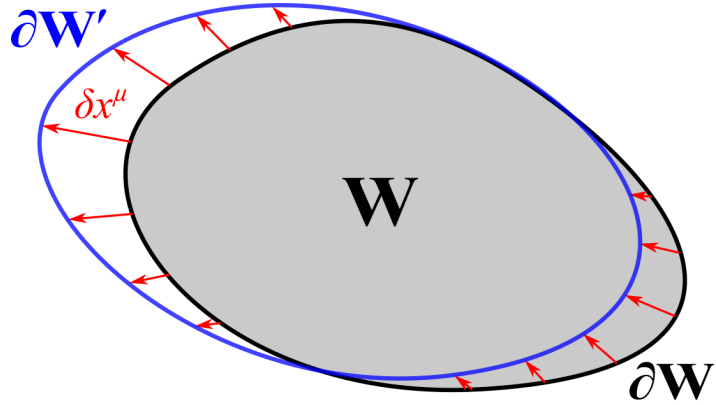}
\end{center}
\caption[Swept Boundary]{This figure illustrates the deformation of the region \(\textbf{W}\) and the infinitesimal displacement of the boundary \(\partial \textbf{W}\) to \(\partial \textbf{W}^\prime\) by the displacement vector \(\delta x^\mu\).} \label{3Fg-BoundarySweep}
\end{figure}

\noindent We establish the convention that if \(\mathscr{L}\) appears without any arguments, it means that \(\mathscr{L}\) has the following arguments: \(\mathscr{L}=\mathscr{L}(\varphi^A,\partial_\mu \varphi^A,x^\mu)\). The first order expansion of the action is:
\begin{equation} \label{3F-ActionExpandedNC}
\begin{aligned}
S[\varphi^A+\delta \varphi^A]&=S[\varphi^A] +\int_{\textbf{W}^\prime} \left(\frac{\partial \mathscr{L}}{\partial \varphi^A} \, \delta \varphi^A + \frac{\partial \mathscr{L}}{\partial (\partial_\mu \varphi^A)} \, \partial_\mu \delta \varphi^A\right) \, d^4 x + \int_{\partial\textbf{W}} \mathscr{L}(\varphi^A,\partial_\mu \varphi^A) \, \delta x^\mu\> d^3\bar{\Sigma}_\mu\\
&=S[\varphi^A] +\int_{\textbf{W}} \left(\frac{\partial \mathscr{L}}{\partial \varphi^A} \, \delta \varphi^A + \frac{\partial \mathscr{L}}{\partial (\partial_\mu \varphi^A)} \, \partial_\mu \delta \varphi^A\right) d^4 x + \int_{\partial\textbf{W}} \left(\mathscr{L} + \frac{\partial \mathscr{L}}{\partial \varphi^A} \, \delta \varphi^A + \frac{\partial \mathscr{L}}{\partial (\partial_\nu \varphi^A)} \, \partial_\nu \delta \varphi^A \right) \delta x^\mu\, d^3\bar{\Sigma}_\mu
\end{aligned}
\end{equation}

\noindent where \(d^3\bar{\Sigma}_\mu\) is the directed surface element on \(\partial \textbf{W}\). The directed surface element has the explicit expression:
\begin{equation} \label{3F-DirectedSurfElement}
\begin{aligned}
d^3\bar{\Sigma}_\mu=\frac{1}{3!}\>\underline{\epsilon}_{\mu \alpha \beta \gamma} \> \frac{\partial x^\alpha}{\partial y^i} \frac{\partial x^\beta}{\partial y^j} \frac{\partial x^\gamma}{\partial y^k} \> dy^i \wedge dy^j \wedge dy^k
\end{aligned}
\end{equation}

\noindent where \(\underline{\epsilon}_{\mu \alpha \beta \gamma}\) is the Levi-Civita symbol. The boundary integrals in (\ref{3F-ActionExpandedNC}) may be justified by noting that under an infinitesimal displacement \(\delta x^\mu\) of the boundary \(\partial \textbf{W}\), the boundary \(\partial \textbf{W}\) sweeps out a volume \(\delta V \approx \delta x^\mu \> d{^3}\bar{\Sigma}_\mu\) (see figure (\ref{3Fg-BoundarySweep})).

The expansion of the action (\ref{3F-ActionExpandedNC}) may be further simplified by noting that terms containing \(\delta \varphi^A \>\delta x^\mu \) and  \(\partial_\nu \delta \varphi^A \>\delta x^\mu\) are second order in infinitesimal variations. If all variations are proportional to a single infinitesimal parameter, these terms may be ignored if we only seek the variation of the action to first order in the variations. The first-order variation of the action is then:
\begin{equation} \label{3F-VariationActionNC}
\begin{aligned}
\delta S&=S[\varphi^A+\delta \varphi^A]-S[\varphi^A] = \int_{\textbf{W}} \left(\frac{\partial \mathscr{L}}{\partial \varphi^A} \> \delta \varphi^A + \frac{\partial \mathscr{L}}{\partial (\partial_\mu \varphi^A)} \> \partial_\mu \delta \varphi^A\right) \> d^4 x+\int_{\partial\textbf{W}} \mathscr{L} \> \delta x^\mu\> d^3\bar{\Sigma}_\mu
\end{aligned}
\end{equation}

\noindent The action must be reworked so that the integral over \(\textbf{W}\) has an integrand proportional to \(\delta \varphi^A\), so that a functional derivative may be defined. We do this by performing a generalized ``integration by parts''; the term containing \(\partial_\mu \delta \varphi^A\) may be converted to a term proportional to \(\delta \varphi^A\) and a divergence term by way of the Leibniz rule:
\begin{equation} \label{3F-LeibnizRuleNC}
\begin{aligned}
\partial_\mu \left(\frac{\partial \mathscr{L}}{\partial (\partial_\mu \varphi^A)} \> \delta \varphi^A \right)=\frac{\partial \mathscr{L}}{\partial (\partial_\mu \varphi^A)} \> \partial_\mu \delta \varphi^A + \partial_\mu \left(\frac{\partial \mathscr{L}}{\partial (\partial_\mu \varphi^A)} \>\right) \delta \varphi^A
\end{aligned}
\end{equation}

\noindent The divergence theorem in \(\textbf{W} \subset \mathbb{R}^4\) takes the following form:
\begin{equation} \label{3F-DivTheoremR4} 
\int_{\textbf{W}} \partial_{\mu}W^\mu \> d^4x=\int_{\partial \textbf{W}} W^\mu \> d^3\bar{\Sigma}_\mu.
\end{equation}

\noindent which may be used to rewrite (\ref{3F-VariationActionNC}):
\begin{equation} \label{3F-VariationActionNC2}
\begin{aligned}
\delta S&=\int_{\textbf{W}} \left(\frac{\partial \mathscr{L}}{\partial \varphi^A} - \partial_\mu \left(\frac{\partial \mathscr{L}}{\partial (\partial_\mu \varphi^A)}\right)\right) \delta \varphi^A \> d^4 x + \int_{\partial\textbf{W}} \left( \mathscr{L} \> \delta x^\mu+\frac{\partial \mathscr{L}}{\partial (\partial_\mu \varphi^A)} \> \delta \varphi^A \right) d^3\bar{\Sigma}_\mu
\end{aligned}
\end{equation}

If we require that \(\delta x^\mu |_{\partial \textbf{W}}=0\) and \(\delta \varphi^A |_{\partial \textbf{W}}=0\), the boundary terms in (\ref{3F-VariationActionNC2}) vanish, and the variation of the action \(\delta S\) is an integral over \(\textbf{W}\) with an integrand proportional \(\delta \varphi^A\). The functional derivative of the action functional \(S[\varphi^A]\) is defined by the following formula:
\begin{equation} \label{3F-FunctionalDerivative}\index{Functional Derivative!Functional Derivative (Fields)}
\begin{aligned}
\delta S &:=\int_{\textbf{W}} \frac{\delta S}{\delta \varphi^A} \> \delta \varphi^A(x) \> d^4 x\\
\end{aligned}
\end{equation}

\noindent The functional derivative may be identified as:
\begin{equation} \label{3F-FunctionalDerivativeNC}
\begin{aligned}
\frac{\delta S}{\delta \varphi^A}=\frac{\partial \mathscr{L}}{\partial \varphi^A} - \partial_\mu \left(\frac{\partial \mathscr{L}}{\partial (\partial_\mu \varphi^A)}\right)
\end{aligned}
\end{equation}

\noindent If we require that the functional derivative vanishes, we obtain the \textit{Euler-Lagrange field equations}\index{Euler-Lagrange Field Equations}:
\begin{equation} \label{3F-EulerLagrangeFieldEqnsNC2}
\begin{aligned}
\frac{\partial \mathscr{L}}{\partial \varphi^A} = \partial_\mu \left(\frac{\partial \mathscr{L}}{\partial (\partial_\mu \varphi^A)}\right)
\end{aligned}
\end{equation}

We now write (\ref{3F-VariationActionNC}) in the local Weiss form--we obtain the field theory generalization of the Weiss variation for mechanics (\ref{3M-WeissVariation}). To do this, we define the total change in the field values at the boundary:
\begin{equation} \label{3F-ChangeinField}
\begin{aligned}
\Delta \varphi^A:&=(\varphi^A+\delta \varphi^A)|_{\partial \textbf{W}^\prime} \> - \> \varphi^A|_{\partial \textbf{W}}\\
&=\left(\delta \varphi^A+\partial_\mu \varphi^A \> \delta x^\mu \right) |_{\partial \textbf{W}}
\end{aligned}
\end{equation}

\noindent The variation of the action (\ref{3F-VariationActionNC}) in the local Weiss form becomes:
\begin{equation} \label{3F-VariationAction3A}
\begin{aligned}
\delta S &:=\int_{\textbf{W}} \left(\frac{\partial \mathscr{L}}{\partial \varphi^A}  - \partial_\mu \left(\frac{\partial \mathscr{L}}{\partial (\partial_\mu \varphi^A)}\right) \right)\delta \varphi^A \> d^4 x + \int_{\partial\textbf{W}} \left(P_A^\mu \> \Delta \varphi^A - \mathscr{H}{^\mu}{_\nu}\> \delta x^\nu\right)  d^3 \bar{\Sigma}_\mu
\end{aligned}
\end{equation}

\noindent where we have identified the canonical energy-momentum ``tensor'' \(\mathscr{H}{^\mu}{_\nu}\> \delta x^\nu\) (this is sometimes called the \textit{Hamiltonian Complex} or \textit{Hamiltonian tensor} \cite{LovelockandRund,RundVarHJ}):\footnote{If we work in Minkowski spacetime and choose \(\delta x^\mu\) so that it corresponds to a translation in spacetime, then we recover the well-known result that \(\mathscr{H}{^\mu}{_\nu}\> \delta x^\nu\) is the N{\"o}ether current for spacetime translation symmetry.}
\begin{equation} \label{3F-HamiltonianComplex}
\begin{aligned}
\mathscr{H}{^\mu}{_\nu}:= P_A^\mu \> \partial_\nu \varphi^A  - \delta^\mu_\nu \>\mathscr{L}
\end{aligned}
\end{equation}

\noindent and the following quantity, which we call the \textit{polymomentum}\index{Polymomentum}:
\begin{equation} \label{3F-CovCovariantFieldMomentumA}
\begin{aligned}
P_A^\mu :=\frac{\partial \mathscr{L}}{\partial (\partial_\mu \varphi^A)}
\end{aligned}
\end{equation}

\noindent This result demonstrates that the canonical energy-momentum ``tensor'' plays a role analogous to the Hamiltonian in mechanics; it may in fact be used to define a Hamiltonian for the field. The \textit{Hamiltonian density}\index{Hamiltonian!Hamiltonian Density} may be defined as the \(\mathscr{H}^0{_0}\) component of the Hamiltonian tensor, which takes the explicit form:
\begin{equation} \label{3F-CovHamiltonianDensityA}
\begin{aligned}
\mathscr{H} := \mathscr{H}^0_0 = \pi_A \> \dot{\varphi}^A -\mathscr{L}
\end{aligned}
\end{equation}

\noindent where \(x^0=t\), the overdot denotes the time derivatives \(\dot{\varphi}^A :=\partial_0 \varphi^A\) and the \textit{conjugate field momentum}\index{Conjugate Field Momentum} \(\pi_A\) is defined as:
\begin{equation} \label{3F-CovConjugateFieldMomentum}
\begin{aligned}
\pi_A:=P_A^0 =\frac{\partial \mathscr{L}}{\partial \dot{\varphi}^A}.
\end{aligned}
\end{equation}

\noindent We assume that one may invert the above to obtain an expression for the function \(\dot{\varphi}^A=\dot{\varphi}^A(\pi_A,\varphi^A,\partial_i \varphi^A)\), and the lowercase index \(i\) (which appears in the partial derivatives \(\partial_i\)) corresponds to the coordinates \(y^i\) for surfaces of constant \(t\).

The Hamiltonian density has a form similar to that of the Hamiltonian in mechanics, and using \(\dot{\varphi}^A=\dot{\varphi}^A(\pi_A,\varphi^A,\partial_i \varphi^A)\), can itself be written as a function of \(\pi_A\), \(\varphi^A\)  and \(\partial_i \varphi^A\).  The Hamiltonian density \(\mathscr{H}\) is not strictly a Hamiltonian, since it is defined at a single point in space, and does not include the degrees of freedom present at other points in space. The Hamiltonian for a field theory is the ``sum'' of the Hamiltonian densities over all points in space; to construct the Hamiltonian, we integrate the Hamiltonian density \(\mathscr{H}\) over a hypersurface \(\Sigma_t\) of constant \(t\):
\begin{equation} \label{3F-FieldHamiltonian}
\begin{aligned}
H[\varphi^A,\pi_A;\Sigma_t]:&=\int_{\Sigma_t} \mathscr{H}(\pi_A,\varphi^A,\partial_i \varphi^A,y^i,t) \> d^3y =\int_{\Sigma_t} (\pi_A \> \dot{\varphi}^A \> -\mathscr{L}) \> d^3y
\end{aligned}
\end{equation}

\noindent which may be rewritten as the Legendre transformation of the field Lagrangian:
\begin{equation} \label{3F-FieldHamiltonianLT}
\begin{aligned}
H[\varphi^A,\pi_A;\Sigma_t]&=\int_{\Sigma_t} \pi_A \> \dot{\varphi}^A \> d^3y - L[\varphi^A,\dot{\varphi}^A;\Sigma_t]\\
\end{aligned}
\end{equation}

\noindent As in the case of the field Lagrangian (\ref{3F-FieldActionNC}), the Hamiltonian \(H[\varphi^A,\pi_A;\Sigma_t]\) is a functional of functions defined on \(\Sigma_t\), in particular the functions \(\varphi^A|_{\Sigma_t}(y)\) and \(\pi_A|_{\Sigma_t}(y)\).

We may now obtain the Weiss form of the variation. To do this, we choose the boundary to consist of two surfaces of constant \(t\), \(\Sigma_{t_1}\) (for \(t=t_1\)) and \(\Sigma_{t_2}\) (for \(t=t_2\)), and a surface at spatial infinity, which we call the spatial boundary. We require vacuum boundary conditions \(\mathscr{H}^\mu{_\nu}=0\) at the spatial boundary, and set \(\delta x^\mu=\delta^\mu_0 \, \Delta t\). The variation of the field action in Weiss form is then:
\begin{equation} \label{3F-VariationAction3WF}
\begin{aligned}
\delta S &:=\int_{\textbf{W}} \left(\frac{\partial \mathscr{L}}{\partial \varphi^A}  - \partial_\mu \left(\frac{\partial \mathscr{L}}{\partial (\partial_\mu \varphi^A)}\right) \right)\delta \varphi^A \> d^4 x + \left( \int_{\Sigma_{t}} \left(P_A^0 \> \Delta \varphi^A\right)  d^3 y - H[\varphi^A,\pi_A;\Sigma_t] \> \Delta t\right) \biggr |^{t_2}_{t_1}
\end{aligned}
\end{equation}

We have shown how the Weiss variation may be carried out in a generic classical field theory. In the next several sections, we lay the groundwork and carry out the Weiss variation for the action of GR (the gravitational action). The GHY boundary term in the gravitational action will introduce additional technical elements to the Weiss variation, so the results in this current section cannot be directly applied to the gravitational action. On the other hand, some of the features of the Weiss variation described in this section will be useful for understanding features of the Weiss variation for the gravitational action.

\section{Geometry, hypersurfaces, and variation of area formulas} 
\subsection{Geometry}
We begin by presenting our definitions for the Riemann curvature tensor, Ricci tensor, and Ricci scalar:
\begin{equation}\label{RiemannCurvatureTensor}
{R^\mu}_{\nu \alpha \beta}:=\partial_\alpha \Gamma^\mu_{\beta \nu}-\partial_\beta \Gamma^\mu_{\alpha \nu}+\Gamma^\mu_{\alpha \sigma} \>\Gamma^\sigma_{\beta \nu}- \Gamma^\mu_{\beta \sigma} \>\Gamma^\sigma_{\alpha \nu}.
\end{equation}
\begin{equation}\label{RicciTensor}
R_{\mu \nu} := {R^\sigma}_{\mu \sigma \nu}
\end{equation}
\begin{equation}\label{RicciScalar}
R:= g^{\mu \nu} \, R_{\mu \nu}
\end{equation}

\noindent which are defined for a Lorentzian manifold \(\mathcal{M}\) endowed with a metric tensor \(g_{\mu \nu}\), and a metric-compatible connection \(\nabla_\mu\) with connection coefficients (Christoffel symbols) \(\Gamma^\alpha_{\mu \nu}\):
\begin{equation}\label{ChristoffelSymbols}
\Gamma^\alpha_{\mu \nu}=\frac{1}{2} g^{\alpha \sigma} (\partial_\mu g_{\sigma \nu}+\partial_\nu g_{\mu \sigma}-\partial_\sigma g_{\mu \nu}) .
\end{equation}

\subsection{Foliations and hypersurfaces} \label{SecFoliations}
It is necessary to discuss the formalism used to describe the geometry of hypersurfaces embedded in 4d bulk spaces, since we take an approach that is slightly different than that present in much of the literature. That literature makes use of the abstract index notation \cite{Wald,Bojowald}, or coordinate bases on hypersurfaces \cite{Poisson,LovelockandRund} (\cite{Gourgolhoun3+1} is an exception, as some key results are worked out in the coordinate basis). In our approach, we obtain many of our results in the \textit{bulk} coordinates, which still requires that we assume the existence of a foliation. We will indicate which results are foliation dependent, and which are not.

Place a foliation on \(\mathcal{M}\), with non-null hypersurfaces \(\Sigma_S\) distinguished by values of some real parameter \(S\). More precisely, the foliation may be defined by an appropriate foliation function \(\phi(x)\), with a hypersurface \(\Sigma_S\) being a level surface defined by the constraint: \(\phi(x)=S\). We define a normal vector field:
\begin{equation} \label{1-GradientVectorField}
\begin{aligned}
 \mathfrak{n}^{\mu}(x)&:=g^{\mu \nu} \> \nabla_{\nu} \phi(x)
\end{aligned}
\end{equation}

\noindent From the above, we may define a unit normal vector field \(n^\mu(x)\) for the foliation:
\begin{equation} \label{1-UnitNormalVector}
n^\mu= \varepsilon \> \alpha \> \mathfrak{n}^{\mu}
\end{equation}

\noindent where \(\varepsilon=+1\) if the unit normal vector is spacelike, and \(\varepsilon=-1\) if the unit normal vector is timelike. The quantity \(\varepsilon\) ensures that the unit normal vector \(n^\mu\) points in the direction of increasing \(\phi\).\footnote{Note that if \(\mathfrak{n}^{\mu}\) is timelike, it points in the direction of \textit{decreasing} \(\phi\).} The quantity \(\alpha=\alpha(x)\) is the ``lapse function'', defined as:\footnote{This is equivalent to the lapse function in the ADM formalism.}
\begin{equation} \label{1-Lapse}
\alpha:=\frac{1}{| \mathfrak{n}_{\mu} \> \mathfrak{n}^{\mu}|^{1/2}}
\end{equation}

\noindent The unit normal vector field \(n^\mu(x)\) allows us to construct the induced metric/projection tensor:
\begin{equation} \label{1-InducedMetric}
\begin{aligned}
\gamma_{\mu \nu}&:=g_{\mu \nu} - \varepsilon \> n_\mu n_\nu
\end{aligned}
\end{equation}

\noindent It is not difficult to show that if coordinates \(y^i\) are placed on the hypersurface \(\Sigma_S\), \(\gamma_{\mu \nu}\) may be expressed in basis of the tangent vectors \(\partial/\partial y^i\) to the hypersurface \(\Sigma_S\). To do this, we define the components \(E^\mu_i\) of the tangent vectors in the following way:
\begin{equation} \label{1-CoordinateBasis}
\begin{aligned}
&\frac{\partial}{\partial y^i}=E^\mu_i \> \frac{\partial}{\partial x^\mu} \>\>\>\>\>\>\>\> \>\>\>\>\>\>\>\> E^\mu_i:=\frac{\partial x^\mu}{\partial y^i}
\end{aligned}
\end{equation}

\noindent We then write the induced metric \(\gamma_{\mu \nu}\) in terms of the basis vectors:
\begin{equation} \label{1-InducedMetricCoordinateBasis}
\gamma_{ij}=E^\mu_i \> E^\nu_j \> \gamma_{\mu \nu} = E^\mu_i \> E^\nu_j \> g_{\mu \nu}
\end{equation}

\noindent We use the above to define the metric-compatible, torsion-free covariant derivative \(D_i\) for the hypersurface \(\Sigma_S\) in the usual manner, with connection coefficients \(\bar{\Gamma}^i_{jk}\) being the Christoffel symbols corresponding to \(\gamma_{ij}\). In the bulk coordinate basis, the metric-compatible, torsion-free covariant derivative for a tensor \({T^{\mu_1 ... \mu_r}}_{\nu_1 ... \nu_s}\) with indices tangent to the hypersurface \(\Sigma_S\) may be written as:
\begin{equation} \label{1-InducedCovariantDerivative}
D_{\sigma} {T^{\mu_1 ... \mu_r}}_{\nu_1 ... \nu_s}=\left(\gamma_{\alpha_1 }^{\mu_1} ... \gamma_{\alpha_r}^{\mu_r}\right)\left(\gamma^{\beta_1 }_{\nu_1} ... \gamma^{\beta_s}_{\nu_s}\right)  \gamma^{\rho}_{\sigma} \> \nabla_\rho  {T^{\alpha_1 ... \alpha_r}}_{\beta_1 ... \beta_s}
\end{equation}

\noindent With some work, one may show that the above definition is equivalent to the connection \(D_i\) with the definition \(\bar{\Gamma}^i_{jk}:=E^{\sigma}_{j} E^{i}_{\mu}  \> \nabla_{\sigma}  E_{k}^{\mu}\) for the connection coefficients. The covariant derivative may be used to construct the intrinsic Riemann curvature tensor \({\bar{R}^{\alpha}}{_{\beta \mu \nu}}\) of the hypersurface \(\Sigma_S\) from the commutator of the derivative \(D_\mu\) defined above:
\begin{equation} \label{1-HypersurfaceRiemannCommutator}
{} [D_{\mu},D_{\nu}] \> X^\alpha = {\bar{R}^{\alpha}}{_{\beta \mu \nu}} \>  X^\beta
\end{equation}

\noindent for a vector \(X^\mu\) tangent to the hypersurface \(\Sigma_S\). It should be straightforward to infer that \({\bar{R}^{\alpha}}_{\> \> \> \beta \mu \nu}= E_{a}^{\alpha} \> E_{\beta}^{b} \> E_{\mu}^{i} \> E_{\nu}^{j} \>{\bar{R}^{a}}_{\> \> \> b i j}\), where \({\bar{R}^{a}}_{\> \> \> b i j}\) is given by the following expression:
\begin{equation}\label{2-CurvatureTensorHypersurface}
{\bar{R}^a}_{\>\>\>bij}=\partial_a \bar{\Gamma}^a_{jb}-\partial_j \bar{\Gamma}^a_{ib}+\bar{\Gamma}^a_{is} \>\bar{\Gamma}^s_{jb}- \bar{\Gamma}^a_{js} \> \bar{\Gamma}^s_{ib}.
\end{equation}

It is natural at this point to ask how the hypersurface curvature tensor \({\bar{R}^{\alpha}}_{\> \> \> \beta \mu \nu}\) is related to the curvature tensor \({R^{\alpha}}_{\beta \mu \nu}\) in the bulk manifold \(\mathcal{M}\). The derivation below (and those of Appendix \ref{AppendixGaussCodazziRicci}) is well known, but we carry it through explicitly to emphasize the importance of the sign, \(\varepsilon\). Before we can discuss this relationship, we must first introduce another notion of curvature, the extrinsic curvature. 

The extrinsic curvature tensor \(K_{\mu \nu }\) of a hypersurface \(\Sigma_S\) may be defined by the following expression:
\begin{equation}\label{1-ExtrinsicCurvatureTensorDefn}
\begin{aligned}
K_{\mu \nu } \, X^\mu \, Y^\nu=-n_\nu \, X^\mu \, \nabla_\mu Y^\nu
\end{aligned}
\end{equation}

\noindent for two vectors \(X^\mu\) and \(Y^\mu\) tangent to \(\Sigma_S\): \(X^\mu\, n_\mu=0\) and \(Y^\mu\, n_\mu=0\). This definition is independent of the foliation, and depends only on the manner in which the surface \(\Sigma_S\) is embedded in the bulk manifold \(\mathcal{M}\) and does not depend on any other surface in the foliation. On the other hand, if a foliation exists, the extrinsic curvature tensor may be written in three different ways:
\begin{equation}\label{1-ExtrinsicCurvatureTensor}
\begin{aligned}
K_{\mu \nu }&=\frac{1}{2}\pounds_n \gamma_{\mu \nu}=\gamma_{\mu }^{\sigma }\gamma _{\mu }^{\tau }\nabla _{\sigma } n_{\tau }=\nabla_{\mu} n_{\nu }-\varepsilon \> n_{\mu} \> a_{\nu}
\end{aligned}
\end{equation}

\noindent where \(\pounds_n\) is the Lie derivative, which acts on \(\gamma_{\mu \nu}\) in the following way:
\begin{equation}\label{1-LieDerivInducedMetric}
\begin{aligned}
\pounds_n \gamma_{\mu \nu}&=n^\alpha \partial_\alpha \gamma_{\mu \nu}+ \gamma_{\alpha \nu} \partial_\mu n^\alpha+ \gamma_{\mu \alpha} \partial_\nu n^\alpha=n^\alpha \nabla_\alpha \gamma_{\mu \nu}+ \gamma_{\alpha \nu} \nabla_\mu n^\alpha+ \gamma_{\mu \alpha} \nabla_\nu n^\alpha\\
\end{aligned}
\end{equation}

\noindent The last equality in (\ref{1-ExtrinsicCurvatureTensor}) makes use of the acceleration \(a_\mu\) for the integral curves of the unit normal vector field, which may be written as:
\begin{equation} \label{1-Acceleration}
a_\nu =n^\mu \> \nabla_\mu  n_\nu=-\varepsilon \> D_\nu(\ln{\alpha})
\end{equation}

\noindent where the last equality in the above comes from the torsion-free property of the covariant derivative \(\nabla_\alpha\).

The trace of the extrinsic curvature tensor, the mean curvature, is given by the following expression:
\begin{equation}\label{1-MeanCurvatureA}
K:=\gamma^{\mu \nu} \> K_{\mu \nu}=g^{\mu \nu} \> K_{\mu \nu}=\nabla_\alpha n^\alpha
\end{equation}

\noindent where the second equality comes from the fact that \(K_{\mu \nu} n^\nu=0\), and the last equality comes from the properties \(\gamma^{\mu \nu} \> \gamma^\alpha_\mu=\gamma^{\alpha \nu}\) and \(n^\nu \nabla_\alpha n_\nu=0\).

There are three formulas which relate the bulk Riemann curvature tensor \({R}{^\mu}_{\nu \alpha \beta}\) for \(\mathcal{M}\) to the extrinsic curvature \(K_{\mu \nu}\) of a hypersurface \(\Sigma\), and the Riemann tensor \(\bar{R}{^\mu}_{\nu \alpha \beta}\) for the surface \(\Sigma\). The derivation of these formulas is provided in Appendix \ref{AppendixGaussCodazziRicci}.\footnote{Again, these derivations are well-known. We carry it through explicitly to emphasize the importance of the sign, \(\varepsilon\).} The first is the Gauss equation:
\begin{equation} \label{1-GaussEquation}
\gamma^{\rho}_{\lambda} \gamma^{\tau}_{\sigma} \gamma^{\alpha}_{\mu} \gamma^{\beta}_{\nu} \>  {R^\lambda}_{\tau \alpha \beta} ={\bar{R}^{\rho}}_{\> \> \> \sigma \mu \nu} + \varepsilon (K^{\rho}_{\nu} \> K_{\mu \sigma}-K^{\rho}_{\mu} \> K_{\nu \sigma})
\end{equation}

\noindent the second is the Codazzi equation:
\begin{equation} \label{2-CodazziEquation}
\gamma _{\kappa }^{\tau }\gamma _{\alpha }^{\mu }\gamma _{\beta }^{\nu } \> {R^{\kappa }}_{\varepsilon \mu \nu } \> n^{\varepsilon }=D_{\alpha }{K_{\beta}}^{\tau}-D_{\beta }{K_{\alpha}}^{\tau}
\end{equation}

\noindent and the third equation is the Ricci equation:
\begin{equation} \label{1-LieDerivExtrinsicCurvC2}
\begin{aligned}
\pounds _n K_{\mu \nu }&=-R_{ \alpha \mu \beta \nu} \> n^{\alpha} n^\beta+K_{\mu \alpha}K_{\nu }{}^{\alpha }-\varepsilon \> a_{\mu }a_{\nu }+D_\mu a_\nu\\
&=-R_{ \alpha \mu \beta \nu} \> n^{\alpha} n^\beta+K_{\mu \alpha}K_{\nu }{}^{\alpha }-\frac{\varepsilon}{\alpha} \> D_{\mu }D_{\nu }\alpha \\
\end{aligned}
\end{equation}

\noindent Note that the right hand side of (\ref{1-LieDerivExtrinsicCurvC2}) is tangent to the hypersurface; if we contract any index with the unit normal vector, the right hand side vanishes. While the Gauss and Codazzi equations do not explicitly refer to the foliation, the Ricci equation depends explicitly on the foliation through the lapse function \(\alpha\), and via the Lie derivative of the extrinsic curvature. Finally, we present the contracted forms of the Gauss and Codazzi equations:
\begin{equation} \label{1-ContractedGaussEquation}
R-2 \> \varepsilon \> n^\mu \> n^\nu \> R_{\mu \nu}={\bar{R}} + \varepsilon (K^{\mu \nu} \> K_{\mu \nu}-K^2)
\end{equation}
\begin{equation} \label{1-CodazziEquationContracted}
\gamma _{\beta }^{\nu }\>R_{\mu \nu } \> n^{\mu }=D_{\alpha }\left({K_{\beta}}^{\alpha}-\gamma^{\alpha}_{\beta } \> K \right)
\end{equation}

\noindent which are typically used to obtain the 3+1 split of the Einstein field equations.

\subsection{The variation of area formulas}
We now introduce the variation of area formulas, which describe the change in the volume (or ``area'') of a hypersurface under infinitesimal displacements. Define the volume \(A\) for some region \(\textbf{Q}\) of a hypersurface \(\Sigma_S\) (\(\textbf{Q} \subset \Sigma_S\)) to be the following:
\begin{equation} \label{1-HypersurfaceArea} 
A := \int_{\textbf{Q}} d\Sigma
\end{equation}

\noindent where \(d\Sigma\) is the hypersurface volume element, which may be written as (\ref{A-ISurfaceElementC2}):
\begin{equation} \label{1-HypersurfaceAreaElement} 
 d\Sigma = \varepsilon \, \sqrt{|\gamma|} \,  d^{3}y
\end{equation}

\noindent with \(\gamma:=\det(\gamma_{ij})\). Now consider an infinitesimal displacement of the surface \(\textbf{Q}\). If the surface \(\textbf{Q}\) is parameterized by the functions \(x^\mu(y)\) (\(y^i\) being coordinates on \(\textbf{Q}\)), then we may characterize the displacement of the surface by adding \(\delta x^\mu(y)\), so that the parameterization of the displaced surface may be described by the functions:
\begin{equation} \label{1-DisplacedHypersurfaceParameterization} 
x{^\prime}{^\mu}(y)=x^\mu(y)+\delta x^\mu(y) .
\end{equation}

\noindent It is helpful to decompose the displacement \(\delta x^\mu(y)\) in the following manner:
\begin{equation} \label{1-DisplacementVectorDecomposition} 
\begin{aligned}
\delta x^\mu &=\delta a \> n^\mu + \delta b^\mu\\
\delta a &:=\varepsilon \, \delta x^\alpha n_\alpha\\
\delta b^\mu &:=\gamma^\mu_\alpha \> \delta x^\alpha
\end{aligned}
\end{equation}

\noindent Under the displacement \(\delta x^\mu\) of the boundary surface, the first variation of area formula may be written as \cite{Frankel} (see Appendix \ref{AppendixVariationAreas} for the derivation):
\begin{equation} \label{1-VariationArea2} 
\delta A = \int_{\textbf{Q}} \delta a \> K  \> d\Sigma+\int_{\partial \textbf{Q}} \delta b^i \> d\sigma_i
\end{equation}

\noindent where \(d\sigma_i\) is the directed surface element on \(\partial \textbf{Q}\). Explicitly, we may write:
\begin{equation} \label{1-VariationArea3} 
\delta A = \int_{\textbf{Q}} \delta x^\mu n_\mu \> K  \> \sqrt{|\gamma|} \, d^3 y+\int_{\partial \textbf{Q}} \> \delta x^\nu \> \gamma^\mu_\nu \left(\frac{\partial y^i}{\partial x^\mu}\right) \> r_i \> \varepsilon_r \> \sqrt{\text{det}|\sigma_{AB}|} d^{2} z
\end{equation}

\noindent where \(\sigma_{AB}\) is the induced metric on \(\partial \textbf{Q}\) with respect to coordinates \(z^A\), and \(r_i\) is the unit normal to \(\partial \textbf{Q}\) tangent to \(\textbf{Q}\), with \(\varepsilon_r=r^i r_i=\pm 1\).

To obtain the second variation of area formula, we introduce a displacement \(\delta \tilde{x}^\mu\) which is in general independent of the displacement \(\delta x^\mu\). We decompose \(\delta \tilde{x}^\mu\) in a manner similar to the decomposition in (\ref{1-DisplacementVectorDecomposition}):
\begin{equation} \label{1-DisplacementVectorDecompositionb} 
\begin{aligned}
\delta \tilde{x}^\mu &=\delta \tilde{a} \> n^\mu + \delta \tilde{b}^\mu\\
\delta \tilde{a} &:=\varepsilon \, \delta \tilde{x}^\alpha n_\alpha\\
\delta \tilde{b}^\mu &:=\gamma^\mu_\alpha \> \delta \tilde{x}^\alpha
\end{aligned}
\end{equation}

\noindent The \textit{second variation of area} is the change in \(\delta A\) with respect to the displacement  \(\delta \tilde{x}^\mu\) (see Appendix \ref{AppendixVariationAreas} for the derivation):
\begin{equation} \label{1-2ndVariationofAreaA2}
\begin{aligned}
\delta_{\tilde{x}} \left(\delta_{x} A\right) =\int_{\textbf{Q}} \biggl(&\delta a \> \delta \tilde{a} \> \varepsilon (1/2)\left(\bar{R}+\varepsilon(K^2-K_{\mu \nu} \> K^{\mu \nu})-R\right) + D_j (\delta \tilde{b}^j \> D_i\delta b^i + \delta a  \> \delta \tilde{b}^j  \>  K)  + \delta \tilde{a} \> D_i \delta b^i \> K \biggr)d\Sigma\\
\end{aligned}
\end{equation}

\noindent where \(\delta_x\) denotes a variation with respect to the displacement \(\delta x^\mu\), and \(\delta_{\tilde{x}}\) denotes a variation with respect to the displacement \(\delta \tilde{x}^\mu\). If we choose \(\delta \tilde{x}^\mu=\delta {x}^\mu\), the the second variation of area formula reduces to:
\begin{equation} \label{1-2ndVariationofAreaB}
\begin{aligned}
\delta^2 A =\int_{\textbf{Q}} \biggl(&(\delta a)^2 \> \varepsilon (1/2)\left(\bar{R}+\varepsilon(K^2-K_{\mu \nu} \> K^{\mu \nu})-R\right) + D_j (\delta {b}^j \> D_i\delta b^i + \delta a  \> \delta {b}^j  \>  K)  + \delta  {a} \> D_i \delta b^i \> K \biggr)d\Sigma\\
\end{aligned}
\end{equation}

\noindent Note that both of these formulas for the second variation of area depend only on the properties of a hypersurface and the way it is embedded in the bulk manifold; they are foliation-independent.

The usefulness of the variation of area formulas will become apparent when we perform the variation of the Gravitational action. In particular, the GHY boundary term is a special case of the \textit{first} variation of area (\ref{1-VariationArea3}) (for \(\delta a=\varepsilon\) and \(\delta b^i=0\)) so that the variation of the GHY term may be written in terms of the second variation of area (\ref{1-2ndVariationofAreaA2}). 

\section{Variation of the gravitational action: Spacetimes wth no spatial boundary}
As stated in the introduction, the reader familiar with the ADM canonical formalism \cite{ADM62,Poisson,Gourgolhoun3+1} should be able to infer the Weiss variation of the gravitational action. In this section, we \textit{explicitly} derive the Weiss variation of the gravitational action in a geometric manner by making use of the first and second variation of area formulas. To simplify the derivation, we first consider the case of spacetimes \textit{without} spatial boundary.
\subsection{The gravitational action}
To simplify the derivation, we consider a globally hyperbolic spacetime \(\mathcal{M}\) that is spatially compact. By this, we mean that \(\mathcal{M}\) has the topology \(\mathbb{R} \times \Sigma\), where \(\Sigma\) is a three dimensional manifold without boundary. Let \(\textbf{U}\subset \mathcal{M}\) be a region of spacetime with the boundary \(\partial \textbf{U}=\Sigma_{I} \cup \Sigma_{F}\) consisting of the smooth, boundaryless spacelike surfaces \(\Sigma_{I}\) and \(\Sigma_{F}\), with \(\Sigma_{I}\) being a surface at early time and \(\Sigma_{F}\) being a surface at late time. The gravitational action on \(\textbf{U}\) is given by:
\begin{equation} \label{3GR-GravitationalAction}
\begin{aligned}
S_{GR}[g^{\mu \nu}]&:=S_{EH}[g^{\mu \nu}]+S_{GHY}\\
\end{aligned}
\end{equation}

\noindent where \(S_{EH}[g^{\mu \nu}]\) is the Einstein-Hilbert action:
\begin{equation} \label{3GR-EinsteinHilbertActionCovCopy}
S_{EH}[g^{\mu \nu}]:=\frac{1}{2 \kappa}\int_{\textbf{U}} \> R \>\sqrt{|g|} \> d^4 x 
\end{equation}

\noindent and \(S_{GHY}\) is the GHY Boundary term:
\begin{equation} \label{3GR-GHYBoundaryTerm}
\begin{aligned}
{S}_{GHY}:=\frac{1}{\kappa}\int_{\partial \textbf{U}} K \> \varepsilon \> \sqrt{|\gamma|}  \> d^3 y .
\end{aligned}
\end{equation}

\noindent At this point, we may recognize the GHY boundary term \({S}_{GHY}\) as a variation of area for the choice \(\delta a=1\) and \(\delta b^i=0\) (cf. equation (\ref{1-DisplacementVectorDecomposition})). This allows one to write the variation of \({S}_{GHY}\) under boundary displacements in terms of the second variation of area formula. This is the key observation that allows us to obtain the variation of \({S}_{GHY}\) under boundary displacements in a geometric manner.

\subsection{Variation of the Einstein-Hilbert action}
We now review the variation of the Einstein-Hilbert action. To obtain the variation of \(S_{EH}[g^{\mu \nu}]\), we begin by adding an infinitesimal, symmetric, rank-2 tensor \(\delta g^{\mu \nu}\) to the inverse metric \(g^{\mu \nu}\). It is convenient to define the following:
\begin{equation} \label{3GR-VariedMetric}
\begin{aligned}
\tilde{g}^{\mu \nu}&:=g^{\mu \nu}+\delta g^{\mu \nu}\\
\tilde{g}_{\mu \nu}&:=g_{\mu \nu}+\delta g_{\mu \nu}\\
\end{aligned}
\end{equation}

\noindent where \(\delta g_{\mu \nu}\) is defined by the following condition:
\begin{equation} \label{3GR-InverseConditionVariation}
\begin{aligned}
&\tilde{g}^{\mu \sigma} \, \tilde{g}_{\sigma \nu}=(g^{\mu \sigma} + \delta g^{\mu \sigma}) (g_{\sigma \nu}+\delta g_{\sigma \nu})=\delta^\mu_\nu \\
& \Rightarrow \>\>\>\>\>\>\>\> g_{\sigma \nu} \>  \delta g^{\mu \sigma} + g^{\mu \sigma} \> \delta g_{\sigma \nu} +  \delta g^{\mu \sigma} \> \delta g_{\sigma \nu} =0
\end{aligned}
\end{equation}

\noindent To first order in \(\delta g^{\mu \nu}\) and \(\delta g_{\mu \nu}\), we obtain the following result:
\begin{equation} \label{3GR-InverseConditionVariationA}
\begin{aligned}
g_{\sigma \nu} \>  \delta g^{\mu \sigma} + g^{\mu \sigma} \> \delta g_{\sigma \nu} \approx 0 \>\>\>\>\>\>\>\> \Rightarrow \>\>\>\>\>\>\>\>  \delta g_{\mu \nu} = - g_{\sigma \mu} \> g_{\tau \nu} \> \delta g^{\sigma \tau}
\end{aligned}
\end{equation}

\noindent Given the above expressions, we can obtain the Taylor expansion of the volume element \(\sqrt{|g|}\) to first order in \(\delta g_{\mu \nu}\):
\begin{equation} \label{3GR-VolumeElementTaylor}
\begin{aligned}
(\sqrt{|g|})|_{g_{\mu \nu}+\delta g_{\mu \nu}}=(\sqrt{|g|})\>|_{g_{\mu \nu}}+\left(\frac{\partial \sqrt{|g|}}{\partial g_{\mu \nu}}\right)\biggr|_{g_{\mu \nu}} \> \delta g_{\mu \nu}+O((\delta g_{\mu \nu}){^2})
\end{aligned}
\end{equation}

\noindent Using the Jacobi formula for the derivative of the determinant, we obtain the following:
\begin{equation} \label{3GR-JacobiFormulaMetricDeterminant}
\begin{aligned}
&\frac{\partial g}{\partial s} =g \> g^{\beta \alpha} \> \frac{\partial g_{\alpha \beta}}{\partial s} \>\>\>\>\> \Rightarrow \>\>\>\>\> \frac{\partial \sqrt{|g|}}{\partial g_{\mu \nu}} = \frac{sgn(g)}{2 \sqrt{|g|}}\frac{\partial g}{\partial g_{\mu \nu}}= \frac{|g| \, g^{\beta \alpha}}{2 \sqrt{|g|}} \> \frac{\partial g_{\alpha \beta}}{\partial g_{\mu \nu}} = \frac{1}{2} \sqrt{|g|} \> g^{\nu \mu}
\end{aligned}
\end{equation}

\noindent where \(\text{sgn}(g)=g/|g|\) picks\footnote{Alternately, we may rewrite this as \(|g| \> = g\, \text{sgn}(g) \).} out the sign of \(g\); \(\text{sgn}(g)=-1\) for a 4 dimensional Lorentzian spacetime. To simplify our expressions, we provide the following definition and expressions for a quantity which we call the variation of the volume element as:
\begin{equation} \label{3GR-VolumeElementVariationA}
\begin{aligned}
\delta \sqrt{|g|} :=\left(\frac{\partial \sqrt{|g|}}{\partial g_{\mu \nu}}\right)\biggr|_{g_{\mu \nu}} \> \delta g_{\mu \nu}= \frac{1}{2} \sqrt{|g|} \> g^{\nu \mu} \> \delta g_{\mu \nu}=-\frac{1}{2} \sqrt{|g|} \> g_{\mu \nu} \> \delta g^{\mu \nu}
\end{aligned}
\end{equation}

\noindent Where (\ref{3GR-InverseConditionVariationA}) has been used in the last equality. To first order, (\ref{3GR-VolumeElementTaylor}) becomes:
\begin{equation} \label{3GR-VolumeElementTaylor2}
\begin{aligned}
(\sqrt{|g|})|_{g_{\mu \nu}+\delta g_{\mu \nu}}=(\sqrt{|g|})\>|_{g_{\mu \nu}}-\frac{1}{2} \sqrt{|g|} \> g_{\mu \nu} \> \delta g^{\mu \nu}+O((\delta g_{\mu \nu}){^2})
\end{aligned}
\end{equation}

We now write the varied Einstein-Hilbert action \(S_{EH}[g^{\mu \nu}]\) to first order in the variations of the inverse metric \(\delta g^{\mu \nu}\) and boundary displacements \(\delta x^\mu\):
\begin{equation} \label{3GR-EinsteinHilbertActionTaylor}
\begin{aligned}
S_{EH}[g^{\mu \nu}+\delta g^{\mu \nu}]&=\frac{1}{2 \kappa}\int_{\textbf{U}^\prime} \> (R \> \sqrt{|g|})|_{g_{\mu \nu}+\delta g_{\mu \nu}} \> d^4 x\\
&=\frac{1}{2 \kappa}\int_{\textbf{U}} \> \left(R \>  \sqrt{|g|}+\delta R \>  \sqrt{|g|}+R \> \delta \sqrt{|g|} +O((\delta g^{\mu \nu}){^2})\right) d^4 x + \frac{1}{2 \kappa}\int_{\partial \textbf{U}} R \> \delta x^\mu \> d^3 \bar{\Sigma}_\mu\\
&=S_{EH}[g^{\mu \nu}] + \frac{1}{2 \kappa}\int_{\textbf{U}} \> \left(\delta R \>  \sqrt{|g|}+R \> \delta \sqrt{|g|} +O((\delta g^{\mu \nu}){^2})\right) d^4 x + \frac{1}{2 \kappa}\int_{\partial \textbf{U}} R \> \delta x^\mu \> d^3 \bar{\Sigma}_\mu
\end{aligned}
\end{equation}

\noindent where the covariant directed surface element \(d^3 \bar{\Sigma}_\mu\) is given by the following expression:
\begin{equation} \label{3GR-CovariantDirectedSurfaceElement}
d^3 \bar{\Sigma}_\mu:=\varepsilon \> n_\mu \> \sqrt{|\gamma|}  \> d^3 y 
\end{equation}

\noindent where \(n^\mu\) is the unit normal vector to the boundary \(\partial \textbf{U}\), \(\varepsilon=n^\mu \> n_\nu= \pm 1\), \(y^i\) are coordinates on \(\partial \textbf{U}\), and \(\gamma\) is the determinant of the induced metric \(\gamma_{ij}\) of the boundary \(\partial \textbf{U}\). Note that the variation of the connection coefficients, \(\delta \Gamma^\alpha_{\mu \nu}\), being defined as a difference between two different connection coefficients (one constructed using the metric \(g_{\mu \nu}\) and the other constructed from the metric \(\tilde{g}_{\mu \nu}\)), transforms as a tensor. One may use this to show that the first order variation of the Ricci scalar is:
\begin{equation}\label{3GR-RicciScalarVariation}
\begin{aligned}
\delta {R}&:=\tilde{g}^{\mu \nu}\>\tilde{R}_{\mu \nu}-g^{\mu \nu}\>{R}_{\mu \nu}\\
&\;=\nabla_\mu (g^{\alpha \beta}\delta \Gamma^\mu_{\beta \alpha}-g^{\alpha \mu}\delta \Gamma^\beta_{\beta \alpha})+{R}_{\mu \nu}\> \delta g^{\mu \nu}\\
&\;=\nabla_\mu ((g^{\alpha \beta}\> \delta^\mu_\nu-g^{\alpha \mu} \> \delta^\beta_\nu )\> \delta \Gamma^\nu_{\beta \alpha})+{R}_{\mu \nu}\> \delta g^{\mu \nu}
\end{aligned}
\end{equation}

\noindent where \(\tilde{g}^{\alpha \beta}\) is the inverse of the metric \(\tilde{g}_{\alpha \beta}:=g_{\alpha \beta}+\delta g_{\alpha \beta}\), and \(\tilde{R}_{\alpha \beta}\) is the Ricci tensor calculated from \(\tilde{g}_{\alpha \beta}\).

The variation of the action, to first order in the variations of the inverse metric \(\delta g^{\mu \nu}\), is given by the following expression:
\begin{equation} \label{3GR-EinsteinHilbertActionVariation}
\begin{aligned}
\delta S_{EH}&:=S_{EH}[g^{\mu \nu}+\delta g^{\mu \nu}]-S_{EH}[g^{\mu \nu}]\\
&\;=\frac{1}{2 \kappa}\int_{\textbf{U}} \biggl(\nabla_\mu ((g^{\alpha \beta}\> \delta^\mu_\nu-g^{\alpha \mu} \> \delta^\beta_\nu )\> \delta \Gamma^\nu_{\beta \alpha})+{R}_{\mu \nu}\> \delta g^{\mu \nu} - \frac{1}{2} \>R \> g_{\mu \nu} \> \delta g^{\mu \nu} \biggr) \sqrt{|g|} \>  d^4 x + \frac{1}{2 \kappa}\int_{\partial \textbf{U}} R \> \delta x^\mu \> d^3 \bar{\Sigma}_\mu\\
\end{aligned}
\end{equation}

\noindent Upon applying the covariant divergence theorem, (\ref{3GR-EinsteinHilbertActionVariation}) becomes:
\begin{equation} \label{3GR-EinsteinHilbertActionVariation2}
\begin{aligned}
\delta S_{EH}&=\frac{1}{2 \kappa}\int_{\textbf{U}} \,{G}_{\mu \nu} \, \delta g^{\mu \nu} \sqrt{|g|} \>  d^4 x + \frac{1}{2 \kappa}\int_{\partial \textbf{U}} \left((g^{\alpha \beta}\> \delta^\mu_\nu-g^{\alpha \mu} \> \delta^\beta_\nu )\> \delta \Gamma^\nu_{\beta \alpha}+R \> \delta x^\mu \right) \> d^3 \bar{\Sigma}_\mu\\
\end{aligned}
\end{equation}

\noindent where \({G}_{\mu \nu}:={R}_{\mu \nu}-\tfrac{1}{2} \, R \, {g}_{\mu \nu} \) is the Einstein tensor.

We now attach a geometric meaning to the boundary term; in doing so, we motivate the use of the GHY boundary term in the gravitational action. First, we place foliations near the boundary surfaces \(\Sigma_I\) and \(\Sigma_F\) (the early time and late time spacelike boundary surfaces) such that the boundary surfaces are contained in the foliation. This allows us to define a unit normal vector field \(n^\mu(x)\) near the boundary surface, so that the covariant derivatives of the unit normal vector field \(n^\mu\) are well-defined. We may choose a coordinate system adapted to the foliation, so that the foliation surfaces correspond to the value of a coordinate \(r\). In the ADM formalism, the unit normal vector and its dual may be written in terms of a lapse function \(\alpha=|g^{00}|^{-1/2}\) and a shift vector \(\beta^i=-\varepsilon \, \alpha^2 \, g^{0i}\):
\begin{equation} \label{3GR-UnitNormalVector}
\begin{aligned}
{} [n^\mu]&=(1/\alpha,[-\beta^i/\alpha])=(1/\alpha,-\beta^1/\alpha,-\beta^2/\alpha,-\beta^3/\alpha)\\
{} [n_\mu]&=(\varepsilon \> \alpha,0,0,0)\\
\end{aligned}
\end{equation}

\noindent Though the respective lapse and shift, \(\alpha\) and \(\beta^i\), form parts of the bulk inverse metric tensor \(g^{\mu \nu}\), they are not physical degrees of freedom--specifying \(\alpha\) and \(\beta^i\) is equivalent to specifying the coordinate system on the spacetime manifold. We may take advantage of this, and impose a coordinate/gauge condition in the neighborhood of the boundary surfaces \(\Sigma_{I}\) and \(\Sigma_F\) so that \(\alpha\) and \(\beta^i\) are unchanged under the variation. Furthermore, we may impose the gauge conditions \(\partial_\mu n^\mu=0\) and \(\partial_\mu \alpha=0\), and require that the variations preserve these conditions. From equation (\ref{3GR-UnitNormalVector}), the requirement that these coordinate conditions be enforced when the variation is carried out may be summarized by following statements:
\begin{equation} \label{3GR-UnitNormalVectorVar}
\begin{aligned}
&\delta n^\mu=0\\
&\delta n_\mu=0\\
\end{aligned}
\end{equation}

\noindent which is equivalent to requiring that \(\delta \alpha=0\) and \(\delta \beta^i=0\). 


We turn to the mean curvature, which may be written in the following manner:
\begin{equation}\label{3GR-MeanCurvature}
\begin{aligned}
K&=\nabla_{\nu} n^\nu=\partial_\nu n^\nu+\Gamma^\nu_{\nu \sigma} n^\sigma\\
&=g^{\alpha \beta} \> \nabla_{\alpha} n_\beta=g^{\alpha \beta} \> \partial_\alpha n_\beta - g^{\alpha \beta} \> \Gamma^\nu_{\alpha \beta} n_\nu
\end{aligned}
\end{equation}

\noindent Under the coordinate conditions (\ref{3GR-UnitNormalVectorVar}), the normal vector and its partial derivatives are unchanged under the variation, so that the variation of the mean curvature takes the form:
\begin{equation}\label{3GR-MeanCurvatureVariation}
\begin{aligned}
\delta K&=\delta \Gamma^\nu_{\nu \sigma} n^\sigma\\
&=- \delta g^{\alpha \beta} \> \Gamma^\nu_{\alpha \beta} n_\nu- g^{\alpha \beta} \> \delta \Gamma^\nu_{\alpha \beta} n_\nu
\end{aligned}
\end{equation}

\noindent We stress that the above formula is only valid if the condition (\ref{3GR-UnitNormalVectorVar}) is satisfied, which is equivalent to requirement that the coordinate/gauge condition \(\partial_\mu n^\mu=0\) and \(\partial_\mu \alpha=0\) is preserved by the variation. Note that if \(\partial_\mu \alpha=0\), then \(a_\mu=-\varepsilon \> D_\nu (\ln{\alpha})=0\) (recall (\ref{1-Acceleration})), the extrinsic curvature tensor, \(K_{\alpha \beta}\) takes the form (equation (\ref{1-ExtrinsicCurvatureTensor})):
\begin{equation}\label{3GR-ExtrinsicCurvatureTensorNormalCoords}
\begin{aligned}
K_{\alpha \beta}=\nabla_\alpha n_\beta=-\Gamma^\nu_{\alpha \beta}\> n_{\nu}
\end{aligned}
\end{equation}

With these conditions in mind, we now examine the boundary terms in the variation of the action (\ref{3GR-EinsteinHilbertActionVariation2}), which takes the form:
\begin{equation} \label{3GR-EinsteinHilbertActionVariationBoundaryTerms}
\begin{aligned}
\delta_{\partial} {S}_{EH} &:=\frac{1}{2 \kappa}\int_{\partial \textbf{U}} \left((g^{\alpha \beta}\> \delta^\mu_\nu-g^{\alpha \mu} \> \delta^\beta_\nu )\> \delta \Gamma^\nu_{\beta \alpha}+R \> \delta x^\mu \right) \varepsilon \> n_\mu \> \sqrt{|\gamma|}  \> d^3 y \\
&\;=\frac{\varepsilon}{2 \kappa}\int_{\partial \textbf{U}} \left(g^{\alpha \beta}\> n_\nu \> \delta \Gamma^\nu_{\beta \alpha}  -  n^{\alpha} \> \delta \Gamma^\sigma_{\sigma \alpha}+R \> \delta x^\mu \> n_\mu \right) \sqrt{|\gamma|}  \> d^3 y \\
\end{aligned}
\end{equation}

\noindent where we have made use of equation (\ref{3GR-CovariantDirectedSurfaceElement}) for \(d^3 \bar{\Sigma}_\mu\), and use the notation \(\delta_{\partial} S\) to pick out boundary terms in the variation \(\delta S\). The boundary terms (\ref{3GR-EinsteinHilbertActionVariationBoundaryTerms}) can be rewritten:
\begin{equation} \label{3GR-EinsteinHilbertActionVariationBoundaryTerms2}
\begin{aligned}
\delta_{\partial} {S}_{EH}&=\frac{\varepsilon}{2 \kappa}\int_{\partial \textbf{U}} \left(-\delta g^{\alpha \beta} \> \Gamma^\nu_{\alpha \beta} n_\nu -  2 \> \delta K+R \> \delta x^\mu \> n_\mu \right) \sqrt{|\gamma|}  \> d^3 y \\
&=\frac{\varepsilon}{2 \kappa}\int_{\partial \textbf{U}} \left(\delta g^{\alpha \beta} \> K_{\alpha \beta}-  2 \> \delta K+R \> \delta x^\mu \> n_\mu \right) \sqrt{|\gamma|}  \> d^3 y \\
&=\frac{\varepsilon}{2 \kappa}\int_{\partial \textbf{U}} \left(\delta \gamma^{\alpha \beta} \> K_{\alpha \beta}-  2 \> \delta K+R \> \delta x^\mu \> n_\mu \right) \sqrt{|\gamma|}  \> d^3 y \\
\end{aligned}
\end{equation}

\noindent where the last equality comes from the definition of the projection tensor \(\gamma^{\alpha \beta}:=g^{\alpha \beta}- \varepsilon \> n^\alpha \> n^\beta\) and the fact that under the gauge condition, the unit normal vectors \(n^\mu\) are held fixed at the boundaries.

If the boundaries are held fixed (if we set \(\delta x^\mu=0\)) and if all the components of the metric tensor are held fixed at the boundary so that \(\delta g^{\mu \nu}|_{\partial \textbf{U}}=0\), the boundary terms reduce to:
\begin{equation} \label{3GR-EinsteinHilbertActionVariationBoundaryTerms3}
\begin{aligned}
\delta_{\partial} {S}_{EH}&=-\frac{\varepsilon}{\kappa}\int_{\partial \textbf{U}} \delta K \> \sqrt{|\gamma|}  \> d^3 y \\
\end{aligned}
\end{equation}

\noindent This shows that if the metric is held fixed at the boundary, and the boundary itself is also held fixed (no boundary displacements), the variation of the GHY boundary term \(\delta S_{GHY}\) cancels out the remaining boundary term in \(\delta_{\partial} {S}_{EH}\).

\subsection{Variation of the GHY boundary term: No boundary displacements}
One might infer from equation (\ref{3GR-EinsteinHilbertActionVariationBoundaryTerms3}) the following expression for the variation of the GHY boundary term:
\begin{equation} \label{3GR-GravitationalActionVariationGHYBoundaryTerm}
\begin{aligned}
\delta {S}_{GHY}&=\frac{\varepsilon}{\kappa}\int_{\partial \textbf{U}} \delta K \> \sqrt{|\gamma|}  \> d^3 y\\
\end{aligned}
\end{equation}

\noindent under the condition that the induced metric \(\gamma_{\mu \nu}\) and its inverse \(\gamma^{\mu \nu}\) are held fixed (\(\delta \gamma_{\mu \nu}=0\) and \(\delta \gamma^{\mu \nu}=0\)). However, the above expression for \(\delta {S}_{GHY}\) will not suffice for the Weiss variation, since the Weiss variation will include variations in \(\gamma_{\mu \nu}\), so that \(\delta \gamma_{\mu \nu} \neq 0\) and \(\delta \gamma^{\mu \nu} \neq 0\).

In this section, we derive the variation of the GHY boundary term for the case where the induced metric (of the boundary \(\partial \textbf{U}\)) \(\gamma_{ij}\) and its inverse \(\gamma^{ij}\) is allowed to vary. We ignore boundary displacements (the boundary \(\partial \textbf{U}\) is held fixed with respect to the coordinates on the spacetime manifold \(\mathcal{M}\)), and compute the variation \(\delta S_{GHY}\) due to changes in the induced metric \(\gamma_{\mu \nu}\) on \(\textbf{U}\); the resulting variation will be denoted by \(\delta_g S_{GHY}\). The variation \(\delta_g S_{GHY}\) takes the form:
\begin{equation} \label{3GR-GHYVariationMetric}
\begin{aligned}
\delta_g S_{GHY}=\frac{ \varepsilon}{\kappa}\int_{\partial \textbf{U}} \left(K{^\prime}  \> \sqrt{|\gamma{^\prime}|} -  K \> \sqrt{|\gamma|}\right) d^3 y
\end{aligned}
\end{equation}

\noindent To first order, we may make use of (\ref{3GR-JacobiFormulaMetricDeterminant}) to write the volume element \(\sqrt{|\gamma{^\prime}|}\) in terms of the induced metric \(\gamma_{ij}\) (which depends on the bulk metric \(g_{\mu \nu}\)) and its variation \(\delta \gamma_{ij}\):
\begin{equation} \label{3GR-GHYVolumeElementExpansion}
\begin{aligned}
\sqrt{|\gamma{^\prime}|}=\sqrt{|\gamma|}-\frac{1}{2} \sqrt{|\gamma|} \> \gamma_{ij} \> \delta \gamma^{ij} .
\end{aligned}
\end{equation}

\noindent Next, we expand \(K{^\prime}\) in the following manner:
\begin{equation} \label{3GR-GHYMeanCurvatureExpansion}
\begin{aligned}
K{^\prime}=K+\delta K
\end{aligned}
\end{equation}

\noindent where \(\delta K\) is given by (\ref{3GR-MeanCurvatureVariation}) (again, we impose the gauge condition: \(\partial_\mu n^\nu=0\) and \(\partial_\mu \alpha=0\)). To first order, the variation \(\delta_g S_{GHY}\) takes the form:
\begin{equation} \label{3GR-GHYVariationMetric2}
\begin{aligned}
\delta_g S_{GHY}&=\frac{ \varepsilon}{2\kappa}\int_{\partial \textbf{U}} \left(2\> \delta K  -  K \> \gamma_{ij} \> \delta \gamma^{ij}\right)\sqrt{|\gamma|} \> d^3 y\\
&=\frac{ \varepsilon}{2\kappa}\int_{\partial \textbf{U}} \left(2\> \delta K  -  K \> \gamma_{\mu \nu} \> \delta \gamma^{\mu \nu}\right)\sqrt{|\gamma|} \> d^3 y\\
\end{aligned}
\end{equation}

\noindent where the last equality comes from making use of the fact that the induced metric \(\gamma_{ij}\) and projection tensor \(\gamma_{\mu \nu}=g_{\mu \nu}-\varepsilon \> n_{\mu}\>n_{\nu}\) are related by a change of basis. 

Note the differences between (\ref{3GR-GHYVariationMetric2}) for \(\delta_g S_{GHY}\) and (\ref{3GR-EinsteinHilbertActionVariationBoundaryTerms2}) for \(\delta_{\partial} {S}_{EH}\). If the boundary \(\partial \textbf{U}\) and its inverse induced metric  \(\gamma^{\mu \nu}\) are held fixed (\(\delta x^\mu=0\) and \(\delta \gamma^{\mu \nu}=0\)), then \(\delta_g S_{GHY}\) and \(\delta_{\partial} {S}_{EH}\) cancel. However, if \(\gamma^{\mu \nu}\) is allowed to vary (\(\delta \gamma^{\mu \nu} \neq 0\)), then \(\delta_g S_{GHY}\) and \(\delta_{\partial} {S}_{EH}\) \textit{do not} cancel.

\subsection{The Weiss variation}
We now discuss the variation induced by the displacement of the boundary \(\partial \textbf{U}\), which we denote by \(\delta_x S_{GHY}\). Earlier, it was remarked after equation (\ref{3GR-GHYBoundaryTerm}) that \({S}_{GHY}\) may be interpreted as the first variation of area (\ref{1-VariationArea2}) for the choices \(\delta {a}= 1\) and \(\delta{b}^i=0\). In particular, compare the following expressions (equations (\ref{1-VariationArea2}) and (\ref{3GR-GHYBoundaryTerm})):
\begin{equation} \label{3GR-GHYBoundaryTermComparison}
\begin{aligned}
\delta A &=\int_{\textbf{Q}} \delta {a} \> K  \> d\Sigma+\int_{\partial \textbf{Q}} \delta {b}^i \> d\sigma_i= \int_{\textbf{Q}}  K  \> \sqrt{|\gamma|} \, d^3 y \\
 {S}_{GHY} &=\frac{1}{\kappa}\int_{\partial \textbf{U}} K \> \varepsilon \> \sqrt{|\gamma|}  \> d^3 y
\end{aligned}
\end{equation}

\noindent where we have set\footnote{We may also get rid of the boundary integral over \(\partial \textbf{Q}\) by requiring that \(\textbf{Q}\) be boundaryless. For instance, if \(\textbf{Q}=\partial \textbf{U}\), then this is indeed the case by the boundary of a boundary principle: \(\partial \partial=0\).} \(\delta {a} = 1\) and \(\delta {b}^i=0\) in the expression for \(\delta A\) (we use \(\textbf{Q}\) to denote general 3d surfaces in \(\mathcal{M}\)). The generalized second variation of area formula (\ref{1-2ndVariationofAreaA2}) describes the change in the first variation of area for a hypersurface under an arbitrary displacement of the hypersurface. Using the generalized second variation of area formula (\ref{1-2ndVariationofAreaA2}), we obtain the following result for the variation of \(\delta A\) (as given in (\ref{3GR-GHYBoundaryTermComparison})) under a displacement\footnote{The displacement \(\delta x^\mu\) corresponds to the \textit{second} variation (\ref{1-DisplacementVectorDecompositionb}) in the variation of area formulas.} \(\delta x^\mu\):
\begin{equation} \label{3GR-VariationGHYDisplacementwithBdy}
\begin{aligned}
\delta_{x} (\delta A)&=\delta_{x}\int_{\textbf{Q}} K \> \sqrt{|\gamma|} \> d^3 y\\
&=\frac{1}{2}\int_{\textbf{Q}} \delta x^\mu \> n_\mu \left[ {^3}{R}+\varepsilon(K^2-K_{\alpha \beta} \> K^{\alpha \beta})-R \> \right] \sqrt{|\gamma|} \> d^3 y + \varepsilon \int_{\partial \textbf{Q}} K \, \delta \tilde{b}^i \, r_i \sqrt{|\sigma|}d^2z\\
\end{aligned}
\end{equation}

\noindent where \(\delta \tilde{b}^i=E^i_\mu \, \delta x^\mu\), \(r^i\) is the unit normal vector to \(\partial \textbf{Q}\), and \(\sqrt{|\sigma|}d^2z\) is the volume element for \(\partial \textbf{Q}\). If \(\textbf{Q}\) has no boundary, then the boundary integral vanishes; this is indeed the case for the surfaces \(\Sigma_I\) and \(\Sigma_F\) that form \(\partial \textbf{U}\). We may make use of the general expression (\ref{3GR-VariationGHYDisplacementwithBdy}) to obtain the variation of the GHY term under displacements of the boundary \(\partial \textbf{U}\):
\begin{equation} \label{3GR-VariationGHYDisplacement}
\begin{aligned}
\kappa \> \delta_x S_{GHY} &=\delta_{x} \int_{\partial \textbf{U}} K \> \varepsilon \> \sqrt{|\gamma|} \> d^3 y =\frac{\varepsilon}{2}\int_{\partial \textbf{U}} \delta x^\mu \> n_\mu \left[ {^3}{R}+\varepsilon(K^2-K_{\alpha \beta} \> K^{\alpha \beta})-R \> \right] \sqrt{|\gamma|} \> d^3 y\\
\end{aligned}
\end{equation}

\noindent The total variation of the GHY boundary term takes the following form:
\begin{equation} \label{3GR-VariationGHYTotalA}
\begin{aligned}
\delta S_{GHY}&=\delta_g S_{GHY}+\delta_x S_{GHY}\\
&=\frac{ \varepsilon}{2\kappa}\int_{\partial \textbf{U}} \left(2\> \delta K  -  K \> \gamma_{\mu \nu} \> \delta \gamma^{\mu \nu}\right)\sqrt{|\gamma|} \> d^3 y + \frac{\varepsilon}{2 \kappa}\int_{\partial \textbf{U}} \delta x^\mu \> n_\mu \left[{^3}{R}+\varepsilon(K^2-K_{\alpha \beta} \> K^{\alpha \beta})-R \> \right] \sqrt{|\gamma|} \> d^3 y\\
\end{aligned}
\end{equation}

\noindent We now combine the expression for \(\delta S_{GHY}\) (\ref{3GR-VariationGHYTotalA}) with \(\delta_\partial S_{EH}\) in (\ref{3GR-EinsteinHilbertActionVariationBoundaryTerms2}) to obtain the full boundary term for \(\delta S_{GR}\), which we denote \(\delta_\partial S_{GR}\):\footnote{Recall that we use the notation \(\delta_{\partial} S\) to mean the boundary terms that appear in the variation \(\delta S\).}
\begin{equation} \label{3GR-VariationGHYTotalB}
\begin{aligned}
\delta_{\partial} S_{GR}&:=\delta_\partial S_{EH}+\delta S_{GHY}\\
&\;=\frac{\varepsilon}{2 \kappa}\int_{\partial \textbf{U}} \left(\delta \gamma^{\alpha \beta} \> K_{\alpha \beta}-  2 \> \delta K+R \> \delta x^\mu \> n_\mu \right) \sqrt{|\gamma|}  \> d^3 y + \frac{ \varepsilon}{2\kappa}\int_{\partial \textbf{U}} \left(2\> \delta K  -  K \> \gamma_{\mu \nu} \> \delta \gamma^{\mu \nu}\right)\sqrt{|\gamma|} \> d^3 y\\
&\>\>\>\>\> \>\>+\frac{\varepsilon}{2 \kappa}\int_{\partial \textbf{U}} \delta x^\mu \> n_\mu \left[{^3}{R}+\varepsilon(K^2-K_{\alpha \beta} \> K^{\alpha \beta})-R \> \right] \sqrt{|\gamma|} \> d^3 y\\
&\;=\frac{ \varepsilon}{2\kappa}\int_{\partial \textbf{U}} \biggl((K_{\mu \nu} - K \> \gamma_{\mu \nu} )\delta \gamma^{\mu \nu} +\delta x^\mu \> n_\mu ({^3}{R}+\varepsilon(K^2-K_{\alpha \beta} \> K^{\alpha \beta}))\biggr) \> \sqrt{|\gamma|} \> d^3 y\\
\end{aligned}
\end{equation}

\noindent The full variation of the gravitational action takes the form:
\begin{equation} \label{3GR-GravitationalActionVariationFull}
\begin{aligned}
\delta S_{GR}&=\frac{1}{2 \kappa}\int_{\textbf{U}} \, G_{\mu \nu} \,  \delta g^{\mu \nu} \sqrt{|g|} \>  d^4 x+\frac{ \varepsilon}{2\kappa}\int_{\partial \textbf{U}} \left((K_{\mu \nu} - K \> \gamma_{\mu \nu} ) \, \delta \gamma^{\mu \nu}\right) \sqrt{|\gamma|} \> d^3 y\\
&\>\>\>\>\> +\frac{ \varepsilon}{2\kappa}\int_{\partial \textbf{U}} \,\delta x^\mu \> n_\mu ({^3}{R}+\varepsilon(K^2-K_{\alpha \beta} \> K^{\alpha \beta}))  \> \sqrt{|\gamma|} \> d^3 y\\
\end{aligned}
\end{equation}

\noindent At this point, we note that if the boundary \(\partial \textbf{U}\) is held fixed, then the projection tensor/induced metric \(\gamma^{\mu \nu}\) must be held fixed in order to obtain a functional derivative of \(S_{GR}[g^{\mu \nu}]\) (see \cite{York1986}); this suggests that the induced metric for hypersurfaces in a foliation of spacetime forms the degrees of freedom for the gravitational field. 

To convert the above result (\ref{3GR-GravitationalActionVariationFull}) to the Weiss form, we define the total change in \(\gamma^{\mu \nu}\) and provide a first-order expression:
\begin{equation} \label{3GR-ChangeInducedMetric}
\begin{aligned}
\Delta \gamma^{\mu \nu}&:= \gamma^{\mu \nu}|_{\partial \textbf{U}{^\prime}} - \gamma^{\mu \nu}|_{\partial \textbf{U}}\\
&\;\approx \delta \gamma^{\mu \nu} + \pounds_{\delta x^\alpha} \gamma^{\mu \nu}\\
\end{aligned}
\end{equation}

\noindent where \(\delta x^\mu\) is the displacement for the boundary. To simplify calculations, we decompose the displacement \(\delta x^\mu\) in the following manner:\footnote{Again, we remind the reader that the displacement \(\delta x^\mu\) corresponds to the \textit{second} variation (\ref{1-DisplacementVectorDecompositionb}) in the variation of area formulas.}
\begin{equation} \label{3GR-DisplacementDecomp}
\begin{aligned}
\delta x^\mu &:= \tilde{a} \> n^\mu + \tilde{b}^\mu\\
\tilde{a}&:=\varepsilon \> \delta x^\mu \> n_\mu\\
\tilde{b}^\mu&:= \gamma^\mu_\nu \> \delta x^\nu=(\delta^\mu_\nu - \varepsilon \> n^\mu \> n_\nu) \delta x^\nu
\end{aligned}
\end{equation}

\noindent The Lie derivative of the induced metric with respect to \(\delta x^\mu\) takes the following form:
\begin{equation} \label{3GR-tPartialDerivativeRank2Tensor}
\begin{aligned}
\pounds_{\delta x^\alpha} \gamma^{\mu \nu} &=\delta x^\alpha \nabla_\alpha \gamma^{\mu \nu} - \gamma^{\alpha \nu}  \> \nabla_\alpha \delta x^\mu - \gamma^{\mu \alpha}  \> \nabla_\alpha \delta x^\nu\\
&=\tilde{a} \pounds_{n} \gamma^{\mu \nu}+\pounds_{\tilde{b}} \gamma^{\mu \nu}+ (n^\mu \> \gamma^{\alpha \nu} \nabla_\alpha \tilde{a} + n^\nu \> \gamma^{\mu \alpha} \nabla_\alpha \tilde{a})
\end{aligned}
\end{equation}

\noindent We combine (\ref{3GR-ChangeInducedMetric}) and (\ref{3GR-tPartialDerivativeRank2Tensor}) to obtain the expression:
\begin{equation} \label{3GR-ChangeInducedMetricB}
\begin{aligned}
\delta \gamma^{\mu \nu} &=\Delta \gamma^{\mu \nu}- \pounds_{\delta x^\alpha} \gamma^{\mu \nu}\\
&=\Delta \gamma^{\mu \nu} - \tilde{a} \pounds_{n} \gamma^{\mu \nu} - \pounds_{\tilde{b}} \gamma^{\mu \nu} - (n^\mu \> \gamma^{\alpha \nu} \nabla_\alpha \tilde{a} + n^\nu \> \gamma^{\mu \alpha} \nabla_\alpha \tilde{a})
\end{aligned}
\end{equation}

\noindent Contracting with \(K_{\mu \nu} - K \> \gamma_{\mu \nu}\), we obtain:
\begin{equation} \label{3GR-ChangeInducedMetric2}
\begin{aligned}
(K_{\mu \nu} - K \> \gamma_{\mu \nu}) \delta \gamma^{\mu \nu} &=(K_{\mu \nu} - K \> \gamma_{\mu \nu})(\Delta \gamma^{\mu \nu} - \tilde{a} \pounds_{n} \gamma^{\mu \nu} - \pounds_{\tilde{b}} \gamma^{\mu \nu})\\
&=(K_{\mu \nu} - K \> \gamma_{\mu \nu})(\Delta \gamma^{\mu \nu} + 2 \tilde{a} \> K^{\mu \nu})-(K_{ij} - K \> \gamma_{ij})  \pounds_{\tilde{b}} \gamma^{ij}\\
&=(K_{\mu \nu} - K \> \gamma_{\mu \nu})\Delta \gamma^{\mu \nu} + 2 \tilde{a} \> (K_{ij} \> K^{ij} - K^2) - (K_{ij} - K \> \gamma_{ij})  \pounds_{\tilde{b}} \gamma^{ij}
\end{aligned}
\end{equation}

\noindent where we have used \((K_{\mu \nu} - K \> \gamma_{\mu \nu})n^\mu=0\) in the first line (since both \(K_{\mu \nu}\) and \(\gamma_{\mu \nu}\) are both tangent to \(\partial \textbf{U}\)), and we have used the expression \(K^{\mu \nu }=-\tfrac{1}{2} \gamma^{\mu}_{\alpha}  \gamma^{\nu}_{\beta} \pounds_n \gamma^{\alpha \beta}\) in the second line. A change of basis has been performed in the third line, and  the last term is justified by the gauge we have chosen, in which the boundary is a surface of a constant value of some coordinate \(r\); this allows us to rewrite the Lie derivative in terms of the coordinate basis \(y^i\).

The boundary term becomes:
\begin{equation} \label{3GR-VariationGHYTotal2}
\begin{aligned}
\delta_{\partial} S_{GR}&=\frac{ \varepsilon}{2\kappa}\int_{\partial \textbf{U}} \biggl((K_{\mu \nu} - K \> \gamma_{\mu \nu} )\Delta \gamma^{\mu \nu} + 2 \varepsilon \> \delta x^\mu \> n_\mu \> (K_{ij} \> K^{ij} - K^2)-(K_{ij} - K \, \gamma_{ij})  \pounds_{\tilde{b}} \gamma^{ij}\\
& \>\>\>\>\> \>\>\>\>\> \>\>\>\>\> \>\>\>\>\> \>\>\> +\delta x^\mu \, n_\mu ({^3}{R}+\varepsilon(K^2-K_{ij} \, K^{ij}))\biggr) \sqrt{|\gamma|} \, d^3 y\\
&=\frac{ \varepsilon}{2\kappa}\int_{\partial \textbf{U}} \biggl((K_{\mu \nu} - K \> \gamma_{\mu \nu} )\Delta \gamma^{\mu \nu} -(K_{ij} - K \, \gamma_{ij})  \pounds_{\tilde{b}} \gamma^{ij} +\delta x^\mu \> n_\mu ({^3}{R}-\varepsilon(K^2-K_{ij} \, K^{ij}))\biggr) \sqrt{|\gamma|} \> d^3 y\\
\end{aligned}
\end{equation}

\noindent We may use the formula\footnote{Note that the covariant derivative \(D^i:=\gamma^{ij} D_j\) on \(\partial \textbf{U}\) satisfies metric compatibility: \(D^k \gamma^{ij}=0\).} \(\pounds_{\tilde{b}} \gamma^{ij}=-D^i \tilde{b}^j-D^j \tilde{b}^i\) with the divergence theorem to rewrite the term containing \(\pounds_{\tilde{b}} \gamma^{ij}\) (this result will be used later on):
\begin{equation} \label{3GR-DivergenceThmMomb}
\begin{aligned}
-\int_{\Sigma} \biggl((K_{ij} - K \> \gamma_{ij})  \pounds_{\tilde{b}} \gamma^{ij} \biggr) \> \sqrt{|\gamma|} \> d^3 y &=2 \int_{\Sigma} \biggl((K_{ij} - K \> \gamma_{ij})  D^i \tilde{b}^j \biggr) \> \sqrt{|\gamma|} \> d^3 y\\
&=-2 \int_{\Sigma} \biggl(D^i(K_{ij} - K \> \gamma_{ij})   \tilde{b}^j \biggr) \> \sqrt{|\gamma|} \> d^3 y + 2 \int_{\partial\Sigma} \biggl((K_{ij} - K \> \gamma_{ij})   \tilde{b}^j \biggr) r^i \> \sqrt{|\sigma|} \> d^2 z\\
\end{aligned}
\end{equation}

\noindent where \(r^i\) is a vector tangent to a hypersurface \(\Sigma\) that forms the unit normal to a 2-surface \(\partial\Sigma\), and \(\sqrt{|\sigma|} \> d^2 z\) is the volume element for \(\partial\Sigma\). Since the surfaces in \(\partial \textbf{U}\) have no boundary, the second term vanishes, so that:
\begin{equation} \label{3GR-VariationGHYTotal3}
\begin{aligned}
\delta_{\partial} S_{GR}=\frac{ \varepsilon}{2\kappa}\int_{\partial \textbf{U}} \biggl(&(K_{\mu \nu} - K \> \gamma_{\mu \nu} )\Delta \gamma^{\mu \nu} - 2 D^i (K_{ij} - K \> \gamma_{ij})  \> \tilde{b}^j + \delta x^\mu \> n_\mu ({^3}{R}-\varepsilon(K^2-K_{ij} \> K^{ij}))\biggr) \> \sqrt{|\gamma|} \> d^3 y
\end{aligned}
\end{equation}

\noindent At this point, we note that if \(\gamma^{\mu \nu}\) form the degrees of freedom for the gravitational field, then we may define the quantity \(P_{\mu \nu}\) to be its conjugate momentum:
\begin{equation} \label{3GR-ConjugateMomentum}
\begin{aligned}
P_{\mu \nu}&:=\frac{\varepsilon}{2 \kappa}\, p_{\mu \nu} \, \sqrt{|\gamma|}\\
\end{aligned}
\end{equation}

\noindent where:
\begin{equation} \label{3GR-ConjugateMomentumTensor}
\begin{aligned}
p_{\mu \nu}&:=K_{\mu \nu} - K \> \gamma_{\mu \nu}\\
\end{aligned}
\end{equation}

\noindent It is straightforward to invert these formulas to obtain the following expression for \(K_{\mu \nu}\):
\begin{equation} \label{3GR-ExtrinsicCurvatureInverted}
\begin{aligned}
K_{\mu \nu}=-2\, \kappa \,  \frac{1}{\sqrt{\gamma}} \left({P}_{\mu \nu} + \frac{1}{2} \, {\gamma}_{\mu \nu} \, {\gamma}^{\alpha \beta} \, {P}_{\alpha \beta }\right)
\end{aligned}
\end{equation}

\noindent Finally, we write out the full variation of the gravitational action in Weiss form:
\begin{equation} \label{3GR-GravitationalActionVariationFullWeiss}
\begin{aligned}
\delta S_{GR}&=\frac{1}{2 \kappa}\int_{\textbf{U}}  \, {G}_{\mu \nu} \,  \delta g^{\mu \nu} \sqrt{|g|} \,  d^4 x+ \frac{ \varepsilon}{2\kappa}\int_{\partial \textbf{U}} \biggl(p_{\mu \nu} \> \Delta \gamma^{\mu \nu} + \left[ n_\mu \left({^3}{R}-\varepsilon(K^2-K_{ij} \, K^{ij})\right)- 2 \, D_\alpha p^{\alpha \beta} \, \gamma_{\mu \beta} \right]\delta x^\mu \biggr) \sqrt{|\gamma|} \, d^3 y\\
\end{aligned}
\end{equation}

\noindent where we reintroduce the definition \(\tilde{b}^\mu = \gamma^\mu_\nu \> \delta x^\nu\). Upon comparing the Weiss form of the variation (\ref{3GR-GravitationalActionVariationFullWeiss}) with that for a generic classical field theory (\ref{3F-VariationAction3A}), we may identify the canonical energy-momentum ``tensor'' for the gravitational field (valid only when evaluated at \(\partial \textbf{U}\)):
\begin{equation} \label{3GR-CanonicalEnergyMomentumTensorGravity}
\begin{aligned}
\mathscr{H}^\mu{_\nu}= \left(\left[({^3}{R}-\varepsilon(K^2-K_{ij} \> K^{ij})) \> n_\nu -2 \> D_\alpha p{^\alpha}{_\nu} \right]n^\mu \,\sqrt{|\gamma|} \right)|_{\partial \textbf{U}}
\end{aligned}
\end{equation}


\noindent We recognize the terms appearing in \(\mathscr{H}^\mu{_\nu}\) as the geometrical parts of the momentum and Hamiltonian constraints:
\begin{equation} \label{3GR-MomConstraint}
\begin{aligned}
&D_\alpha p{^\alpha}{_\nu}=D_{\alpha}\left(K{^\alpha}{_\nu}-\gamma^{\alpha}_{\nu} \> K \right)=\kappa \> \gamma^\mu_\alpha \, T_{\mu \nu}\, n^\nu\\
\end{aligned}
\end{equation}
\begin{equation} \label{3GR-HamConstraint}
\begin{aligned}
&-\varepsilon({^3}{R} - \varepsilon (K^2 - K_{ij} \> K^{ij}))=2 \kappa \> T_{\mu \nu} \, n^\mu \, n^\nu\\
\end{aligned}
\end{equation}

\noindent where \(T_{\mu \nu}\) is the energy-momentum tensor (the source term for the Einstein field equations), which vanishes in the absence of matter. If the above constraints are satisfied (as they should for solutions of the vacuum Einstein field equations), what we would regard as the canonical energy-momentum tensor for the gravitational field vanishes. In the literature, this is often attributed to the reparameterization invariance of GR;\footnote{See, for instance, \cite{Kiefer2007QG}, which contains a detailed discussion of reparameterization invariance in GR. A more general discussion of reparameterization-invariance may be found in \cite{RundVarHJ}; one should keep in mind the distinction between reparameterization invariance and the invariance of the action under coordinate transformations. The difference is that under coordinate transformations, tensors pick up transformation matrices while reparameterizations do not generate transformation matrices; reparameterizations only affect the functional form of the fields, so that their effects only show up in the derivatives of the field.} however, the canonical energy-momentum tensor vanishes \textit{identically} for a reparameterization-invariant theory \cite{RundVarHJ}, while the constraints (\ref{3GR-MomConstraint}) and (\ref{3GR-HamConstraint}) do not. This is because the gravitational action \(S_{GR}\) is not written in a reparameterization-invariant form; one may easily verify that the Lagrangian density does not transform as a reparameterization-invariant Lagrangian density. Furthermore, the presence of the Hamiltonian and momentum constraints suggest that the variables we have chosen to describe the gravitational field (the 10 components of the metric tensor \(g_{\mu \nu}\)) are greater in number than the physical degrees of freedom for the gravitational field; it is well-known that there are only two physical degrees of freedom.\footnote{To see that there are only two physical degrees of freedom, note that the constraints (\ref{3GR-MomConstraint}) and (\ref{3GR-HamConstraint}) consist of four independent equations, which may in principle be used to fix four components of the metric tensor. Specifying the spacetime coordinates (there are four in number) fixes another four components of the metric tensor (the lapse function and shift vector), leaving two components.} If one can identify the physical degrees of freedom (which we define as those that identically satisfy the constraints (\ref{3GR-MomConstraint}) and (\ref{3GR-HamConstraint})), the gravitational action written in terms of the physical degrees of freedom will have a canonical energy momentum tensor that vanishes identically for pure gravity. Thus, if the physical degrees of freedom for the gravitational field are identified,\footnote{The identification of the \textit{physical} degrees of freedom for the gravitational field is a highly nontrivial problem, and to our knowledge, remains an open problem.} one may expect the resulting action to be reparameterization invariant.

We now attempt to construct a Hamilton-Jacobi formulation for gravity. We begin by simplifying \(\delta S_{GR}\) (\ref{3GR-GravitationalActionVariationFullWeiss}); if we choose the variation \(\delta x^\mu\) to be proportional to the unit normal vector \(n^\mu\):
\begin{equation} \label{4QG-NormalBoundaryDisplacement}
\begin{aligned}
\delta x^\mu =n^\mu \, \Delta \tau(y) ,
\end{aligned}
\end{equation}

\noindent we may interpret \(\Delta \tau (y)\) as the amount (measured in proper time) by which the boundary \(\partial \textbf{U}\) is displaced in the normal direction. Upon performing a change of basis to write \(P_{\mu \nu} \> \Delta \gamma^{\mu \nu}=P_{ij} \> \Delta \gamma^{ij}\), the variation (\ref{3GR-GravitationalActionVariationFullWeiss}) simplifies to:
\begin{equation} \label{3GR-GravitationalActionVariationFullWeissB}
\begin{aligned}
\delta S_{GR}&=\frac{1}{2 \kappa}\int_{\textbf{U}}  \, {G}_{\mu \nu} \,  \delta g^{\mu \nu} \sqrt{|g|} \>  d^4 x+ \int_{\partial \textbf{U}} \biggl(P_{ij} \, \Delta \gamma^{ij} -\mathscr{H}_{gf}\,\Delta \tau \biggr) \> d^3 y\\
\end{aligned}
\end{equation}

\noindent where we make use of \(\gamma_{\mu \beta}\, n^\mu=0\), and we define the ``gauge fixed'' Hamiltonian density:
\begin{equation} \label{3GR-HamiltonianDensityGaugeFixed}
\begin{aligned}
\mathscr{H}_{gf}(P_{ij},\gamma^{ij}):=-\frac{1}{2\kappa} \, \biggl[{^3}{R}-\varepsilon\, (K^2-K_{ij} \> K^{ij}) \biggr]\sqrt{|\gamma|}
\end{aligned}
\end{equation}

\noindent where \(K_{ij}\) depends on \(P_{ij}\) via formula (\ref{3GR-ExtrinsicCurvatureInverted}). This coincides with the ADM Hamiltonian\footnote{To obtain the full ADM Hamiltonian, we choose the variation to take the form \(\delta x^\mu=(\alpha \, n^\mu+\beta^\mu) \Delta t\).} in Gaussian normal coordinates, where the spacetime metric \(g_{\mu \nu}\) and its inverse \(g^{\mu \nu}\) satisfies the following on a surface \(\Sigma_t\) of constant \(x^0=t\):
\begin{equation} \label{4QG-GaussianNormalCoordinateMetric}
\begin{aligned}
&g_{00}|_{\Sigma_t}=\varepsilon \>\>\>&\>\>\>& g^{00}|_{\Sigma_t}=\varepsilon \\
&g_{0i}|_{\Sigma_t}=0   \>\>\>&\>\>\>& g^{0i}|_{\Sigma_t}=0\\
&g_{ij}|_{\Sigma_t}=\gamma_{ij} \>\>\>&\>\>\>& g^{ij}|_{\Sigma_t}=\gamma^{ij}
\end{aligned}
\end{equation}

\noindent with \(\Sigma_t\) having no boundary. 

We now attempt to obtain the Hamilton-Jacobi equation for the gravitational field by defining a classical action \(S^{c}_{GR} \llbracket \gamma^{ij}_1,\tau_1;\gamma^{kl}_2,\tau_2 \rrbracket \), where the brackets \(\llbracket \, \rrbracket\) denote functionals over boundary surfaces. The quantities \(\gamma^{ij}_1(y_1)\) and \(\tau_1(y_1)\) are functions over the surfaces \(\Sigma_I\), and the quantities \(\gamma^{ij}_2(y_2)\) and \(\tau_2(y_2)\) are functions over the surfaces \(\Sigma_F\). The quantities \(\gamma^{ij}_1(y_1)\) and \(\gamma^{ij}_2(y_2)\) are the inverse induced metrics on the respective boundary surfaces \(\Sigma_I\) and  \(\Sigma_F\), and the values of \(\tau_1\) and \(\tau_2\) correspond to the time coordinate in Gaussian normal coordinates constructed at the boundary surfaces \(\Sigma_I\) and  \(\Sigma_F\). If \(\tau_1\) is held fixed, the differential of \(S^{c}_{GR} \llbracket \gamma^{ij}_1,\tau_1;\gamma^{kl}_2,\tau_2 \rrbracket \) takes the following form:
\begin{equation} \label{4GR-ClassicalActionDifferential}
\begin{aligned}
\delta S^{c}_{GR}= \frac{\delta S^{c}_{GR}}{\delta \gamma^{ij}_1} \, \delta \gamma^{ij}_1+\frac{\delta S^{c}_{GR}}{\delta \gamma^{kl}_2} \, \delta \gamma^{kl}_2+\frac{\delta S^{c}_{GR}}{\delta \tau_2} \, \delta \tau_2
\end{aligned}
\end{equation}

\noindent The value of the classical action \(S^{c}_{GR}\) coincides with the usual action \(S_{GR}\) for solutions of the vacuum Einstein field equations \(G_{\mu \nu}=0\). Furthermore, the Hamiltonian constraint (\ref{3GR-HamConstraint}) for such solutions suggests that \(\mathscr{H}_{gf}=0\). When comparing the differential (\ref{4GR-ClassicalActionDifferential}) with the variation (\ref{3GR-GravitationalActionVariationFullWeissB}), we obtain the following expressions:
\begin{equation} \label{4GR-HamJacMom}
\begin{aligned}
&\frac{\delta S^{c}_{GR}}{\delta \gamma^{kl}_2} =P_{kl}|_{\Sigma_F}\\
\end{aligned}
\end{equation}

\begin{equation} \label{4GR-HamJac}
\begin{aligned}
&\frac{\delta S^{c}_{GR}}{\delta \tau_2} = \mathscr{H}_{gf} \left(P_{ij}|_{\Sigma_F}  , \, \gamma^{ij}_2\right) = \mathscr{H}_{gf} \left({\delta S^{c}_{GR}}/{\delta \gamma^{ij}_2}, \, \gamma^{ij}_2\right)=0\\
\end{aligned}
\end{equation}

\noindent where we have made a substitution in (\ref{4GR-HamJac}) using (\ref{4GR-HamJacMom}). It is common to identify (\ref{4GR-HamJac}) as the Hamilton-Jacobi equation for (vacuum) GR, as is often done in the literature \cite{Peres1962,RovelliLQG,Rovelli2004}, and it may be shown that (\ref{4GR-HamJac}) define the dynamics for (vacuum) GR \cite{Gerlach1969}. We note that equations (\ref{4GR-HamJac}) do not form the Hamilton-Jacobi equation for GR in the same sense as the Hamilton-Jacobi equation in mechanics; \(\mathscr{H}_{gf}\) is a Hamiltonian \textit{density}, not a Hamiltonian, so (\ref{4GR-HamJac}) should be viewed as a set of local constraints. The Hamilton-Jacobi equation for GR is the following:
\begin{equation} \label{4GR-HamiltonJacobi}
\begin{aligned}
&H_{gf}\left\llbracket {\delta S^{c}_{GR}}/{\delta \gamma^{ij}_2}, \, \gamma^{ij}_2 \right\rrbracket=\int_{\Sigma_F}\mathscr{H}_{gf} \left({\delta S^{c}_{GR}}/{\delta \gamma^{ij}_2}, \, \gamma^{ij}_2\right) d^3y=0\\
\end{aligned}
\end{equation}

\noindent which forms a functional differential equation for \(S^{c}_{GR}\).

\section{Variation of the gravitational action: Spacetimes with spatial boundary}
\subsection{Cylindrical boundaries and the action} \label{SectionGRVar-Boundaries}

\begin{figure} 
\begin{center}
\includegraphics[scale=1.25]{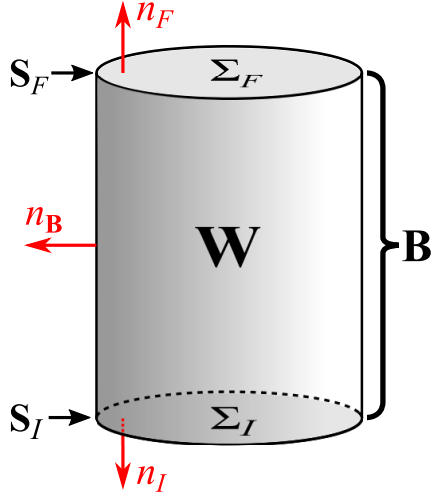}
\end{center}
\caption[Notation for Spacetime Boundary]{An illustration of a cylindrical boundary for a spacetime region \(\textbf{W}\), with boundary \(\partial \textbf{W}=\Sigma_I \cup \textbf{B} \cup \Sigma_F\). The vertical direction is timelike, so that \(\Sigma_I\) and \(\Sigma_F\) are spacelike surfaces of codimension one, and \(\textbf{B}\) is a timelike surface of codimension one. The surfaces \(\textbf{S}_I\) and  \(\textbf{S}_F\) are two dimensional surfaces (assumed to have exclusively spacelike tangent vectors) that form boundaries between \(\Sigma_I\), \(\textbf{B}\) and \(\Sigma_F\). The unit normal vectors (shown in red) are defined to be outward pointing; \(n_I=[n^\mu_I]\) is the unit normal to \(\Sigma_I\), \(n_{\textbf{B}}=[n^\mu_{\textbf{B}}]\) is the unit normal to \(\textbf{B}\), and \(n_F=[n^\mu_F]\) is the unit normal to \(\Sigma_F\).} \label{3GR-BoundaryNotation}
\end{figure}
We now consider a compact region of spacetime \(\textbf{W}\) with a boundary \(\partial \textbf{W}\) that has the cylindrical topology indicated in figure \ref{3GR-BoundaryNotation}. In particular, we choose the boundary \(\partial \textbf{W}\) so that the spacelike portions \(\Sigma_I\) and \(\Sigma_F\) have the topology of a solid 3-sphere,\footnote{In particular, a solid 3-sphere is a subset of \(\mathbb{R}^3\) defined by the condition \(x^2+y^2+z^2 \leq C\), where \(C\) is a constant.} and the timelike portion \(\textbf{B}\) has the topology of the manifold \(\bar{\mathbb{R}} \times S^2\), where \(\bar{\mathbb{R}}\) is a compact subset of \(\mathbb{R}\).\footnote{It must be mentioned that we must \textit{choose} the region \(\textbf{W}\) of a dimensional spacetime so that admits a boundary with such a topology.} For the remainder of this paper, we shall (unless otherwise stated) require that the spacetime boundary has such a topology, and the the surfaces \(\Sigma_I\),  \(\Sigma_F\) and \(\textbf{B}\) maintain their respective signatures. We shall also require that the unit normal vectors be outward pointing, and use the notation established in figure \ref{3GR-BoundaryNotation}; \(n^\mu_I\) is the unit normal to \(\Sigma_I\), \(n^\mu_{\textbf{B}}\) is the unit normal to \(\textbf{B}\), and \(n^\mu_F\) is the unit normal to \(\Sigma_F\). We shall also require that the variations are such that the boundary and the metric \(g_{\mu \nu}\) is held fixed at the 2-surfaces \(\textbf{S}_I\) and \(\textbf{S}_F\), and that the inner products of the unit normal vectors on either side of the 2-surfaces \(\textbf{S}_I\) and \(\textbf{S}_F\) are held fixed; in particular, we hold fixed the following quantities:
\begin{equation} \label{3GR-InnerProductBoundaryVectors}
\begin{aligned}
 \langle n_I,n_{\textbf{B}} \rangle|_{{\textbf{S}}_I}&:=g_{\mu \nu}|_{{\textbf{S}}_I}\> (n^\mu_I)|_{{\textbf{S}}_I} \>(n^\nu_{\textbf{B}})|_{{\textbf{S}}_I}\\
 \langle n_F,n_{\textbf{B}} \rangle|_{{\textbf{S}}_F}&:=g_{\mu \nu}|_{{\textbf{S}}_F}\> (n^\mu_F)|_{{\textbf{S}}_F} \>(n^\nu_{\textbf{B}})|_{{\textbf{S}}_F}\\
\end{aligned}
\end{equation}

\noindent It is convenient to introduce some additional notation for quantities defined on the different portions of the boundary \(\partial \textbf{W}\). For the induced metric \(\gamma_{ij}\) of the boundary \(\partial \textbf{W}\), we write:
\begin{equation} \label{3GR-BoundaryMetric}
\begin{aligned}
h^{ij}_I&:=\gamma^{ij}|_{\Sigma_I}\\
q^{ab}&:=\gamma^{ab}|_{\textbf{B}}\\
h^{ij}_F&:=\gamma^{ij}|_{\Sigma_F}\\
\end{aligned}
\end{equation}

\noindent where lowercase Latin indices \(i,j,...\) from the middle of the alphabet correspond to coordinates on \(\Sigma_I\) and \(\Sigma_F\), and lowercase Latin indices \(a,b,...\) from the beginning of the alphabet correspond to coordinates on \(\textbf{B}\). An underline will be used to indicate quantities defined on the spatial boundary \(\textbf{B}\); for instance, \(\underline{K}_{ij}\) and \(\underline{K}\) denote the respective extrinsic curvature and mean curvature for the spatial boundary surface \(\textbf{B}\).

The gravitational action over the region \(\textbf{W}\) is given by:
\begin{equation} \label{3GR-GravitationalAction2}
\begin{aligned}
S_{GR,\textbf{B}}[g^{\mu \nu}]&:=S_{EH}[g^{\mu \nu}]+S_{GHY}+S_{C}\\
\end{aligned}
\end{equation}

\noindent where \(S_{EH}[g^{\mu \nu}]\) is the Einstein-Hilbert action:
\begin{equation} \label{3GR-EinsteinHilbertActionCovCopy2}
S_{EH}[g^{\mu \nu}]:=\frac{1}{2 \kappa}\int_{\textbf{W}} \> R \>\sqrt{|g|} \> d^4 x .
\end{equation}

\noindent As before, \(S_{GHY}\) is the GHY Boundary term, but it now takes the form:
\begin{equation} \label{3GR-GHYBoundaryTerm2}
\begin{aligned}
{S}_{GHY}=-\frac{1}{\kappa}\int_{\Sigma_F} K \> \sqrt{|h|}  \> d^3 y+\frac{1}{\kappa}\int_{\textbf{B}} \underline{K} \> \sqrt{|q|}  \> d^3 y-\frac{1}{\kappa}\int_{\Sigma_I} K \> \sqrt{|h|}  \> d^3 y.
\end{aligned}
\end{equation}

\noindent The quantity \(S_{C}\) is the ``\textit{corner term},'' which one must include if the boundary \(\partial \textbf{W}\) is nonsmooth \cite{Sorkin1975,Sorkin1981,HartleSorkin1981,Hayward1993,BrillHayward1994}.\footnote{Our construction assumes the 2-surfaces \(\textbf{S}_I\), \(\textbf{S}_F\), and the 3-surfaces \(\Sigma_I\), and \(\Sigma_F\) are all spacelike (in the sense that they have spacelike tangent vectors), and that the surface \(\textbf{B}\) is timelike (in the sense that it has a Lorentzian signature for the induced metric). The boundary terms for the more general case, where the boundaries are nonsmooth and contain null surfaces, may be found in \cite{Parattu2016a,Parattu2016b,Lehner2016,Jubb2017,Chakraborty2017}.} For the boundary surface in figure \ref{3GR-BoundaryNotation}, the corner term takes the following form:
\begin{equation} \label{3GR-HaywardTerm}
\begin{aligned}
{S}_{C}:=\frac{1}{\kappa}\int_{\textbf{S}_I} \eta_I \sqrt{|\sigma|}  \> d^2 z+\frac{1}{\kappa}\int_{\textbf{S}_F} \eta_F \sqrt{|\sigma|}  \> d^2 z
\end{aligned}
\end{equation}

\noindent where:
\begin{equation} \label{3GR-HaywardIntegrand}
\begin{aligned}
 &\eta_I:=\text{arcsinh}\left(\langle n_I,n_{\textbf{B}} \rangle|_{{\textbf{S}}_I}\right)\\
 &\eta_F:=\text{arcsinh}\left(\langle n_F,n_{\textbf{B}} \rangle|_{{\textbf{S}}_F}\right)\\
\end{aligned}
\end{equation}

\noindent with \(\langle n_I,n_{\textbf{B}} \rangle|_{{\textbf{S}}_I}\) and \(\langle n_F,n_{\textbf{B}} \rangle|_{{\textbf{S}}_F}\) being defined by (\ref{3GR-InnerProductBoundaryVectors}). Note that if the unit normal vectors are orthogonal, the corner term vanishes.

\subsection{The Weiss variation}
We now write down the Weiss variation of the action (\ref{3GR-GravitationalAction2}). The earlier result (\ref{3GR-GravitationalActionVariationFullWeiss}) for the variation of the gravitational action may be carried over if we demand that the variations do not reach the 2-surfaces \(\textbf{S}_I\) and \(\textbf{S}_F\), where the boundary \(\partial \textbf{W}\) becomes nonsmooth. In particular, we require that \(\delta x^\mu|_{{\textbf{S}}_I}=0\), \(\delta x^\mu|_{{\textbf{S}}_F}=0\), \(\delta g_{\mu \nu}|_{{\textbf{S}}_I}=0\) and \(\delta g_{\mu \nu}|_{{\textbf{S}}_F}=0\). We also require that \(\langle n_I,n_{\textbf{B}} \rangle|_{{\textbf{S}}_I}\) and \(\langle n_F,n_{\textbf{B}} \rangle|_{{\textbf{S}}_F}\) are held fixed under the variations so that the variation of the corner term vanishes: \(\delta S_C=0\). Under these conditions, the variation of the gravitational action has the same form as (\ref{3GR-GravitationalActionVariationFullWeiss}):
\begin{equation} \label{3GR-GravitationalActionVariationFullWeiss2}
\begin{aligned}
\delta S_{GR,\textbf{B}}&=\frac{1}{2 \kappa}\int_{\textbf{W}} \, {G}_{\mu \nu} \,  \delta g^{\mu \nu} \sqrt{|g|} \>  d^4 x + \frac{ \varepsilon}{2\kappa}\int_{\partial \textbf{W}} \, p_{ij} \> \Delta \gamma^{ij} \, \sqrt{|\gamma|} \, d^3 y\\
&\>\>\>\>\>+ \frac{ \varepsilon}{2\kappa}\int_{\partial \textbf{W}} \biggl(\left({^3}{R}-\varepsilon(K^2-K_{ij} \> K^{ij})\right)  n_\mu  - 2 \> D_\alpha p^{\alpha \beta} \> \gamma_{\mu \beta} \biggr) \delta x^\mu \sqrt{|\gamma|} \, d^3 y\\
\end{aligned}
\end{equation}

We may simplify the above expression by choosing a boundary displacement \(\delta x^\mu\) of the following form:
\begin{equation} \label{3GR-NormalBoundaryDisplacementSpatialPiecewise1}
\begin{aligned}
\delta x^\mu|_{\Sigma_I} &=n^\mu \, \Delta \tau_i(y) &\>\>\>\>\>\>\>\>\>\>&\text{for \(y \in {\Sigma_I}\)}\\
\delta x^\mu|_{\textbf{B}} &=n^\mu \, \Delta r(y) &\>\>\>\>\>\>\>\>\>\>&\text{for \(y \in \textbf{B}\)}\\
\delta x^\mu|_{\Sigma_F} &=n^\mu \, \Delta \tau_f(y) &\>\>\>\>\>\>\>\>\>\>&\text{for \(y \in {\Sigma_F}\)}\\
\end{aligned}
\end{equation}

\noindent where \(\delta x^\mu\) is assumed to be continuous, and vanishes at the 2-surfaces \(\textbf{S}_I\) and \(\textbf{S}_F\):
\begin{equation} \label{4QG-NormalBoundaryDisplacementSpatialPiecewise2}
\begin{aligned}
\delta x^\mu|_{\textbf{S}_I} &=0\\
\delta x^\mu|_{\textbf{S}_F} &=0\\
\end{aligned}
\end{equation}

\noindent The variation of the action may then be written as (\(I\) and \(F\) are labels--they are \textit{not} indices to be summed over):
\begin{equation} \label{3GR-GravitationalActionVariationFullWeiss2b}
\begin{aligned}
\delta S_{GR,\textbf{B}}&=\frac{1}{2 \kappa}\int_{\textbf{W}} \, {G}_{\mu \nu} \, \delta g^{\mu \nu} \sqrt{|g|} \,  d^4 x + \int_{\Sigma_I} \biggl(P^I_{ij} \, \Delta h^{ij}_I  - \mathscr{H}_I \, \Delta \tau_I \biggr)d^3 y\\
&\>\>\>\>\> + \int_{ \textbf{B}} \biggl(\underline{P}_{ab} \> \Delta q^{ab}  - \mathscr{H}_{\textbf{B}} \, \Delta s \biggr) \, d^3 y  + \int_{\Sigma_F} \biggl(P^F_{ij} \> \Delta h^{ij}_F  - \mathscr{H}_F \, \Delta \tau_F \biggr) \, d^3 y\\
\end{aligned}
\end{equation}

\noindent where we have defined the momentum densities (recall that an underline denotes quantities defined on the boundary \(\textbf{B}\)):
\begin{equation} \label{3GR-MomentumDensitySpatial}
\begin{aligned}
{P}^I_{ij}&:=-\frac{1}{2 \kappa}\,({K}_{ij}-{K} \> h^I_{ij})  \, \sqrt{|h_I|}\\
\underline{P}_{ab}&:=\frac{1}{2 \kappa}\,(\underline{K}_{ab}-\underline{K} \> q_{ab})  \, \sqrt{|q|}\\
{P}^F_{ij}&:=-\frac{1}{2 \kappa}\,({K}_{ij}-{K} \> h^F_{ij})  \, \sqrt{|h_F|}\\
\end{aligned}
\end{equation}

\noindent and the Hamiltonian densities:
\begin{equation} \label{3GR-HamiltonianDensitySpatial}
\begin{aligned}
\mathscr{H}_{I}&:=-\frac{1}{2\kappa} \, \biggl[{^3}{R}+ (K^2-K_{ij} \> K^{ij}) \biggr]\sqrt{|h_I|}\\
\mathscr{H}_{\textbf{B}}&:=-\frac{1}{2\kappa} \, \biggl[{^3}\underline{R}-(\underline{K}^2-\underline{K}_{ab} \> \underline{K}^{ab}) \biggr]\sqrt{|q|}\\
\mathscr{H}_{F}&:=-\frac{1}{2\kappa} \, \biggl[{^3}{R}+ (K^2-K_{ij} \> K^{ij}) \biggr]\sqrt{|h_F|}\\
\end{aligned}
\end{equation}

\noindent where \(\mathscr{H}_{I}\) is defined on \(\Sigma_I\) and \(\mathscr{H}_{F}\) is defined on \(\Sigma_F\).

\subsection{Time evolution and the Brown-York quasilocal energy}
We conclude this paper with a brief discussion of time evolution, and a derivation of the Brown-York quasilocal energy. One might imagine time evolution as a displacement of the boundary \(\Sigma_F\) in the future time direction, with an accompanying stretch of the boundary \(\textbf{B}\). To see this, consider the classical action \(S^{c}_{GR,\textbf{B}}\llbracket h_I^{ij};q^{ab};h_F^{ij} \rrbracket\) which has the value of the action functional \(S_{GR,\textbf{B}}[g^{\mu \nu}]\) evaluated on solutions of the vacuum Einstein field equations \(G_{\mu \nu}=0\). The Ricci scalar \(R\) for these solutions vanishes, so that the action takes the following form:
\begin{equation} \label{3GR-ClassicalActionGRSpb}
\begin{aligned}
S^{c}_{GR,\textbf{B}}\llbracket h_I^{ij};q^{ab};h_F^{ij} \rrbracket=-\frac{1}{\kappa}\int_{\Sigma_F} K \sqrt{|h|}  \, d^3 y+\frac{1}{\kappa}\int_{\textbf{B}} \underline{K} \sqrt{|q|}  \, d^3 y-\frac{1}{\kappa}\int_{\Sigma_I} K \sqrt{|h|}  \, d^3 y
\end{aligned}
\end{equation}

\noindent where the extrinsic curvatures for the boundary \(K\) and \(\underline{K}\) are obtained from vacuum solutions of the Einstein field equations consistent with boundary conditions given by the induced boundary metrics \(h_I^{ij}\), \(q^{ab}\) and \(h_F^{ij}\). We note that \(\mathscr{H}_{I}=0\), \(\mathscr{H}_{\textbf{B}}=0\), and \(\mathscr{H}_{F}=0\) on vacuum solutions of the Einstein field equations; from equation (\ref{3GR-GravitationalActionVariationFullWeiss2b}), it follows that the variation of the classical action \(S^{c}_{GR,\textbf{B}}\) satisfies:
\begin{equation} \label{3GR-GravitationalActionVariationFullWeissClassical}
\begin{aligned}
\delta S^{c}_{GR,\textbf{B}}&= \int_{\Sigma_I} P^I_{ij} \, \Delta h^{ij}_I \, d^3 y+ \int_{ \textbf{B}} \underline{P}_{ab} \> \Delta q^{ab} \, d^3 y  + \int_{\Sigma_F} P^F_{ij} \> \Delta h^{ij}_F \, d^3 y\\
\end{aligned}
\end{equation}

\noindent We therefore find that the classical action \(S^{c}_{GR,\textbf{B}}\) is independent of displacements \(\delta x^\mu\) of the boundary in the normal direction (under the condition (\ref{4QG-NormalBoundaryDisplacementSpatialPiecewise2})). A stretch in the boundary \(\textbf{B}\), on the other hand, does affect the value of the classical action \(S^{c}_{GR,\textbf{B}}\); a stretching of the boundary \(\textbf{B}\) corresponds to an increase in its 3-volume, which will affect the integral over \(\textbf{B}\) in (\ref{3GR-GravitationalActionVariationFullWeissClassical}).

To obtain the Brown-York quasilocal energy, we perform a \(3+1\) decomposition of the boundary metric \(q_{ab}\):
\begin{equation} \label{3GR-BoundaryLineElementGeneral}
\begin{aligned}
ds^2=q_{ab} \, dy^a \, dy^b=- (\underline{\alpha}^2+ \sigma_{A B}\, \underline{\beta}^A \, \underline{\beta}^B) \, dt^2 + \sigma_{A B} \, \underline{\beta}^A \, dz^B \, dt + \sigma_{A B}\, dz^A \, dz^B
\end{aligned}
\end{equation}

\noindent where \(\sigma_{A B}\) is the induced metric on constant \(t\) hypersurfaces of \(\textbf{B}\). The volume element may be written as \(\underline{\alpha} \sqrt{|\det(\sigma_{AB})|}\); we may characterize the stretching of the boundary with a change in the lapse function \(\underline{\alpha}\). The inverse metric components \(q^{ab}\) may be written as: 
\begin{equation} \label{3GR-InverseBoundaryMetricADM}
\begin{aligned}
q^{00}&=-\underline{\alpha}^{-2}\\
q^{0A}&= \underline{\alpha}^{-2} \> \underline{\beta}^A\\
q^{AB}&=-\underline{\alpha}^{-2} \> \underline{\beta}^A \> \underline{\beta}^{B}+\sigma^{AB} \\
\end{aligned}
\end{equation}

\noindent In mechanics, the Hamiltonian in Hamilton-Jacobi theory is the derivative of the action with respect to a change in the time parameter \(t\). In the classical action \(S^{c}_{GR,\textbf{B}}\), the lapse function \(\underline{\alpha}\) characterizes the stretch in the boundary, so it plays the role of a time parameter. The analogue to the Hamiltonian is the following functional derivative:
\begin{equation} \label{3GR-GravitationalActionDerivativeq00}
\begin{aligned}
\frac{\underline{\alpha}}{\sqrt{|q|}}\frac{\delta S_{GR,\textbf{B}}}{\delta \underline{\alpha} }&=2\frac{1}{\underline{\alpha}^2\sqrt{|q|}}\left(\frac{\delta S_{GR,\textbf{B}}}{\delta q^{00}}-2 \frac{\delta S_{GR,\textbf{B}}}{\delta q^{0A}} \underline{\beta}^A+\frac{\delta S_{GR,\textbf{B}}}{\delta q^{AB}} \underline{\beta}^A \, \underline{\beta}^B\right)\\
&=2\frac{1}{\underline{\alpha}^2\sqrt{|q|}}\left(P_{00}-2 P_{0A} \underline{\beta}^A+P_{AB} \underline{\beta}^A \, \underline{\beta}^B\right)\\
\end{aligned}
\end{equation}

\noindent We define a unit vector \([\bar{n}^a]:=(1/\underline{\alpha},-\beta^A/\underline{\alpha})\) that is normal to the constant \(t\) surfaces, and tangent to the boundary \(\textbf{B}\). Equation (\ref{3GR-GravitationalActionDerivativeq00}) may then be rewritten:
\begin{equation} \label{3GR-GravitationalActionDerivativeq00B}
\begin{aligned}
\frac{\underline{\alpha}}{\sqrt{|q|}}\frac{\delta S_{GR,\textbf{B}}}{\delta \underline{\alpha} }&=\frac{1}{\kappa} \, \bar{n}^a \, \bar{n}^b \, \underline{P}_{ab}=\frac{1}{\kappa} \, \bar{n}^a \, \bar{n}^b \left(\underline{K}_{ab}-\underline{K} \> q_{ab}\right)
\end{aligned}
\end{equation}

\noindent We may integrate this over a constant \(t\) surface on \(\textbf{B}\) to obtain the following expression for the energy \cite{BrownYork1993}:
\begin{equation} \label{3GR-Brown-YorkQLE}
\begin{aligned}
E:=\frac{1}{\kappa} \int_{\textbf{S}_t} \bar{n}^a \, \bar{n}^b \left(\underline{K}_{ab}-\underline{K} \> q_{ab}\right) \sqrt{|\det(\sigma_{AB})|} \, d^2 z
\end{aligned}
\end{equation}

\noindent This expression is equivalent to the Brown-York quasilocal energy, up to a subtraction term. One may, following \cite{BrownYork1993}, obtain similar integral expressions for a momentum-like and a stress-like quantity from functional derivatives with respect to \(\underline{\beta}^A\) and \(\sigma_{AB}\). Note that, unlike the original result in \cite{BrownYork1993} our expression (\ref{3GR-Brown-YorkQLE}) is independent of the foliation in the bulk manifold \(\textbf{W}\); we do not require that the time coordinate in the bulk be the same as the time coordinate \(t\) on the boundary \(\textbf{B}\), and while the result in \cite{BrownYork1993} requires the condition that the foliation in the bulk consist of hypersurfaces that are orthogonal to the boundary \(\textbf{B}\), our expression (\ref{3GR-Brown-YorkQLE}) for quasilocal energy does not require such a condition.

\begin{acknowledgements}
This article is based on the dissertation work of J. C. Feng. We thank Mr. Mark Selover, Prof. E. C. G. Sudarshan and Prof. G. Bhamathi for their comments and encouragement. J. C. Feng also thanks Prof. Austin Gleeson, Prof. Philip J. Morrison, Prof. Richard D. Hazeltine, and Prof. Robert E. Gompf for their guidance and service as members of his dissertation committee. This work was partially supported by the National Science Foundation under Grant Number PHY-1620610.
\end{acknowledgements}

\bibliography{bibStd,bibGRQG,bibGRADM,bibSptop,bibOurs}

\appendix
\section{The Gauss, Codazzi and Ricci Equations in the bulk coordinate basis}\label{AppendixGaussCodazziRicci}

In this appendix, we establish some formulas (the Gauss, Codazzi and Ricci equations) relating the curvature of a hypersurface \(\Sigma_S\) to the curvature of the bulk manifold \(\mathcal{U}\) (assumed to be Lorentzian). While these equations are well-known, the derivations we have encountered in the physics literature were obtained with the 3+1 formalism in mind; in particular, they assume a spacelike surface embedded in a Lorentzian spacetime. Our formulas apply to both spacelike and timelike surfaces; the distinction is governed by the quantity \(\varepsilon=n^\mu n_\mu=\pm 1\). We assume the existence of a foliation in \(\mathcal{U}\); all foliation-dependent quantities are defined with respect to this foliation. The definitions in section \ref{SecFoliations} will be used here.

\subsection{Some preliminary results}
We begin by deriving a few results that will be useful for us later. Given a vector field \(V^\alpha\) tangent to the hypersurfaces \(\Sigma_S\), i.e. one that satisfies \(V^\alpha n_\alpha=0\) for the unit normal vector field \(n^\alpha\), we infer that \(\nabla_\mu (V^\alpha n_\alpha)=0\). By the product rule, we may obtain the following:
\begin{equation}\label{1-DerivTangentVectDefn}
n_\alpha \nabla_\mu V^\alpha + V^\alpha \nabla_\mu n_\alpha=0 \> \> \> \> \Rightarrow \> \> \> \> n_\alpha \nabla_\mu V^\alpha = - V^\alpha \nabla_\mu n_\alpha
\end{equation}

\noindent If we project the free index of the above onto the hypersurface, we obtain:
\begin{equation}\label{1-DerivTangentVectDefnProj}
\begin{aligned}
\gamma^{\mu}_{\nu}  n_\alpha \nabla_\mu V^\alpha &= - V^\alpha \gamma^{\mu}_{\nu} \nabla_\mu n_\alpha\\
&= - V^\alpha K_{\nu \alpha}\\
\end{aligned}
\end{equation}

Next, work out the expression for the covariant derivative of the induced metric:
\begin{equation}\label{1-CovDerivInducedMetricA}
\begin{aligned}
\nabla_\alpha \gamma_{\mu \nu }&=\nabla_\alpha g_{\mu \nu } - \varepsilon \> \nabla_\alpha (n_{\mu} \> n_{\nu})\\
&=- \varepsilon \> n_{\nu} \> \nabla_\alpha n_{\mu}- \varepsilon \> n_{\mu} \> \nabla_\alpha n_{\nu}
\end{aligned}
\end{equation}

\noindent From equation (\ref{1-ExtrinsicCurvatureTensor}), we have \(\nabla _{\mu} n_{\nu }=K_{\mu \nu }+\varepsilon \> n_{\mu} \> a_{\nu}\), so that we obtain the following expressions for the covariant derivatives of the induced metric and projection tensor:
\begin{equation}\label{1-CovDerivInducedMetric}
\begin{aligned}
\nabla_\alpha \gamma_{\mu \nu }&=- \varepsilon \> n_{\nu} \> K_{\alpha \mu}- n_{\nu} \> n_{\alpha} \> a_{\mu} - \varepsilon \> n_{\mu} \> K_{\alpha \nu} - n_{\mu} \> n_{\alpha} \> a_{\nu}\\
\nabla_\alpha \gamma^{\mu}_{\nu }&=- \varepsilon \> n_{\nu} \> K^{\mu}_{\alpha }- n_{\nu} \> n_{\alpha} \> a^{\mu} - \varepsilon \> n^{\mu} \> K_{\alpha \nu} - n^{\mu} \> n_{\alpha} \> a_{\nu}.\\
\end{aligned}
\end{equation}

\noindent If we project the derivative index onto the hypersurface, we obtain the following formulas for the covariant derivative of the projection tensor:
\begin{equation}\label{1-CovDerivInducedMetricProj}
\begin{aligned}
\gamma^{\alpha}_{\beta} \nabla_\alpha \gamma_{\mu \nu }&=- \varepsilon \> n_{\nu} \> K_{\beta \mu}- \varepsilon \> n_{\mu} \> K_{\beta \nu}\\
\gamma^{\alpha}_{\beta} \nabla_\alpha \gamma^{\mu}_{\nu }&=- \varepsilon \> n_{\nu} \> K^{\mu}_{\beta} - \varepsilon \> n^{\mu} \> K_{\beta \nu}.\\
\end{aligned}
\end{equation}

Finally, we derive a useful expression for \(\nabla_\mu n^\nu \> \nabla_\nu n^\mu\). Using the expression \(\nabla _{\mu} n_{\nu} =K_{\mu \nu }+\varepsilon \> n_{\mu} \> a_{\nu}\), we may write the following:
\begin{equation}\label{1-QuadraticUnitNormalDerivative1}
\begin{aligned}
\nabla_\mu n^\nu \> \nabla_\nu n^\mu &= ({K_{\mu}}^{\nu}+\varepsilon \> n_{\mu} \> a^{\nu})( {K_{\nu}}^{\mu}+\varepsilon \> n_{\nu} \> a^{\mu})\\
&= {K_{\mu}}^{\nu} \> {K_{\nu}}^{\mu}+\varepsilon \> n_{\mu} \> a^{\nu} \> {K_{\nu}}^{\mu}+\varepsilon \> {K_{\mu}}^{\nu} \> n_{\nu} \> a^{\mu}+\varepsilon^2 \>  n_{\mu} \> a^{\nu} \> n_{\nu} \> a^{\mu}\\
\end{aligned}
\end{equation}

\noindent Since \(n^\mu a_\mu=n_\mu a^\mu=0\),\footnote{One may show this using the formula for acceleration (\ref{1-Acceleration}) and the requirement that \(n^\mu\) have unit norm.} and \(n_\nu {K_\mu}^\nu=0\), the last three terms vanish, and we obtain the result:
\begin{equation}\label{1-QuadraticUnitNormalDerivative2}
\nabla_\mu n^\nu \> \nabla_\nu n^\mu = {K_{\mu}}^{\nu} \> {K_{\nu}}^{\mu}
\end{equation}

\subsection{Deriving the Gauss equation}
We are now in a position to derive the Gauss equation, which relates the intrinsic (Riemann) and extrinsic curvature of a hypersurface \(\Sigma_S\) to the curvature of the bulk manifold \(\mathcal{U}\). For some vector \(V^\alpha\) field tangent to the hypersurfaces \(\Sigma_S\), the Riemann tensor for \(\Sigma_S\) satisfies the commutator relation:
\begin{equation} \label{2-HypersurfaceRiemannCommutatorCopy1}
{} D_{\mu}D_{\nu} V^\alpha -D_{\nu} D_{\mu} V^\alpha= {\bar{R}^{\alpha}}_{\> \> \> \beta \mu \nu} \>  V^\beta.
\end{equation}

\noindent We begin by analyzing the first term in the commutator:
\begin{equation} \label{2-HypersurfaceRiemannCommutatorTerm1}
\begin{aligned}
{} D_{\mu}D_{\nu} V^\rho &=\gamma^{\alpha}_{\mu}   \gamma^{\sigma}_{\nu} \gamma^{\rho}_{\tau} \nabla_{\alpha} (\gamma^{\beta}_{\sigma}  \gamma^{\tau}_{\lambda} \nabla_\beta V^\lambda)\\
&=\gamma^{\alpha}_{\mu}   \gamma^{\sigma}_{\nu} \gamma^{\rho}_{\tau} \gamma^{\tau}_{\lambda} \nabla_{\alpha} \gamma^{\beta}_{\sigma} \nabla_\beta V^\lambda+\gamma^{\alpha}_{\mu}   \gamma^{\sigma}_{\nu} \gamma^{\rho}_{\tau} \gamma^{\beta}_{\sigma}  \nabla_{\alpha} \gamma^{\tau}_{\lambda} \nabla_\beta V^\lambda+\gamma^{\alpha}_{\mu}   \gamma^{\sigma}_{\nu} \gamma^{\rho}_{\tau} \gamma^{\beta}_{\sigma}  \gamma^{\tau}_{\lambda} \nabla_{\alpha} \nabla_\beta V^\lambda\\
&=\gamma^{\alpha}_{\mu} \gamma^{\sigma}_{\nu}  \gamma^{\rho}_{\lambda} \nabla_{\alpha} \gamma^{\beta}_{\sigma} \nabla_\beta V^\lambda+\gamma^{\alpha}_{\mu} \gamma^{\rho}_{\tau}\gamma^{\beta}_{\nu}  \nabla_{\alpha} \gamma^{\tau}_{\lambda} \nabla_\beta V^\lambda+\gamma^{\alpha}_{\mu} \gamma^{\rho}_{\lambda} \gamma^{\beta}_{\nu}  \nabla_{\alpha} \nabla_\beta V^\lambda\\
&=\gamma^{\alpha}_{\mu} \gamma^{\beta}_{\nu}  \gamma^{\rho}_{\lambda} \nabla_{\alpha} \gamma^{\sigma}_{\beta} \nabla_\sigma V^\lambda+\gamma^{\alpha}_{\mu} \gamma^{\beta}_{\nu}  \gamma^{\rho}_{\lambda} \nabla_{\alpha} \gamma^{\lambda}_{\tau} \nabla_\beta V^\tau+\gamma^{\alpha}_{\mu} \gamma^{\beta}_{\nu} \gamma^{\rho}_{\lambda}   \nabla_{\alpha} \nabla_\beta V^\lambda\\
\end{aligned}
\end{equation}

\noindent Using equation (\ref{1-CovDerivInducedMetricProj}), we have:
\begin{equation} \label{2-HypersurfaceRiemannCommutatorTerm2}
\begin{aligned}
{} D_{\mu}D_{\nu} V^\rho &=-\varepsilon \> \gamma^{\beta}_{\nu}  \gamma^{\rho}_{\lambda} (n_{\beta} \> K^{\sigma}_{\mu}+ n^{\sigma} \> K_{\mu \beta}) \nabla_\sigma V^\lambda - \varepsilon \> \gamma^{\beta}_{\nu}  \gamma^{\rho}_{\lambda} (n_{\tau} \> K^{\lambda}_{\mu} + n^{\lambda} \> K_{\mu \tau}) \nabla_\beta V^\tau\\
&\> \> \> \> +\gamma^{\alpha}_{\mu} \gamma^{\beta}_{\nu} \gamma^{\rho}_{\lambda} \nabla_{\alpha} \nabla_\beta V^\lambda\\
&=-\varepsilon\>  \gamma^{\beta}_{\nu}  \gamma^{\rho}_{\lambda} \> K_{\mu \beta} \> n^{\sigma} \nabla_\sigma V^\lambda - \varepsilon \> \gamma^{\beta}_{\nu}  \gamma^{\rho}_{\lambda} \> K^{\lambda}_{\mu} \>n_{\tau} \nabla_\beta V^\tau+ \gamma^{\alpha}_{\mu} \gamma^{\beta}_{\nu} \gamma^{\rho}_{\lambda}   \nabla_{\alpha} \nabla_\beta V^\lambda\\
&=-\varepsilon\>K_{\mu \nu} \>  \gamma^{\rho}_{\lambda} \> n^{\sigma} \nabla_\sigma V^\lambda - \varepsilon \> K^{\rho}_{\mu} (\gamma^{\beta}_{\nu} \> n_{\tau} \nabla_\beta V^\tau) + \gamma^{\rho}_{\lambda}  \gamma^{\alpha}_{\mu} \gamma^{\beta}_{\nu} \nabla_{\alpha} \nabla_\beta V^\lambda\\
\end{aligned}
\end{equation}

\noindent and using equation (\ref{1-DerivTangentVectDefnProj}) in the second term on the last line, we have the result:
\begin{equation} \label{2-HypersurfaceRiemannCommutatorTerm3}
\begin{aligned}
{} D_{\mu}D_{\nu} V^\rho &=-\varepsilon\>K_{\mu \nu} \>  \gamma^{\rho}_{\beta} \> n^{\sigma} \nabla_\sigma V^\beta + \varepsilon \> K^{\rho}_{\mu} \> K_{\nu \tau} \> V^\tau +\gamma^{\rho}_{\lambda}  ( \gamma^{\alpha}_{\mu} \gamma^{\beta}_{\nu}  \nabla_{\alpha} \nabla_\beta V^\lambda)\\
\end{aligned}
\end{equation}

\noindent Finally, we plug this result back into the commutator to obtain the result:
\begin{equation} \label{2-HypersurfaceRiemannCommutatorExplicit}
{} [D_{\mu},D_{\nu}] V^\rho=\varepsilon (K^{\rho}_{\mu} \> K_{\nu \tau}-K^{\rho}_{\nu} \> K_{\mu \tau}) V^\tau +\gamma^{\rho}_{\lambda}  ( \gamma^{\alpha}_{\mu} \gamma^{\beta}_{\nu}  [\nabla_{\alpha}, \nabla_\beta] V^\lambda)
\end{equation}

\noindent Upon comparison of the above with equation (\ref{2-HypersurfaceRiemannCommutatorCopy1}) and the expression\([\nabla_\nu,\nabla_\nu] V^\rho={R^\rho}_{\sigma \mu \nu} V^\sigma\), we obtain:
\begin{equation} \label{2-HypersurfaceRiemannComparisonBulk}
\begin{aligned}
{} {\bar{R}^{\rho}}_{\> \> \> \sigma \mu \nu} \>  V^\sigma &=\varepsilon (K^{\rho}_{\mu} \> K_{\nu \tau}-K^{\rho}_{\nu} \> K_{\mu \tau}) V^\tau  +\gamma^{\rho}_{\lambda}  ( \gamma^{\alpha}_{\mu} \gamma^{\beta}_{\nu}  {R^\lambda}_{\tau \alpha \beta} V^\tau)\\
&=-\varepsilon (K^{\rho}_{\nu} \> K_{\mu \sigma}-K^{\rho}_{\mu} \> K_{\nu \sigma}) V^\sigma  +\gamma^{\rho}_{\lambda}  ( \gamma^{\alpha}_{\mu} \gamma^{\beta}_{\nu}  {R^\lambda}_{\tau \alpha \beta } \gamma^{\tau}_{\sigma} V^\sigma)\\
&=-\left(\varepsilon (K^{\rho}_{\nu} \> K_{\mu \sigma}-K^{\rho}_{\mu} \> K_{\nu \sigma}) +\gamma^{\rho}_{\lambda} \gamma^{\tau}_{\sigma}\gamma^{\alpha}_{\mu} \gamma^{\beta}_{\nu}  {R^\lambda}_{\tau \alpha \beta}  \right)V^\sigma\\
\end{aligned}
\end{equation}

\noindent The above expression must hold for any vector field \(V^\alpha\) tangent to the hypersurfaces \(\Sigma_S\), which implies the following:
\begin{equation} \label{2-GaussEquation}
\fbox{$ \gamma^{\rho}_{\lambda} \gamma^{\tau}_{\sigma} \gamma^{\alpha}_{\mu} \gamma^{\beta}_{\nu} \>  {R^\lambda}_{\tau \alpha \beta} ={\bar{R}^{\rho}}_{\> \> \> \sigma \mu \nu} + \varepsilon (K^{\rho}_{\nu} \> K_{\mu \sigma}-K^{\rho}_{\mu} \> K_{\nu \sigma}) $}
\end{equation}

\noindent The formula above is called the \textit{Gauss Equation},\index{Gauss Equation} which establishes an algebraic relationship between the Riemann curvature tensor \({R^\lambda}_{\tau \alpha \beta}\) in the bulk manifold \(\mathcal{U}\) to the respective intrinsic (Riemann) and extrinsic curvature tensors \({\bar{R}^{\rho}}_{\> \> \> \sigma \mu \nu}\) and \(K_{\mu \nu}\) for the hypersurface \(\Sigma_S\).

\subsection{Useful contractions of the Gauss equation}
We now discuss some contractions of the Gauss equation that will appear often in this article. First, we contract the indices \(\rho\) and  \(\sigma\) of the Gauss equation (\ref{2-GaussEquation}) to obtain an expression for the hypersurface Ricci tensor \(\bar{R}_{\mu \nu}:={\bar{R}^{\sigma}}_{\> \> \> \mu \sigma \nu}\):
\begin{equation} \label{2-GaussEquationRicciA1}
\gamma^{\sigma}_{\lambda} \gamma^{\tau}_{\sigma} \gamma^{\alpha}_{\mu} \gamma^{\beta}_{\nu} \>  {R^\lambda}_{\alpha \tau \beta} ={\bar{R}^{\sigma}}_{\> \> \> \mu \sigma \nu} + \varepsilon (K^{\sigma}_{\nu} \> K_{\sigma \mu}-K \> K_{\nu \mu})
\end{equation}

\noindent where we have used the formula for the mean curvature  \(K=\gamma^{\mu \nu} K_{\mu \nu}\). The left-hand side of the above equation becomes:
\begin{equation} \label{2-ProjectedContractedBulkRiemann1}
\begin{aligned}
\gamma^{\sigma}_{\lambda} \gamma^{\tau}_{\sigma} \gamma^{\alpha}_{\mu} \gamma^{\beta}_{\nu} \>  {R^\lambda}_{\alpha \tau \beta} & =\gamma^{\tau}_{\lambda} \gamma^{\alpha}_{\mu} \gamma^{\beta}_{\nu} \>  {R^\lambda}_{\alpha \tau \beta} = (\delta^{\tau}_{\lambda} -\varepsilon \> n^\tau \> n_\lambda) \gamma^{\alpha}_{\mu} \gamma^{\beta}_{\nu} \>  {R^\lambda}_{\alpha \tau \beta}\\
&=\gamma^{\alpha}_{\mu} \gamma^{\beta}_{\nu} \>  R_{\alpha  \beta} -\varepsilon \> n^\tau \> n_\lambda \gamma^{\alpha}_{\mu} \gamma^{\beta}_{\nu} \>  {R^\lambda}_{\alpha \tau \beta}\\
&=\gamma^{\alpha}_{\mu} \gamma^{\beta}_{\nu}( R_{\alpha  \beta} -\varepsilon \>  n^\sigma \> n^\tau \>  {R}_{\sigma \alpha \tau \beta})\\
\end{aligned}
\end{equation}

\noindent It should be noted that \( n^\alpha \> n^\sigma \> n^\tau \> {R}_{\sigma \alpha \tau \beta}=0\) and \(n^\sigma \> n^\tau \> n^\beta \> {R}_{\sigma \alpha \tau \beta}=0\) due to the antisymmetry in the indices \(\alpha\) and \(\sigma\), and the indices \(\tau\) and \(\beta\). It follows that  \(\gamma^\alpha_\mu \> \gamma^\beta_\nu \> n^\sigma \> n^\tau \> {R}_{\sigma \alpha \tau \beta}= \> n^\sigma \> n^\tau \> {R}_{\sigma \mu \tau \nu}\). Equation (\ref{2-GaussEquationRicciA1}) becomes the following:\index{Gauss Equation!\textit{contracted}}
\begin{equation} \label{2-GaussEquationRicciA2}
\fbox{$ \gamma^{\alpha}_{\mu} \gamma^{\beta}_{\nu}  R_{\alpha  \beta} -\varepsilon \>  n^\sigma \> n^\tau \>  {R}_{\sigma \mu \tau \nu} ={\bar{R}}_{\mu \nu} + \varepsilon (K^{\sigma}_{\nu} \> K_{\sigma \mu}-K \> K_{\nu \mu}) $}
\end{equation}

We now contract the remaining two indices (with the induced metric/projection tensor) to get an expression for the Ricci scalar. Again, the contraction of the right hand side of (\ref{2-GaussEquationRicciA2}) is straightforward, but the contraction of the left hand side requires some algebra:
\begin{equation} \label{2-ProjectedContractedBulkRicci1}
\begin{aligned}
\gamma^{\mu \nu} (\gamma^{\alpha}_{\mu} \gamma^{\beta}_{\nu}  R_{\alpha  \beta} -\varepsilon \>  n^\sigma \> n^\tau \> {R}_{\sigma \mu \tau \nu}) &=\gamma^{\mu \nu} \> \gamma^{\alpha}_{\mu} \gamma^{\beta}_{\nu}  R_{\alpha  \beta} -\varepsilon \>  n^\sigma \> n^\tau \> \gamma^{\mu \nu} \>  {R}_{\sigma \mu \tau \nu}\\
&=\gamma^{\alpha \beta} \> R_{\alpha  \beta} -\varepsilon \>  n^\sigma \> n^\tau \> (g^{\mu \nu}-\varepsilon \> n^\mu \> n^\nu) \>  {R}_{\sigma \mu \tau \nu}\\
&=(g^{\alpha \beta}-\varepsilon \> n^\alpha \> n^\beta) \> R_{\alpha  \beta} -\varepsilon \>  n^\sigma \> n^\tau \> {R}_{\sigma \tau}\\
& \> \> \> \> \> +\varepsilon^2 \> n^\sigma \> n^\tau \> n^\mu \> n^\nu \> {R}_{\sigma \mu \tau \nu}\\
&=R-2 \> \varepsilon \> n^\alpha \> n^\beta \> R_{\alpha  \beta}\\
\end{aligned}
\end{equation}

\noindent where we have used \(n^\sigma \> n^\tau \> n^\mu \> n^\nu \> {R}_{\sigma \mu \tau \nu}=0\) (which follows from the antisymmetry of the first two and last two indices) in the third equality. The contracted form of (\ref{2-GaussEquationRicciA2}) is:
\begin{equation} \label{2-GaussEquationRicciB1}
\fbox{$ R-2 \> \varepsilon \> n^\mu \> n^\nu \> R_{\mu \nu}={\bar{R}} + \varepsilon (K^{\mu \nu} \> K_{\mu \nu}-K^2) $}
\end{equation}

\noindent We may obtain an alternate expression by writing \(R_{\mu \nu} \> n^\mu \> n^\nu\) in terms of the extrinsic curvature. From the commutator formula for the Riemann tensor, we have the following expression:
\begin{equation} \label{2-CommutatorRicciUnitNormals}
\begin{aligned}
R_{\mu \nu }n^{\mu }n^{\nu }&=n^{\nu }\left(\nabla _{\sigma }\nabla _{\nu }-\nabla _{\nu }\nabla _{\sigma }\right)n^{\sigma }\\
&=n^{\nu }\nabla _{\sigma }\nabla _{\nu }n^{\sigma }-n^{\nu } \nabla _{\nu }\nabla _{\sigma }n^{\sigma }\\
\end{aligned}
\end{equation}

\noindent Using the Leibniz rule, we may write:
\begin{equation} \label{2-CommutatorRicciUnitNormalsLeibnizTerm}
\begin{aligned}
\nabla _{\sigma }\left(n^{\nu }\nabla _{\nu }n^{\sigma }\right)=\nabla _{\sigma }n^{\nu }\nabla _{\nu }n^{\sigma }+n^{\nu }\nabla _{\sigma }\nabla
_{\nu }n^{\sigma }\\
\nabla _{\nu }\left(n^{\nu }\nabla _{\sigma }n^{\sigma }\right)=\nabla _{\nu }n^{\nu }\nabla _{\sigma }n^{\sigma }+n^{\nu }\nabla _{\nu }\nabla
_{\sigma }n^{\sigma }\\
\end{aligned}
\end{equation}

\noindent The above allows us to rewrite equation (\ref{2-CommutatorRicciUnitNormals}) as:
\begin{equation} \label{2-CommutatorRicciUnitNormals2}
\begin{aligned}
R_{\mu \nu }n^{\mu }n^{\nu }&=\nabla _{\sigma }\left(n^{\nu}\nabla _{\nu }n^{\sigma }\right)-\nabla _{\sigma }n^{\nu }\nabla _{\nu }n^{\sigma }-\nabla _{\nu }\left(n^{\nu }\nabla _{\sigma }n^{\sigma }\right)+\nabla
_{\nu }n^{\nu }\nabla _{\sigma }n^{\sigma }\\
&=\nabla _{\sigma }a^{\sigma }-{K_{\sigma}}^{\nu} \> {K_{\nu}}^{\sigma}-\nabla _{\nu }\left(n^{\nu } \> K\right)+K^2\\
&=K^2-K^{\mu \nu} \> K_{\mu \nu}+\nabla _{\sigma}\left(a^{\sigma }-n^{\sigma } \> K\right)\\
\end{aligned}
\end{equation}

\noindent where we have used the expression \(a^\sigma=n^{\nu}\nabla _{\nu }n^{\sigma }\) for acceleration (\ref{1-Acceleration}), the expression \(K=\nabla_\mu n^\mu\) for mean curvature (\ref{1-MeanCurvatureA}), and equation (\ref{1-QuadraticUnitNormalDerivative2}) in the second line. We may plug this back into equation (\ref{2-GaussEquationRicciB1}) to obtain the following expression for the bulk Ricci scalar:\index{Gauss Equation!\textit{contracted}}
\begin{equation} \label{2-GaussEquationRicciB2a}
\fbox{$ R = {\bar{R}} + \varepsilon (K^2-K^{\mu \nu} \> K_{\mu \nu}) +2 \> \varepsilon \> \nabla _{\sigma}\left(a^{\sigma }-n^{\sigma } \> K\right) $}
\end{equation}

\subsection{The Codazzi equation}
An alternate way of deriving the Gauss equation in the previous section is to project all the indices of the bulk curvature tensor onto the hypersurface; in doing so, we obtain a tensorial equation with all indices tangent to the hypersurface. However, one may choose instead to project some indices of the curvature tensor onto the hypersurface, and to contract the remaining indices with that of the unit normal vector; this procedure also yields tensor equations tangent to the hypersurface, this time of lower rank.

In this section, we obtain a differential relationship between the Riemann curvature tensor \({R^\lambda}_{\tau \alpha \beta}\) in the bulk manifold \(\mathcal{U}\) and the extrinsic curvature tensor \(K_{\mu \nu}\) for the hypersurface \(\Sigma_S\) by contracting one index of \({R^\lambda}_{\tau \alpha \beta}\) with the unit normal vector and applying the projection operator to the remaining indices. Using the commutator formula \([\nabla_\mu,\nabla_\nu]V^\alpha={R^\alpha}_{\beta \mu \nu} \>  V^\beta\) for the bulk Riemann curvature tensor, we may write the following:\footnote{Due to the symmetries of the Riemann tensor, we may write this without loss of generality.}
\begin{equation} \label{2-ProjectionThreeIndicesA}
\begin{aligned}
\gamma _{\kappa }^{\tau }\gamma _{\alpha }^{\mu }\gamma _{\beta }^{\nu } \> {R^{\kappa }}_{\varepsilon \mu \nu } \> n^{\varepsilon }&=\gamma _{\kappa
}^{\tau }\gamma _{\alpha }^{\mu }\gamma _{\beta }^{\nu }\left(\nabla _{\mu }\nabla _{\nu }-\nabla _{\nu }\nabla _{\mu }\right)n^{\kappa }\\
&=\gamma _{\kappa
}^{\tau }\gamma _{\alpha }^{\mu }\gamma _{\beta }^{\nu }\left(\nabla _{\mu }(\nabla _{\nu }n^{\kappa })-\nabla _{\nu }(\nabla _{\mu }n^{\kappa })\right)\\
&=\gamma^{\tau \sigma}\gamma _{\alpha }^{\mu }\gamma _{\beta }^{\nu }\left(\nabla _{\mu }(\nabla _{\nu }n_{\sigma})-\nabla _{\nu }(\nabla _{\mu }n_{\sigma })\right)\\
\end{aligned}
\end{equation}

\noindent where we have made use of metric compatibility \(\nabla_\mu g^{\kappa \sigma}=0\) to lower the index on the normal vector in the last equality. We plug in the expression \(\nabla _{\mu} n_{\nu }=K_{\mu \nu }+\varepsilon \> n_{\mu} \> a_{\nu}\) to obtain the following:
\begin{equation} \label{2-ProjectionThreeIndicesB}
\begin{aligned}
\gamma _{\kappa }^{\tau }\gamma _{\alpha }^{\mu }\gamma _{\beta }^{\nu } \> {R^{\kappa }}_{\varepsilon \mu \nu } \> n^{\varepsilon }&=\gamma^{\tau \sigma}\gamma _{\alpha }^{\mu }\gamma _{\beta }^{\nu }\left(\nabla _{\mu }(K_{\nu \sigma }+\varepsilon \> n_{\nu} \> a_{\sigma})-\nabla _{\nu }(K_{\mu \sigma }+\varepsilon \> n_{\mu} \> a_{\sigma})\right)\\
&=\gamma^{\tau \sigma}\gamma _{\alpha }^{\mu }\gamma _{\beta }^{\nu }(\nabla _{\mu }K_{\nu \sigma }+\varepsilon \>\nabla _{\mu } n_{\nu} \> a_{\sigma}+\varepsilon \> n_{\nu} \>\nabla _{\mu } a_{\sigma}-\nabla _{\nu }K_{\mu \sigma }\\
&\>\>\>\>\>-\varepsilon \> \nabla _{\nu } n_{\mu} \> a_{\sigma}-\varepsilon \> n_{\mu} \>\nabla _{\nu } a_{\sigma})\\
\end{aligned}
\end{equation}

\noindent Using \(\gamma _{\alpha }^{\mu }=0\) and the formula (\ref{1-ExtrinsicCurvatureTensor}) for the extrinsic curvature \(K_{\alpha \beta}=\gamma _{\alpha }^{\mu }\gamma _{\beta }^{\nu }\nabla _{\mu } n_{\nu}\), we have:
\begin{equation} \label{2-ProjectionThreeIndicesC}
\begin{aligned}
\gamma _{\kappa }^{\tau }\gamma _{\alpha }^{\mu }\gamma _{\beta }^{\nu } \> {R^{\kappa }}_{\varepsilon \mu \nu } \> n^{\varepsilon }&=D_{\alpha }{K_{\beta}}^{\tau}-D_{\beta }{K_{\alpha}}^{\tau}+\varepsilon \> a^{\tau} \> (K_{\alpha \beta}- K_{\beta \alpha})
\end{aligned}
\end{equation}

\noindent Where \(D_\mu\) is the hypersurface covariant derivative. Making use of the symmetry \(K_{\mu \nu}=K_{\nu \mu}\), last two terms cancel and we arrive at the Codazzi equation:\index{Codazzi Equation}
\begin{equation} \label{2-CodazziEquation2}
\fbox{$ \gamma _{\kappa }^{\tau }\gamma _{\alpha }^{\mu }\gamma _{\beta }^{\nu } \> {R^{\kappa }}_{\varepsilon \mu \nu } \> n^{\varepsilon }=D_{\alpha }{K_{\beta}}^{\tau}-D_{\beta }{K_{\alpha}}^{\tau} $}
\end{equation}

\noindent We may sum over the indices \(\alpha\) and \(\tau\) to obtain the following expression for the left hand side:
\begin{equation} \label{2-CodazziEquationLHSContract}
\begin{aligned}
\gamma _{\kappa }^{\alpha }\gamma _{\alpha }^{\mu }\gamma _{\beta }^{\nu }\>{R^{\kappa }}_{\varepsilon \mu \nu } \> n^{\varepsilon }&=\gamma _{\kappa }^{\mu }\gamma _{\beta }^{\nu }\>{R^{\kappa }}_{\varepsilon \mu \nu } \> n^{\varepsilon }=\delta _{\kappa }^{\mu }\gamma _{\beta }^{\nu }\>{R^{\kappa }}_{\varepsilon \mu \nu } \> n^{\varepsilon }-\varepsilon \> \gamma _{\beta }^{\nu } \> n^{\mu} \> n_{\kappa }\> {R^{\kappa}}_{\varepsilon \mu \nu } \> n^{\varepsilon }\\
&=\gamma _{\beta }^{\nu }\>{R^{\kappa }}_{\varepsilon \kappa \nu } \> n^{\varepsilon }-\varepsilon \> \gamma _{\beta }^{\nu } \>  R_{\kappa \varepsilon \mu \nu }  \> n^{\kappa }\> n^{\varepsilon } \> n^{\mu} \\
&=\gamma _{\beta }^{\nu }\>R_{\mu \nu } \> n^{\mu }\\
\end{aligned}
\end{equation}

\noindent where we have used \(R_{\kappa \varepsilon \mu \nu } \> n^{\kappa }\> n^{\varepsilon }=0\) in the last line. The contracted Codazzi equation takes the form (using metric compatibility \(D_{\alpha }\gamma^{\alpha}_{\beta }=0\) on the second term):\index{Codazzi Equation!\textit{contracted}}
\begin{equation} \label{2-CodazziEquationContracted}
\fbox{$ \gamma _{\beta }^{\nu }\>R_{\mu \nu } \> n^{\mu }=D_{\alpha }\left({K_{\beta}}^{\alpha}-\gamma^{\alpha}_{\beta } \> K \right) $}
\end{equation}

\subsection{The Lie derivative of extrinsic curvature: The Ricci equation}
We now derive the Ricci equation, which relates the Lie derivative of the extrinsic curvature to the bulk Riemann curvature tensor. We shall take an indirect approach, and begin by computing the Lie derivative of the extrinsic curvature with respect to the unit normal vector field. The Lie derivative of the extrinsic curvature is:
\begin{equation} \label{A-LieDerivExtrinsicCurv}
\pounds _nK_{\mu \nu }=n^{\alpha }\nabla _{\alpha }K_{\mu \nu }+K_{\alpha \nu }\nabla _{\mu }n^{\alpha }+K_{\mu \alpha }\nabla _{\nu }n^{\alpha
}
\end{equation}

\noindent We begin by computing the last two terms:
\begin{equation} \label{A-LastTwoinLieDerivExtrinsicCurvA}
\begin{aligned}
K_{\alpha \nu }\nabla _{\mu }n^{\alpha }+K_{\mu \alpha }\nabla _{\nu }n^{\alpha }&=\left(\nabla _{\alpha }n_{\nu }-\varepsilon \>  n_{\alpha }a_{\nu
}\right)\nabla _{\mu }n^{\alpha }+K_{\mu \alpha }K_{\nu }{}^{\alpha }+\varepsilon \>  K_{\mu \alpha } n_{\nu }a^{\alpha }\\
&=\left(\nabla _{\alpha }n_{\nu }-\varepsilon  \> n_{\alpha }a_{\nu
}\right)\nabla _{\mu }n^{\alpha }+K_{\mu \alpha }K_{\nu }{}^{\alpha }\\
&\>\>\>\>\>+\varepsilon  \left(\nabla _{\mu }n_{\alpha }-\varepsilon  \> n_{\mu }a_{\alpha}\right)n_{\nu }a^{\alpha }\\
\end{aligned}
\end{equation}

\noindent We make use of \(n^{\alpha }\nabla _{\mu }n_{\alpha }=n_{\alpha }\nabla _{\mu }n^{\alpha }=0\) (which follows from \(n^\alpha n_\alpha=\varepsilon=\pm1\)) to obtain the following result:
\begin{equation} \label{A-LastTwoinLieDerivExtrinsicCurvB}
K_{\alpha \nu }\nabla _{\mu }n^{\alpha }+K_{\mu \alpha }\nabla _{\nu }n^{\alpha }=(\nabla _{\alpha }n_{\nu })(\nabla _{\mu }n^{\alpha })+K_{\mu \alpha
}K_{\nu }{}^{\alpha }+\varepsilon  \> n_{\nu } \>a^{\alpha } \> \nabla _{\mu }n_{\alpha }-n_{\mu } \> n_{\nu } \> a_{\alpha } \>a^{\alpha }
\end{equation}

We now turn our attention to the first term in equation (\ref{A-LieDerivExtrinsicCurv}), which is the directional derivative of the extrinsic curvature:
\begin{equation} \label{A-DirectDerivExtrinsicCurvA1}
\begin{aligned}
n^{\alpha }\nabla _{\alpha }K_{\mu \nu }&=n^{\alpha }\nabla _{\alpha }\left(\nabla _{\mu }n_{\nu }-\varepsilon \>  n_{\mu }a_{\nu }\right)=n^{\alpha}\nabla _{\alpha }\nabla _{\mu }n_{\nu }-\varepsilon \> a_{\mu }a_{\nu }-\varepsilon \>  n_{\mu }n^{\alpha }\nabla _{\alpha }a_{\nu }\\
\end{aligned}
\end{equation}

\noindent From the definition of the projection tensor, we may write \(\varepsilon \> n_\mu n^\alpha=\delta^\alpha_\mu -\gamma^\alpha_\mu\). We use this to expand the last term:
\begin{equation} \label{A-DirectDerivExtrinsicCurvA2}
\begin{aligned}
n^{\alpha }\nabla _{\alpha }K_{\mu \nu }&=n^{\alpha}\nabla _{\alpha }\nabla _{\mu }n_{\nu }-\varepsilon \> a_{\mu }a_{\nu }-\delta^\alpha_\mu \> \nabla _{\alpha }a_{\nu } +\gamma^\alpha_\mu \> \nabla _{\alpha }a_{\nu }\\
&=n^{\alpha}\nabla _{\alpha }\nabla _{\mu }n_{\nu }-\varepsilon \> a_{\mu }a_{\nu } - \nabla _{\mu }a_{\nu } +\gamma^\alpha_\mu \> \nabla _{\alpha }a_{\nu }\\
&=n^{\alpha}\nabla _{\alpha }\nabla _{\mu }n_{\nu }-\varepsilon \> a_{\mu }a_{\nu } - \nabla _{\mu } (n^\alpha \> \nabla_{\alpha} n_{\nu}) +\gamma^\alpha_\mu \> \nabla _{\alpha }a_{\nu }\\
&=n^{\alpha}\nabla _{\alpha }\nabla _{\mu }n_{\nu }-\varepsilon \> a_{\mu }a_{\nu } - (\nabla _{\mu } n^\alpha) \> (\nabla_{\alpha} n_{\nu}) -  n^\alpha \>  \nabla _{\mu }\nabla_{\alpha} n_{\nu} +\gamma^\alpha_\mu \> \nabla _{\alpha }a_{\nu }\\
&=n^{\alpha}(\nabla _{\alpha }\nabla _{\mu } - \nabla _{\mu }\nabla_{\alpha})n_{\nu }-\varepsilon \> a_{\mu }a_{\nu } - (\nabla _{\mu } n^\alpha) \> (\nabla_{\alpha} n_{\nu})+\gamma^\alpha_\mu \> \nabla _{\alpha }a_{\nu }\\
\end{aligned}
\end{equation}

\noindent At this point, we recognize the first two terms in the last line as the contraction of the Riemann curvature tensor with two unit normal vectors; it is straightforward to show that \(n^{\alpha}[\nabla _{\alpha },\nabla _{\mu }] n_{\nu }=n^\alpha R_{\nu \beta \alpha \mu} n^\beta\). The directional derivative of the extrinsic curvature becomes:
\begin{equation} \label{A-DirectDerivExtrinsicCurvB}
\begin{aligned}
n^{\alpha }\nabla _{\alpha }K_{\mu \nu }=-R_{ \alpha \mu \beta \nu} \> n^{\alpha} n^\beta-\varepsilon \> a_{\mu }a_{\nu } - (\nabla _{\mu } n^\alpha) \> (\nabla_{\alpha} n_{\nu})+\gamma^\alpha_\mu \> \nabla _{\alpha }a_{\nu }\\
\end{aligned}
\end{equation}

\noindent We plug equations (\ref{A-LastTwoinLieDerivExtrinsicCurvB}) and (\ref{A-DirectDerivExtrinsicCurvB}) into the formula for the Lie Derivative of the extrinsic curvature (\ref{A-LieDerivExtrinsicCurv}) to obtain:
\begin{equation} \label{A-LieDerivExtrinsicCurvB}
\begin{aligned}
\pounds _nK_{\mu \nu }&=-R_{ \alpha \mu \beta \nu} \> n^{\alpha} n^\beta-\varepsilon \> a_{\mu }a_{\nu } - (\nabla _{\mu } n^\alpha) \> (\nabla_{\alpha} n_{\nu})+\gamma^\alpha_\mu \> \nabla _{\alpha }a_{\nu }\\
&\>\>\>\>\>+(\nabla _{\alpha }n_{\nu })(\nabla _{\mu }n^{\alpha })+K_{\mu \alpha}K_{\nu }{}^{\alpha }+\varepsilon  \> n_{\nu } \>a^{\alpha } \> \nabla _{\mu }n_{\alpha }-n_{\mu } \> n_{\nu } \> a_{\alpha } \>a^{\alpha }\\
&=-R_{ \alpha \mu \beta \nu} \> n^{\alpha} n^\beta+K_{\mu \alpha}K_{\nu }{}^{\alpha }-\varepsilon \> a_{\mu }a_{\nu }+\gamma^\alpha_\mu \> \nabla _{\alpha }a_{\nu }+\varepsilon  \> n_{\nu } \>a^{\alpha } \> \nabla _{\mu }n_{\alpha }\\
&\>\>\>\>\>-n_{\mu } \> n_{\nu } \> a_{\alpha } \>a^{\alpha }
\end{aligned}
\end{equation}

\noindent We may simplify this further by working out the hypersurface covariant derivative \(D_\mu a_\nu\) of the acceleration \(a_\nu\), which is tangent to the hypersurfaces \(\Sigma_S\); in doing so, we will recognize that several terms in the above expression (\ref{A-LieDerivExtrinsicCurvB}) combine. Explicitly, we have:
\begin{equation} \label{A-HypersurfaceCovariantAccelerationA1}
\begin{aligned}
D_\mu a_\nu &=\gamma^\alpha_\mu \gamma^\beta_\nu \> \nabla_\alpha a_\beta=\gamma^\alpha_\mu \nabla_\alpha a_\nu-\varepsilon \> n^\beta n_\nu\> \gamma^\alpha_\mu   \nabla_\alpha a_\beta\\
&=\gamma^\alpha_\mu \nabla_\alpha a_\nu-\varepsilon \> n_\nu\> \gamma^\alpha_\mu ( n^\beta \> \nabla_\alpha a_\beta)\\
\end{aligned}
\end{equation}

\noindent Since \(n^\beta \> a_\beta=0\), we may write \(\nabla_\alpha(n^\beta \> a_\beta)=0\), and it follows that \(n^\beta \> \nabla_\alpha a_\beta=- a_\beta \> \nabla_\alpha n^\beta=- a^\beta \> \nabla_\alpha n_\beta \). This allows us to write:
\begin{equation} \label{A-HypersurfaceCovariantAccelerationA2}
\begin{aligned}
D_\mu a_\nu &=\gamma^\alpha_\mu \nabla_\alpha a_\nu+\varepsilon \> n_\nu\> \gamma^\alpha_\mu ( a^\beta \> \nabla_\alpha n_\beta)\\
&=\gamma^\alpha_\mu \nabla_\alpha a_\nu+\varepsilon \> n_\nu \> \delta^\alpha_\mu ( a^\beta \> \nabla_\alpha n_\beta)- \varepsilon^2 \> n_\nu\> n^\alpha \> n_\mu ( a^\beta \> \nabla_\alpha n_\beta)\\
&=\gamma^\alpha_\mu \nabla_\alpha a_\nu+\varepsilon \> n_\nu ( a^\beta \> \nabla_\mu n_\beta)- n_\nu\> n_\mu \>  a^\beta \> n^\alpha \>\nabla_\alpha n_\beta\\
&=\gamma^\alpha_\mu \nabla_\alpha a_\nu+\varepsilon \> n_\nu \> a^\alpha \> \nabla_\mu n_\alpha- n_\nu\> n_\mu \>  a^\alpha \> a_\alpha\\
\end{aligned}
\end{equation}

\noindent The three terms in the above result are the same as the last three terms in equation (\ref{A-LieDerivExtrinsicCurvB}). We may rewrite (\ref{A-LieDerivExtrinsicCurvB}) as:
\begin{equation} \label{A-LieDerivExtrinsicCurvC1}
\pounds _nK_{\mu \nu }=-R_{ \alpha \mu \beta \nu} \> n^{\alpha} n^\beta+K_{\mu \alpha}K_{\nu }{}^{\alpha }-\varepsilon \> a_{\mu }a_{\nu }+D_\mu a_\nu
\end{equation}

\noindent We may simplify this formula once more, using the expression \(a_\nu =-\varepsilon \> D_\nu(\text{ln}(\alpha))\), which is straightforward to derive. Recall that the lapse function is given by \(\alpha={|g^{\mu \nu} \> \mathfrak{n}_{\mu} \> \mathfrak{n}_{\nu}|^{-1/2}}\), where \(\mathfrak{n}_{\mu}=\nabla_\mu \phi\) is the gradient of the foliation function \(\phi\). We obtain:
\begin{equation} \label{A-HypersurfaceCovariantAccelerationB1}
\begin{aligned}
D_{\mu }a_{\nu }&=-\varepsilon \> D_{\mu }D_{\nu }\ln  \alpha=-\varepsilon \>  D_{\mu }\left({\alpha}^{-1} \> D_{\nu }\alpha\right)\\
&=-\varepsilon \> \left({\alpha}^{-1} \> D_{\mu }D_{\nu }\alpha-{\alpha}^{-2} \>D_{\mu }\alpha \> D_{\nu }\alpha\right)\\
&=-\varepsilon \> \left({\alpha}^{-1} \> D_{\mu }D_{\nu }\alpha-(D_{\mu }\ln  \alpha)( D_{\nu }\ln  \alpha)\right)\\
&=-\varepsilon \> {\alpha}^{-1} \> D_{\mu }D_{\nu }\alpha+\varepsilon \>  a_{\mu } \> a_{\nu }
\end{aligned}
\end{equation}

\noindent This may be rewritten as:
\begin{equation} \label{A-HypersurfaceCovariantAccelerationB2}
D_{\mu }a_{\nu }-\varepsilon \>  a_{\mu } \> a_{\nu }=-\varepsilon \> {\alpha}^{-1} \> D_{\mu }D_{\nu }\alpha
\end{equation}

\noindent Finally, we plug this back in to equation (\ref{A-LieDerivExtrinsicCurvC1}) to obtain our result:\index{Ricci Equation}
\begin{equation} \label{A-LieDerivExtrinsicCurvC2}
\fbox{$ \pounds _n K_{\mu \nu }=-R_{ \alpha \mu \beta \nu} \> n^{\alpha} n^\beta+K_{\mu \alpha}K_{\nu }{}^{\alpha }-\frac{\varepsilon}{\alpha} \> D_{\mu }D_{\nu }\alpha $}
\end{equation}

\noindent This formula is called the \textit{Ricci equation}. Note that the right hand side is tangent to the hypersurface; if we contract any index with the unit normal vector, the right hand side vanishes.\footnote{To see that \(R_{\alpha \mu \beta \nu} \>n^\alpha \> n^\beta\) is tangent to the hypersurface, not that the symmetries of the Riemann tensor are such that another contraction of the quantity \(R_{\alpha \mu \beta \nu} \>n^\alpha \> n^\beta\) with the unit normal vector would cause the resulting expression to vanish. One may therefore infer that the quantity \(R_{\alpha \mu \beta \nu} \>n^\alpha \> n^\beta\) is automatically tangent to the hypersurface.} 

\section{Deriving the variation of area formulas} \label{AppendixVariationAreas}


These formulas and portions of their derivation may be found in \cite{Frankel} and \cite{ChowHamilton,Nitsche1989,Dierkesetal2010}, but we choose to derive these formulas in a manner that is less formal than that found in the literature. In particular, we present a derivation of these formulas that is accessible--if still very complicated--to physicists.

\subsection{The first variation of area formula}
In this section, we derive the \textit{first variation of area formula} \cite{Frankel}, which is a formula describing the change in the ``area'' of a hypersurface under infinitesimal displacements. We begin by defining the volume form for the bulk manifold \(\mathcal{M}\):
\begin{equation} \label{A-VolumeForm} 
\Omega:=\frac{1}{N!}\epsilon_{i_1...i_N}  \>  dx^{i_1} \wedge . . . \wedge dx^{i_N}=\frac{1}{N!}\sqrt{|g|} \> \underline{\epsilon}_{i_1...i_N}  \>  dx^{i_1} \wedge . . . \wedge dx^{i_N}
\end{equation}

\noindent where \(\underline{\epsilon}_{i_1...i_N} \) is the Levi-Civita \textit{symbol} and \(\epsilon_{i_1...i_N}:=\sqrt{|g|} \> \underline{\epsilon}_{i_1...i_N}\) is the Levi-Civita \textit{pseudotensor}. It is not difficult to show that on a semi-Riemannian manifold, 
\begin{equation} \label{A-VolumeFormLieDerivative} 
\pounds_V \Omega= d i_V \Omega=\nabla_\mu V^\mu \, \Omega
\end{equation}

\noindent where \(V^\mu\) is a vector field, and the interior product \(i_u \omega\) of some \(p\)-form \(\omega\) and some vector field \(u\) is defined by the expression: 
\begin{equation} \label{A-InteriorProduct} 
\begin{aligned}
i_u \omega:=\frac{1}{(p-1)!} \> u^\mu \> \omega_{\mu \> \alpha_{1} ... \alpha_{p-1}} \> dx^{\alpha_{1}} \wedge ...\wedge dx^{\alpha_{p-1}}
\end{aligned}
\end{equation}

\noindent The formula for the Lie derivative (\ref{A-VolumeFormLieDerivative}) of the volume form \(\Omega\), allows us to write:
\begin{equation} \label{A-LieDerivativeVolumeForm} 
\pounds_n \Omega=\nabla_{\alpha} n^\alpha \> \Omega=K \> \Omega \> \> \> \> \Rightarrow \> \> \> \> K \sim \frac{1}{\delta V}\frac{d (\delta V)}{dS}
\end{equation} 

\noindent We may take this one step further, and demonstrate that the mean curvature also measures the fractional rate of change for the surface element \(d\Sigma\) of some hypersurface \(\Sigma\). The hypersurface volume element \(d\Sigma\) may be defined as:
\begin{equation} \label{A-ISurfaceElementC2} 
\begin{aligned}
d\Sigma &:=i_{n} \Omega=\frac{1}{(N-1)!}\sqrt{|g|} \> n^\mu \> \underline{\epsilon}_{\> \mu \> \alpha_1...\alpha_{N-1}}  \>  dx^{\alpha_1} \wedge. . . \wedge dx^{\alpha_{N-1}}\\
&= \frac{\varepsilon}{(N-1)!}\sqrt{|\gamma|} \> \underline{\epsilon}_{\>i_1...i_{N-1}}  \>  dy^{i_1} \wedge. . . \wedge dy^{i_{N-1}}
\end{aligned}
\end{equation}

\noindent Recall that \(y^i\) are the coordinates on the hypersurface \(\Sigma\), and \(\gamma:=\det(\gamma_{ij})\). From Cartan's formula \(\pounds_u \omega=d i_u \omega + i_u d \omega\), one may obtain the expression \(\pounds_u i_u \omega= i_u \pounds_u \omega\) for a p-form \(\omega\) and some vector field \(u\). Using this result, the Lie derivative of the hypersurface volume element is
\begin{equation} \label{A-LieDerivativeHypersurfaceVolumeForm} 
\begin{aligned}
&\pounds_n d\Sigma=\pounds_n (i_n \Omega)=i_n (\pounds_n \Omega)=i_n (\nabla_{\alpha} n^\alpha \> \Omega)= K \> i_n \Omega\\
& \Rightarrow \>\>\>\>\> \pounds_n d\Sigma =  K \> d\Sigma 
\end{aligned}
\end{equation}

\noindent This expression may also be obtained explicitly by applying the Lie derivative formula for tensors to the components of \(d\Sigma\); since \(\pounds_n n^\mu=0\), it is not surprising that \(\pounds_n \Omega\) and \(\pounds_n d\Sigma\) are both proportional to the mean curvature.

This result in equation (\ref{A-LieDerivativeHypersurfaceVolumeForm}) may be used to obtain a formula for the first variation of area, which is the change in the volume of a hypersurface under an infinitesimal displacement along the flow of some vector field \(v^\mu\). The change in the volume element under an infinitesimal displacement, which we write as \(\delta x^\mu = \delta \lambda \, v^\mu\) (where \(\lambda\) is a parameter along the integral curves of \(v^\mu\)), is:
\begin{equation} \label{A-DisplacementSurfaceElementA} 
\delta d\Sigma=\pounds_{\delta x} d\Sigma=\pounds_{\delta x} (i_n \Omega)
\end{equation}

\noindent We may decompose the displacement ``vector'' \(\delta x^\mu\) into a part normal to the hypersurface and a part tangent to the hypersurface:
\begin{equation} \label{A-DisplacementVectorDecomposition} 
\begin{aligned}
\delta x^\mu &=\delta a \> n^\mu + \delta b^\mu\\
\delta a &:=\varepsilon \, \delta x^\alpha n_\alpha=\varepsilon \, \delta \lambda \> (v^\alpha n_\alpha)|_{\textbf{Q}}\\
\delta b^\mu &:=\gamma^\mu_\alpha \> \delta x^\alpha=\delta \lambda \> \gamma^\mu_\alpha v^\alpha|_{\textbf{Q}}
\end{aligned}
\end{equation}

\noindent Since \(\delta a\) and \(\delta b^\mu\) are only defined on \(\textbf{Q}\), they are functions of points \(y\in \textbf{Q}\), so that derivatives of scalars formed from these quantities in the direction of the unit normal vector must vanish; for instance, \(n^\mu \, \nabla_\mu \delta a=0\). From the properties of the interior product, we note that \(i_{\delta a \> n+\delta b}=\delta a \> i_{n}+i_{\delta b}\), and that \(i_n \> i_n=0\). Using Cartan's formula, we rewrite equation (\ref{A-DisplacementSurfaceElementA}) as:
\begin{equation} \label{A-DisplacementSurfaceElementA2} 
\begin{aligned}
\delta d\Sigma&=d i_{\delta x} i_n \Omega+ i_{\delta x} d(i_n \Omega)\\
&=d (\delta a \> i_{n} \>  i_n \Omega+ i_{\delta b} \> i_n \Omega)+\delta a \> i_{n} \> d(i_n \Omega)+ i_{\delta b} \> d(i_n \Omega)\\
&=d i_{\delta b} (i_n \Omega)+ i_{\delta b} \> d(i_n \Omega)+\delta a \> i_{n} \> d(i_n \Omega)\\
&=\pounds_{\delta b}(i_n \Omega)+\delta a ( i_{n} \> d(i_n \Omega)+d i_n ( i_n \Omega ))\\
&=\pounds_{\delta b}(i_n \Omega)+\delta a (\pounds_{n}(i_n \Omega))\\
\end{aligned}
\end{equation}

\noindent where we have made use of \(i_n \> i_n=0\) in the third and fourth equality (we have added a zero in the latter). We may replace \(i_n \Omega\) with \(\Sigma\), and since the vector \(\delta b\) is tangent to the hypersurface, we may write it in the coordinate basis \(\partial/\partial y^i\) on the hypersurface, so that \(\pounds_{\delta b} d\Sigma=D_i \delta b^i d\Sigma\). The change in the surface element becomes:
\begin{equation} \label{A-DisplacementSurfaceElementA3} 
\begin{aligned}
\delta d\Sigma=\pounds_{\delta x} d\Sigma=(D_i \delta b^i +\delta a \> K )d\Sigma
\end{aligned}
\end{equation}

\noindent If we are given a hypersurface \(\Sigma_S\), then we may obtain the infinitesimal change of the ``area'' (by which we mean the \(N-1\) dimensional volume of the hypersurface \(\Sigma_S\)) by evaluating \(\delta a\), \(\delta b^i\), \(K\) and \(d\Sigma\) at the hypersurface, and integrating (\ref{A-DisplacementSurfaceElementA3}). If \(\textbf{Q} \subset \Sigma_S\) is a region of the hypersurface with boundary  \(\partial \textbf{Q}\), then the variation of area \(\delta A\) is given by:
\begin{equation} \label{A-VariationArea1} 
\delta A = \int_{\textbf{Q}} \delta d\Sigma= \int_{\textbf{Q}} (D_\alpha \delta b^\alpha +\delta a \> K )d\Sigma
\end{equation}

\noindent Using the divergence theorem, we obtain the \textit{first variation of area}\index{Variation of Area!First Variation of Area} formula \cite{Frankel}:
\begin{equation} \label{A-VariationArea2} 
\delta A = \int_{\textbf{Q}} \delta a \> K  \> d\Sigma+\int_{\partial \textbf{Q}} \delta b^i \> d\sigma_i
\end{equation}

\noindent where \(d\sigma_i\) is the directed surface element on \(\partial \textbf{Q}\), and \(\delta a\) and \(\delta b^i=(\partial y^i/\partial x^\mu) \> \delta b^\mu\) are defined in terms of the displacement \(\delta x^\mu\) according to equation (\ref{A-DisplacementVectorDecomposition}). If \(r^i\) is the unit normal vector to \(\partial \textbf{Q}\) (with norm \(\varepsilon_r=r^i r_i\)), \(z^A\) are the coordinates on \(\partial \textbf{Q}\), and \(\sigma_{AB}\) is the induced metric on \(\partial \textbf{Q}\), we may rewrite the first variation of area formula in a more explicit form:
\begin{equation} \label{A-VariationArea3} 
\delta A = \int_{\textbf{Q}} \delta x^\mu n_\mu \> K \> \sqrt{\text{det}|\gamma_{ij}|} d^{N-1} y+\int_{\partial \textbf{Q}} \> \delta x^\nu \> \gamma^\mu_\nu \left(\frac{\partial y^i}{\partial x^\mu}\right) \> r_i \> \varepsilon_r \> \sqrt{\text{det}|\sigma_{AB}|} d^{N-2} z
\end{equation}

\noindent Finally, we note that if \(\Sigma_S\) has no boundary and the integral is performed over the whole of \(\Sigma_S\), the boundary integral over \(\partial \Sigma_S\) vanishes.

\subsection{The second variation of area formula}
We now obtain a formula for the second order change in the volume of the hypersurface due to an infinitesimal displacement, which is called the \textit{second variation of area} formula \cite{ChowHamilton,Nitsche1989,Dierkesetal2010}. For the sake of generality, we will begin by considering two independent displacements of the hypersurface, \(\delta x^\mu\) and \(\delta \tilde{x}^\mu\); and compute \(\pounds_{\delta \tilde{x}^\mu} \pounds_{\delta x^\mu} d\Sigma\), with \(d\Sigma\) being the surface element of the hypersurface. From equation (\ref{A-DisplacementSurfaceElementA3}), we may write:
\begin{equation} \label{A-DisplacementSurfaceElementA3Copy} 
\begin{aligned}
\pounds_{\delta x} d\Sigma=(D_i \delta b^i +\delta a \> K )d\Sigma\\
\pounds_{\delta \tilde{x}} d\Sigma=(D_i \delta \tilde{b}^i +\delta \tilde{a} \> K )d\Sigma\\
\end{aligned}
\end{equation}

\noindent where we employ the decompositions \(\delta x^\mu =\delta a \> n^\mu + \delta b^\mu\) and  \(\delta \tilde{x}^\mu =\delta \tilde{a} \> n^\mu + \delta \tilde{b}^\mu\), with definitions as in equation (\ref{A-DisplacementVectorDecomposition}).
 Again, we stress that the quantities \(\delta {a}\), \(\delta \tilde{a}\), \(\delta b^\mu\) and \(\delta \tilde{b}^\mu\) are functions of \(y \in \textbf{Q}\) only. We note that for some scalar function \(\varphi\), the Leibniz rule yields \(\pounds_V (\varphi d\Sigma)=(\pounds_V \varphi) \> d\Sigma+\varphi \> \pounds_V d\Sigma\). If \((D_i \delta b^i +\delta a \> K )\) is a scalar function, we may use the Leibniz rule to write:
\begin{equation} \label{A-2ndLieDerivSurfaceElemA}
\begin{aligned}
\pounds_{\delta \tilde{x}^\mu} \pounds_{\delta x^\mu} d\Sigma & =\pounds_{\delta \tilde{x}^\mu}( (D_i \delta b^i +\delta a \> K )d\Sigma)\\
& =(\pounds_{\delta \tilde{x}^\mu} (D_i \delta b^i+\delta a \> K )d\Sigma)+ (D_i \delta b^i +\delta a \> K ) \pounds_{\delta \tilde{x}^\mu} \> d\Sigma\\
& =(\pounds_{\delta \tilde{x}^\mu} (D_i \delta b^i+\delta a \> K )+(D_i \delta b^i + \delta a \> K )(D_j \delta \tilde{b}^j +\delta \tilde{a} \> K) )d\Sigma\\
& =(\delta \tilde{x}^{\mu} \nabla_\mu (D_i \delta b^i+\delta a \> K )+ D_i \delta b^i \> D_j \delta \tilde{b}^j\\
& \>\>\>\>\>+\delta a \> D_j \delta \tilde{b}^j  \>  K + \delta \tilde{a} \> D_i \delta b^i \> K +\> \delta a \> \delta \tilde{a} \> K^2 )d\Sigma\\
\end{aligned}
\end{equation}

\noindent where for a scalar function \(\varphi\), \(\pounds_n \varphi=n^\mu \nabla_\mu \varphi\) in the last equality. Expanding further:
\begin{equation} \label{A-2ndLieDerivSurfaceElemB1}
\begin{aligned}
\pounds_{\delta \tilde{x}^\mu} \pounds_{\delta x^\mu} d\Sigma & =(\delta \tilde{a} \> n^\mu \nabla_\mu (D_i \delta b^i+\delta a \> K )+\delta \tilde{b}^j D_j (D_i \delta b^i+\delta a \> K )\\
& \>\>\>\>\> + D_i \delta b^i \> D_j \delta \tilde{b}^j + \delta a \> D_j \delta \tilde{b}^j  \>  K + \delta \tilde{a} \> D_i \delta b^i \> K +\> \delta a \> \delta \tilde{a} \> K^2 ) d\Sigma\\
& =(\delta \tilde{a} \> n^\mu \nabla_\mu (D_i \delta b^i)+\delta \tilde{a} \> n^\mu \nabla_\mu \delta a \> K +\delta a \> \delta \tilde{a} \> n^\mu \nabla_\mu K+\delta \tilde{b}^j D_j D_i \delta b^i \\
& \>\>\>\>\> +\delta \tilde{b}^j D_j \delta a \> K + \delta a \> \delta \tilde{b}^j D_j K + D_i \delta b^i \> D_j \delta \tilde{b}^j + \delta a \> D_j \delta \tilde{b}^j  \>  K \\
& \>\>\>\>\> + \delta \tilde{a} \> D_i \delta b^i \> K +\> \delta a \> \delta \tilde{a} \> K^2 ) d\Sigma\\
& =(\delta a \> \delta \tilde{a} \> n^\mu \nabla_\mu K+(\delta \tilde{b}^j D_j \delta a \> K + \delta a \> \delta \tilde{b}^j D_j K + \delta a \> D_j \delta \tilde{b}^j  \>  K) \\
& \>\>\>\>\>+\delta \tilde{b}^j D_j D_i \delta b^i  + D_i \delta b^i \> D_j \delta \tilde{b}^j  + \delta \tilde{a} \> D_i \delta b^i \> K +\> \delta a \> \delta \tilde{a} \> K^2 ) d\Sigma\\
\end{aligned}
\end{equation}

\noindent where we have eliminated two terms by noting that \(\delta a\) and \(\delta b^i\) are functions of \(y \in \textbf{Q}\) only; normal derivatives of quantities that are purely functions of \(y \in \textbf{Q}\) vanish. In particular, since \(\delta a\) and \(D_i \delta b^i\) are purely functions of \(y \in \textbf{Q}\), we have \(n^\mu \nabla_\mu \delta a=0\) and \(n^\mu \nabla_\mu (D_i \delta b^i)=0\). Three terms in the above expression may be combined into a divergence, so that:
\begin{equation} \label{A-2ndLieDerivSurfaceElemB2}
\begin{aligned}
\pounds_{\delta \tilde{x}^\mu} \pounds_{\delta x^\mu} d\Sigma & =(\delta a \> \delta \tilde{a} \> n^\mu \nabla_\mu K+\delta \tilde{b}^j D_j D_i \delta b^i  + D_i \delta b^i \> D_j \delta \tilde{b}^j + \delta \tilde{a} \> D_i \delta b^i \> K +\> \delta a \> \delta \tilde{a} \> K^2 \\
&\>\>\>\>\>+D_j (\delta a  \> \delta \tilde{b}^j  \>  K)) d\Sigma\\
\end{aligned}
\end{equation}

\noindent Next, we note that \(D_j (\delta \tilde{b}^j \> D_i\delta b^i)= D_j \delta \tilde{b}^j \> D_i \delta b^i + \delta \tilde{b}^j \> D_j D_i \delta b^i \), which allows us to combine another two terms into a divergence:
\begin{equation} \label{A-2ndLieDerivSurfaceElemC}
\begin{aligned}
\pounds_{\delta \tilde{x}^\mu} \pounds_{\delta x^\mu} d\Sigma & =(\delta a \> \delta \tilde{a} \left( n^\mu \nabla_\mu K +K^2 \right)+ \delta \tilde{a} \> D_i \delta b^i \> K  D_j (\delta \tilde{b}^j \> D_i\delta b^i)+  D_j (\delta a  \> \delta \tilde{b}^j  \>  K)) d\Sigma\\
& =(\delta a \> \delta \tilde{a} \left(\pounds_n K +K^2 \right)+ \delta \tilde{a} \> D_i \delta b^i \> K  +D_j (\delta \tilde{b}^j \> D_i\delta b^i + \delta a  \> \delta \tilde{b}^j  \>  K)) d\Sigma\\
\end{aligned}
\end{equation}

\noindent  where we have again made use of the fact that for a scalar \(\varphi\), \(\pounds_n \varphi=n^\mu\nabla_\mu \varphi\) in the last equality. We now evaluate \(\pounds_n K\). To do so, we make use of the expression \(\gamma^\mu_\alpha \gamma^\nu_\beta \> \pounds_n \gamma^{\alpha \beta}= -2 K^{\mu \nu}\) and also the Ricci equation (\ref{A-LieDerivExtrinsicCurvC2}):
\begin{equation} \label{A-LieDerivMeanCurvA}
\begin{aligned}
\pounds_n K &=K_{\mu \nu} \> \pounds_n \gamma^{\mu \nu} + \gamma^{\mu \nu} \> \pounds_n K_{\mu \nu}\\
&=K_{\mu \nu} \> \gamma^\mu_\alpha \gamma^\nu_\beta \> \pounds_n \gamma^{\alpha \beta} - \gamma^{\mu \nu} \> R_{ \alpha \mu \beta \nu} \> n^{\alpha} n^\beta+ \gamma^{\mu \nu} \> K_{\mu \alpha}K_{\nu }^{\alpha }-\varepsilon \> a^2+ \gamma^{\mu \nu} \, D_\mu a_\nu\\
&=-2 K_{\mu \nu} \> K^{\mu \nu}  - g^{\mu \nu} \> R_{ \alpha \mu \beta \nu} \> n^{\alpha} n^\beta + \varepsilon \> n^{\mu} \>  n^{\nu} \> R_{ \alpha \mu \beta \nu} \> n^{\alpha} n^\beta+  K_{\mu \alpha}K^{\mu \alpha} - \varepsilon \> a^2 + D_\mu a^\mu\\
&=- K_{\mu \nu} \> K^{\mu \nu}  -  R_{ \alpha \beta } \> n^{\alpha} n^\beta - \varepsilon \> a^2+ D_\mu a^\mu\\
\end{aligned}
\end{equation}

\noindent where \(a^\mu :=n^\mu \,\nabla_\nu n^\mu\) is the acceleration and \(a^2:=\gamma^{\mu \nu} \, a_\mu \, a_\nu\). In the second line, we made use of \(K_{\mu \nu} \> \gamma^\mu_\alpha \gamma^\nu_\beta=K_{\alpha \beta}\). Plugging the above result back into equation (\ref{A-2ndLieDerivSurfaceElemC}), we obtain:
\begin{equation} \label{A-2ndLieDerivSurfaceElemD}
\begin{aligned}
\pounds_{\delta \tilde{x}^\mu} \pounds_{\delta x^\mu} d\Sigma =\biggl(&\delta a \> \delta \tilde{a} \left( K^2- K_{\mu \nu} \> K^{\mu \nu}  -  R_{ \alpha \beta } \> n^{\alpha} n^\beta - \varepsilon \> a^2+ D_\mu a^\mu \right) \\
&+D_j (\delta \tilde{b}^j \> D_i\delta b^i + \delta a  \> \delta \tilde{b}^j  \>  K)  + \delta \tilde{a} \> D_i \delta b^i \> K \biggr)d\Sigma\\
\end{aligned}
\end{equation}

\noindent We integrate the above formula over a region of the hypersurface  \(\textbf{Q} \subset \Sigma_S\) with boundary \(\partial \textbf{Q}\) to obtain the \textit{generalized second variation of area} formula:
\begin{equation} \label{A-2ndVariationofAreaA1a}
\begin{aligned}
\delta_{\delta \tilde{x}^\mu} \delta_{\delta x^\mu} A &:=\int_{\textbf{Q}} \pounds_{\delta \tilde{x}^\mu} \pounds_{\delta x^\mu} d\Sigma\\
&=\int_{\textbf{Q}} \biggl(\delta a \> \delta \tilde{a} \left( K^2- K_{\mu \nu} \> K^{\mu \nu}  -  R_{ \alpha \beta } \> n^{\alpha} n^\beta - \varepsilon \> a^2+ D_\mu a^\mu \right) \\
&\>\>\>\>\>\>\>\>\>\>\>\>\>\>\>+D_j (\delta \tilde{b}^j \> D_i\delta b^i + \delta a  \> \delta \tilde{b}^j  \>  K)  + \delta \tilde{a} \> D_i \delta b^i \> K \biggr)d\Sigma\\
\end{aligned}
\end{equation}

\noindent From the contracted Gauss equation (\ref{2-GaussEquationRicciB1}), one may obtain:
\begin{equation} \label{A-GaussEquationRicciB2b}
R_{\mu \nu} \> n^\mu \> n^\nu =\frac{\varepsilon}{2} \left(R-{\bar{R}} - \varepsilon \> (K^{\mu \nu} \> K_{\mu \nu}-K^2) \right)
\end{equation}

\noindent where \({\bar{R}}\) is the Ricci scalar for the hypersurface \(\Sigma_S\). Using the above, one may obtain the general second variation of area formula:
\begin{equation} \label{A-2ndVariationofAreaA1b}
\begin{aligned}
\delta_{\delta \tilde{x}^\mu} \delta_{\delta x^\mu} A =\int_{\textbf{Q}} \biggl(&\delta a \> \delta \tilde{a} \> \varepsilon ((1/2)(\bar{R}+\varepsilon(K^2-K_{\mu \nu} \> K^{\mu \nu})-R) -  a^2+ \varepsilon\, D_\mu a^\mu) \\
& +D_j (\delta \tilde{b}^j \> D_i\delta b^i + \delta a  \> \delta \tilde{b}^j  \>  K)  + \delta \tilde{a} \> D_i \delta b^i \> K \biggr)d\Sigma\\
\end{aligned}
\end{equation}

\noindent This formula is foliation dependent due to the presence of the acceleration \(a_\mu\). In the immediate vicinity of \(\textbf{Q}\), we may construct Gaussian normal coordinates, in which the lapse function \(\alpha\) is set to unity, so that by virtue of (\ref{1-Acceleration}), the acceleration \(a_\nu =-\varepsilon \, D_\nu \, \alpha=0\). If our original foliation reduces to that of Gaussian normal coordinates at \(\textbf{Q}\), then we may set \(a_\nu=0\), so that:
\begin{equation} \label{A-2ndVariationofAreaA2}
\begin{aligned}
\delta_{\delta \tilde{x}^\mu} \delta_{\delta x^\mu} A =\int_{\textbf{Q}} \biggl(&\delta a \> \delta \tilde{a} \> \varepsilon (1/2)\left(\bar{R}+\varepsilon(K^2-K_{\mu \nu} \> K^{\mu \nu})-R\right) + D_j (\delta \tilde{b}^j \> D_i\delta b^i + \delta a  \> \delta \tilde{b}^j  \>  K)\\
& + \delta \tilde{a} \> D_i \delta b^i \> K \biggr)d\Sigma\\
\end{aligned}
\end{equation}

\noindent Finally, upon setting \(\delta \tilde{x}^\mu=\delta x^\mu\), we obtain the \textit{second variation of area} formula:
\begin{equation} \label{A-2ndVariationofAreaB}
\begin{aligned}
\delta^2 A =\int_{\textbf{Q}} \biggl(&(\delta a)^2 \> \varepsilon (1/2)\left(\bar{R}+\varepsilon(K^2-K_{\mu \nu} \> K^{\mu \nu})-R\right) + D_j (\delta {b}^j \> D_i\delta b^i + \delta a  \> \delta {b}^j  \>  K)  \\
&+ \delta  {a} \> D_i \delta b^i \> K \biggr)d\Sigma\\
\end{aligned}
\end{equation}

We conclude this appendix by briefly discussing an application of the second variation of area formula in cosmology. If we set \(\delta x^\mu=\delta^\mu_0 \Delta t\), the second variation of area formula may be interpreted as a measure of the acceleration or deceleration for the expansion of the universe. One may also use the integrand of (\ref{A-2ndVariationofAreaB}) as a local measure of whether the expansion of space is accelerating or decelerating. This may be particularly useful in characterizing the inflationary epoch, since the universe must go through a period of accelerating expansion followed by a period of decelerating expansion before the end of inflation. Since the universe is currently in a period of accelerating expansion, a period of rapid inflation requires that the volume of the universe must have at least three inflection points, which may be characterized by the points in time where the second variation of area vanishes.

\end{document}